\documentclass[12pt,english]{article}
\usepackage[OT1]{fontenc}
\usepackage[utf8]{inputenc}
\usepackage[english]{babel}
\usepackage{float}
\usepackage{amsmath}
\usepackage{amsthm}
\usepackage{amssymb}
\usepackage{graphicx}
\usepackage{setspace}
\usepackage{esint}
\usepackage[authoryear]{natbib}
\usepackage[unicode=true,pdfusetitle,
 bookmarks=true,bookmarksnumbered=false,bookmarksopen=false,
 breaklinks=false,pdfborder={0 0 1},colorlinks=false]
 {hyperref}
\hypersetup{
 urlcolor=blue}
\usepackage[a4paper, text={6.2in,9.2in}, centering]{geometry}
\usepackage{abstract}
\usepackage{array}
\usepackage{varioref}
\usepackage{caption}
\usepackage[table]{xcolor}
\usepackage{booktabs}
\usepackage{longtable} 
\usepackage{threeparttable}
\usepackage{rotfloat} 
\usepackage{graphicx}
\usepackage{placeins}
\usepackage{float}
\usepackage{subfigure}
\usepackage[toc,page]{appendix}
\RequirePackage{amsopn} \RequirePackage{amsfonts}
\RequirePackage{amsthm}
\renewcommand{\baselinestretch}{1.36}
\newtheorem{prop}{Proposition}

\newtheorem{cor}{Corollary}

\newtheorem{dfn}{Definition}
\theoremstyle{definition}

\newtheorem{ex}{Example}
\newtheorem{exa}{Example}
\renewcommand{\ex}{{\textbf{Example A\arabic{ex}}}.}
\newcommand{\field}[1]{\mathbb{#1}}
\newcommand{\R}{\field{R}}
\newcommand{\N}{\field{N}}
\newcommand{\E}{\field{E}}
\renewcommand{\Pr}{\field{P}}

\definecolor{light-gray}{gray}{0.80}
\definecolor{yyellow}{gray}{.50}
\definecolor{red-orange}{rgb}{1,0.1,0}
\definecolor{blue}{rgb}{0,0,1}
\definecolor{dark-blue}{rgb}{0,0, 0.55}
\definecolor{dark-red}{rgb}{0.55, 0, 0}

\begin{document}

\renewcommand{\baselinestretch}{1.1}
\title{{Opinion Dynamics via Search Engines}\\{\Large (and other Algorithmic Gatekeepers)\thanks{We thank Larbi Alaoui, Jose Apesteguia, Emilio Calvano, Stefano Colombo, 
Andrea Galeotti, Lisa George, Roberto Imbuzeiro Oliveira, Matthieu Manant, Andrea Mattozzi, Ignacio Monz\'{o}n, Antonio Nicol\`{o}, Nicola Persico, Vaiva Petrikaite, Christian Peukert, Alessandro Riboni, Emanuele Tarantino, Greg Taylor and seminar participants in Athens (EARIE 2018), Barcelona (UPF), Florence (EUI), Geneva (EEA 2016), Hamburg (Economics of Media Bias Workshop 2015), Naples (Media Economics Workshop 2014), Padova (UP), Palma de Mallorca (JEI 2016), Paris (ICT Conference 2017), Petralia (Applied Economics Workshop 2015), Rome (IMEBESS 2016) and Toulouse (Digital Economics Conference 2018) for useful comments and conversations. We are also indebted to Riccardo Boscolo for helpful discussions on the functioning of search engines. Germano acknowledges financial support from grant ECO2017-89240-P (AEI/FEDER, UE), from Fundaci\'{o}n BBVA (grant ``Innovaci\'{o}n e Informaci\'{o}n en la Econom\'{i}a Digital'') and also from the Spanish Ministry of Economy and Competitiveness, through the Severo Ochoa Programme for Centres of Excellence in R\&D (SEV-2015-0563).} }}
\author{Fabrizio Germano\thanks{Department of Economics and Business, Universitat
Pompeu Fabra, and Barcelona Graduate School of Economics,
\textit{fabrizio.germano@upf.edu}} \hspace{.025in} and Francesco
Sobbrio\vspace{15pt}\thanks{Department of Economics and Finance, LUISS  \textquotedblleft G.
Carli", and CESifo. \textit{fsobbrio@luiss.it}}}

\date{September, 2018\\\vspace{15pt}
}
\maketitle

\bigskip
\bigskip

\begin{abstract}
{\footnotesize  
Ranking algorithms are the information gatekeepers of the Internet era. We develop a stylized model to study the effects of ranking algorithms on opinion dynamics. We consider a search engine that uses an algorithm based on popularity and on personalization. We find that popularity-based rankings generate an \textit{advantage of the fewer} effect: fewer websites reporting a given signal attract relatively more traffic overall. This highlights a novel, ranking-driven channel that explains the diffusion of misinformation, as websites reporting incorrect information may attract an amplified amount of traffic precisely because they are few. Furthermore, when individuals provide sufficiently positive feedback to the ranking algorithm, popularity-based rankings tend to aggregate information while personalization acts in the opposite direction.
}

{\footnotesize \noindent\textsc{\textbf{Keywords}}:  Ranking
Algorithm, Information Aggregation, Asymptotic Learning, Popularity Ranking, Personalized Ranking, Misinformation, Fake News.
}
\end{abstract}

\thispagestyle{empty}
\begingroup
\setstretch{1.45}
\newpage

\section{Introduction}

Search engines are among the most important information gatekeepers of the Internet era. Google alone receives over 3.5 billion search queries per day, and, according to some estimates,  80\% of them are informational, dwarfing navigational and transactional searches (Jansen et al., 2008). Individuals increasingly use search engines to look for information on a vast array of topics such as science (Horrigan, 2006), birth control and abortion (Kearney and Levine, 2014), or the pros and cons of alternative electoral outcomes (e.g., the Brexit referendum in the UK; see {\em Google Trends}). Remarkably, what determines the ranking of any website to be displayed for any given search query are automated algorithms. Such algorithms are also used by social media---such as Facebook and Twitter---to rank their posts or tweets. They are so fundamental in establishing what is relevant information or what are relevant information sources, that they necessarily convert search engines and social media platforms into {\em de facto} \textquotedblleft algorithmic gatekeepers\textquotedblright\  (Introna and Nissenbaum, 2000; Rieder, 2005; Granka, 2010; Napoli, 2015; Tufekci, 2015). Despite the importance of such ranking algorithms for a wide variety of online platforms, opinion dynamics via algorithmic rankings is largely understudied. 

This paper aims to fill this gap by  developing a dynamic framework that studies the interaction between individual searches and a stylized ranking algorithm. As actual algorithms used by platforms such as {\em Google} or {\em Facebook} are highly complex (Dean, 2013; MOZ, 2013; Vaughn, 2014), we do not aim to pin down the exact pattern of website traffic or opinion dynamic generated by any specific algorithm, nor do we pursue a mechanism design approach. Rather, our aim is to isolate few essential components of ranking algorithms and to study their interplay with individual clicking behavior. Specifically, we focus on two key aspects of ranking algorithms: $(a)$ rankings may be based on the  \textit{popularity} of the different websites, $(b)$ rankings may be \textit{personalized} and may depend on individuals' characteristics. These two aspects drive the ranking of the available websites provided to individuals by a search engine. 

Individuals use the search engine to look for information on the state of the world (e.g., whether or not to vaccinate a child). Their choices over websites are modeled by means of a random utility model (Luce, 1959; Block and Marschak, 1960; G\"{u}l et al., 2014). This yields probabilities of reading the ranked websites that depend on both website content and ranking. As a result, individuals tend to choose websites that are: $(i)$ higher-ranked (De Cornière and Taylor, 2014; Taylor, 2013; Hagiu and Jullien, 2014;
Burguet et al., 2015), $(ii)$ \textit{ex-ante} more informative, $(iii)$ confirm their prior information (Mullainathan and Shleifer, 2005; Gentzkow and Shapiro, 2010), and $(iv)$ they may trade off content and ranking as a result of their stochastic choices. Website rankings are endogenous meaning that individuals' choices feed back into the search engine's rankings, thereby affecting future searches (Demange, 2014a).  It is important to point out that individuals are na\"{i}ve with respect to the search engine's algorithm in the sense that they do not make any inference from the websites' rankings {\em per se}. This last assumption---apart from making the model tractable---reflects  informational and behavioral limitations of individuals in understanding the working of ranking algorithms (Granka, 2010; Eslami et al., 2016).\footnote{While popularity and personalization are well-established components of ranking algorithms, exact details of the algorithms are typically kept secret (Dean, 2013; MOZ, 2013; Vaughn, 2014; Kulshrestha et al., 2018). As Eslami et al., (2016), p.~1, point out, the ``operation of these algorithms is typically opaque to users.''} 

The model provides three main insights deriving from the interaction of the ranking algorithm with sequential individual searches.  First, we uncover a fundamental property induced by popularity-based rankings, which we call the \textit{advantage of the fewer} ($AOF$).  It says that, all else equal, fewer websites carrying a given signal may attract more traffic overall, than if there were more of them. Popularity-based rankings amplify the static effect of simply concentrating audiences on fewer outlets, since they induce relatively higher rankings for the fewer outlets, which makes them more attractive for subsequent individuals, further raising their ranking and so on, generating a {\em ``few get richer''} dynamic with a potentially sizeable effect in the limit.
To the best of our knowledge, this property of popularity-based rankings has not yet been pointed out in the literature. 
We show that $AOF$ holds well beyond the assumptions of our basic model.   Importantly, $AOF$ highlights 
a further, ranking-driven, channel for why ``alternative-facts'' websites may thrive and gain in authority in the current information environment, dominated by algorithmic gatekeepers,  such as search engines and social media (Allcott and Gentzkow,
2017; Allcott et al., 2018).\footnote{Allcott and Gentzkow (2017) provide an economic model of fake-news and also document that, in the run-up to the 2016 US presidential election, more than 60\% of traffic of fake news websites in the US came from referrals by algorithmic gatekeepers (i.e., search engines and social media). See also Azzimonti and Fernandes (2018) for a model on the diffusion of fake-news in social networks.} 

Second, we provide conditions under which popularity-based rankings can effectively aggregate information. We compare asymptotic learning under popularity-based rankings and under fully random rankings. Not surprisingly, we find that popularity-based rankings do better as long as individuals generate sufficiently positive feedback through their searches (e.g., their private signals are not too noisy, or their preference for confirmatory news is not too strong). 

Third, we study whether personalized rankings can contribute to information aggregation. We compare asymptotic learning under both personalized  and non-personalized rankings. We find that there is a close relationship between the conditions under which non-personalized rankings outperform personalized rankings with the ones under which popularity-based rankings outperform random rankings.  For the common-value searches of our model, personalization limits the feedback among individuals in the opinion dynamic, making it better {\em not} to personalize the ranking when individuals' choices over websites generate a positive feedback. As a consequence, personalized rankings are often dominated in terms of asymptotic learning either by non-personalized rankings or by random rankings. Finally, we also show that personalized  rankings can induce relatively similar individuals to read different websites, which can lead to \textit{belief polarization}.

We conclude with two caveats. 
The results of our model provide some first insights on opinion dynamics via algorithmic rankings with na\"{i}ve 
individuals. A similar model with more sophisticated individuals who can observe the evolution of the ranking sufficiently accurately may well predict that such individuals are likely to always learn the true state of the world in the limit (see Section 6.3). 
As emphasized above, the na\"{i}vit\'{e} in our model reflects informational and behavioral limitations of individuals in assessing the working of the ranking algorithm and provides a natural benchmark for understanding online misinformation. 
 Second, while our results suggest that  personalization may often be sub-optimal in the context of  individuals looking for information on common value issues (e.g., whether or not to vaccinate a child), when individuals care differently about the objects of their searches (e.g., where to have dinner) an appropriately personalized search algorithm might clearly outperform a non-personalized one. 
\color{black}

\bigskip

\noindent
\textbf{Related Literature.} 

To our knowledge, ours is the first paper in economics to analyze opinion dynamics via endogenous algorithmic rankings and the informational gate-keeping role of search engines.\footnote{The existing economics literature on search engines has focused on the important case where search engines have an incentive to distort sponsored and organic search results in order to gain extra profits from advertising and product markets 
(Taylor, 2013; De Cornière and Taylor, 2014; Hagiu and Jullien, 2014; Burguet et al., 2015). See also Grimmelmann (2009) and Hazan (2013) for a legal perspective on the issue.}

The paper is broadly related to the economics literature on the aggregation of information dispersed across various agents (Bikhchandani et al., 1998; Piketty, 1999; Acemoglu and Ozdaglar, 2011; provide surveys on information aggregation, respectively, through observation of behavior of others, through voting, and through learning in social networks). We share with this literature the focus on understanding the conditions under which information that is dispersed among multiple agents might be efficiently aggregated. Our framework differs from this literature in a simple and, yet, crucial aspect: we are interested in investigating the role played by a specific (yet extensively used) \textquotedblleft tool\textquotedblright\ of information diffusion/aggregation, namely, the ranking algorithm, which represents the backbone of many online platforms. Another feature we have in common with a subset of this literature is that we take a non-Bayesian approach and consider individuals who are na\"{i}ve with respect to some key aspects of their choice situation (i.e., the ranking algorithm). In this sense, our model is closer to the papers on non-Bayesian belief formation (DeGroot, 1974; DeMarzo et al., 2003; Acemoglu et al., 2010; Golub and
Jackson, 2010). Moreover, although our individuals perform only one search, ``learning'' occurs through the ranking algorithm that aggregates the information reflected in the clicking behavior and passes it on to subsequent individuals.

Our focus on search engines as information gatekeepers is close in spirit to the economic literature on news media (see DellaVigna and Gentzkow, 2010; Prat and Str\"{o}mberg, 2013, for surveys). At the same time, the presence of an
automated ranking algorithm makes search engines---and other algorithmic gatekeepers---fundamentally different from
news media, where the choice of what information to
gather and disclose is made on a discretionary, case-by-case basis. In the
case of search engines the gate-keeping is unavoidably the result of
automated algorithms (Granka, 2010; Tufekci, 2015).\footnote{Put
differently, \textquotedblleft While humans are certainly responsible for
editorial decisions, these [search engine] decisions are mainly expressed in
the form of software which thoroughly transforms the ways in which procedures
are imagined, discussed, implemented and managed. In a sense, we are closer to
\textit{statistics} than to \textit{journalism} when it comes to bias in Web
search\textquotedblright; Rieder and Sire (2013), p. 2.} Therefore, whatever
\textit{bias} might originate from search engines, its nature is intrinsically
different from one arising in, say, traditional news media. 
As a result, studying the effects of search
engines on the accuracy of individuals' beliefs, requires a different approach from 
the ones used so far in theoretical models of media bias
(e.g., Str\"{o}mberg, 2004; Mullainathan and Shleifer, 2005; Gentzkow and Shapiro, 2006).

Finally, the paper is also
related to the literature outside economics discussing the possible
implications of the search engines' architecture
(or, more generally, of algorithmic gatekeeping) on
democratic outcomes. This literature encompasses communication scholars (Hargittai, 2004; Granka, 2010), legal scholars (Goldman, 2006; Grimmelmann, 2009; Sunstein, 2009), media activists (Pariser, 2011), psychologists (Epstein and Robertson, 2015), political scientists (Putnam, 2001; Hindman, 2009; Lazer, 2015), sociologists (Tufekci, 2015) and, last but not least, computer scientists (Cho et al., 2005; Menczer et al.,
2006; Pan et al., 2007; Glick et al., 2011; Flaxman et al., 2016; Bakshy et al., 2015).

\bigskip
The paper is structured as follows. 
Section~\ref{section: model} describes the framework.
Section~\ref{section:results} presents our central result concerning the effects of the search engine's ranking algorithm on website traffic: the advantage of the fewer. Section~\ref{section:welfare} presents the implications of the model in terms of asymptotic learning, also providing a comparison with a fully randomized ranking. Section~\ref{sect:pers_rank} discusses the effects of personalization of search results on belief polarization and asymptotic learning.
Section~\ref{section: Extensions} discusses some extensions which assess the robustness of the results of the benchmark model and provide further insights. Section~\ref{section: conclusions} concludes. 
Finally, all the proofs and some formal definitions are relegated to the Appendix.



\section{\label{section: model} The Model}

We present a stylized model of a search environment where individuals use a search engine to look for information on a fixed issue (e.g., whether or not to vaccinate a child) that is dispersed across websites.
At the center of the model is a search engine characterized by its ranking algorithm, which ranks and directs individuals to the different websites, using, among other things, the popularity of individuals' choices. To simplify the analysis, we assume that individuals are na\"{i}ve and perform exactly one search, one after the other, without knowing who searched before them and without updating prior beliefs after observing the ranking. We also assume websites simply report their own private signal, assumed to be constant throughout.
We describe the formal environment.


\subsection{Information Structure} \label{sect:infostrct}

There is a binary state of the world $\omega$, which is a $\bigl(\tfrac{1}{2},\tfrac{1}{2}\bigr)$ 
Bernoulli random variable which takes one of two values from the set $\{0,1\}$. 
There are $M$ information sources (websites) and $N$ individuals, 
where, by slight abuse of notation, we let $M = \{ 1, \ldots, M \}$, $N = \{ 1, \ldots, N \}$ 
also denote the set of websites and individuals, respectively. For convenience, we assume $M$ is an odd number (unless otherwise noted).
Each website $m\in M$ receives a private random signal, correlated with the true state $\omega$,
\[
y_{m}\in\{0,1\} \mbox{ with } \Pr(y_{m}=\omega\mid\omega= \xi) = q\in\left(
\tfrac{1}{2},1\right)  , \mbox{  for any } \xi \in\{0,1\} .
\]
This determines the {\em website majority signal}, denoted $y_K \in \{ 0, 1 \}$, which is the signal that is carried by a majority of websites;
let $K=\{ m \in M \, | \, y_m = y_K \}$ denote the set (and number) of websites carrying the signal $y_K$.
Similarly, each individual $n\in N$ receives two private random signals: 
\[
x_{n}\in\{0,1\} \mbox{ with } \Pr(x_{n}=\omega\mid\omega= \xi ) = p\in\left(
\tfrac{1}{2},1\right)  , \mbox{ for any } \xi  \in\{0,1\} ,
\]
which reflects the individual's prior on the true state of the world, and, 
\[
z_{n}\in\{0,1\} \mbox{ with } \Pr(z_{n}= y_K \mid y_K = \zeta ) = \mu \in\left(
\tfrac{1}{2},1\right]  , \mbox{  for any } \zeta \in\{0,1\} ,
\]
which is independent of $x_n$, when conditioned on ($\omega, y_K$), 
and reflects the individual's prior about what the majority of websites (e.g., mainstream or ``authoritative'' websites) are reporting. 

We assume all signals to be conditionally independent across individuals and websites,
when conditioned on ($\omega, y_K$), 
and that $\mu q > p$.\footnote{This condition implies:
\[ \mu \sum_{K > \frac{M}{2}} \binom{M}{K} q^K (1-q)^{M-K}  + (1- \mu ) \sum_{K> \frac{M}{2}} \binom{M}{K} q^{M-K} (1-q)^{K} > p 
\hspace{.2in} \mbox{ and } \hspace{.2in} q > p , \] where the first inequality implies that an individual's signal $z_n$ is ex ante more informative about $\omega$ than the same individual's signal $x_n$; and the second inequality implies that a website's signal $y_m$ is ex-ante more informative about $\omega$ than an individual's signal $x_n$.} 
This very last condition implies not only that, from an ex-ante perspective, the website signals ($y_m$) and hence also the website majority signal ($y_K$) are more informative than the individual private signals on the state of the world ($x_n$), but also that, for a given individual, his private \textit{perceived} website majority signal ($z_{n}$) is also more informative than his other signal ($x_n$).   
Among other things, the signals $z_n$ allow us to introduce a ``rational'' side to individuals' choices as we will discuss further below. 
From the onset, we emphasize that the case $\mu \le 1$ is meant to capture the intrinsic noise faced by individuals in identifying the ex-ante more informative website majority signal ($y_K$).  Accordingly,  $(1-\mu)$ represents the probability that an individual incorrectly identifies the minority signal as the majority one  (e.g., looking at a website reporting ``alternative facts'' believing that it is reporting mainstream or ``authoritative'' news).
All the main conclusions remain unchanged if we assume $\mu=1$, meaning that individuals can perfectly identify the majority signal (e.g., individuals know which are the mainstream websites appearing in the search result pages).


\subsection{Information sources}

Each of the $M$ websites represents an information source. A website is characterized by a signal $y_m \in \{0,1\}$ as described above, which is posted and held constant throughout.  Websites can be seen as articles or documents posted on the web that contains pertinent information to a given search query. 


\subsection{Individuals}

\label{sect: individuals}

Individuals in $N$ enter in a random order, sequentially, such that, at any point in time $t$, there is a unique individual $t \equiv n \in N$, 
who receives random (i.i.d.) signals $(x_n, z_n) \in \{ 0 , 1 \}^2$ as specified above, who performs exactly one search, faces the ranking 
and chooses a website to read.\footnote{For reasons of tractability, we assume that all individuals perform the same search query exactly once.} 
When choosing which website to read, individuals make a stochastic choice derived from a random utility model (Luce, 1959; Block and Marschak, 1960; G\''{u}l et al., 2014; Agranov and Ortoleva, 2017, provide empirical evidence),
\begin{equation} \label{RUM}
\rho_{n,m} = \frac{v_{n,m}}{\sum_{m' \in M} v_{n,m'}}, 
\end{equation}
where $\rho_{n,m}$ is the probability individual $n$ clicks on website $m$, and where $v_{n,m}$ is the value derived from clicking on website $m$. The values $v_{n,m}$ are to be interpreted as measuring desirability in the sense of a stochastic preference, which, as we will see, reflect a ranking aspect and a content aspect of website $m$. We now define the $v_{n,m}$'s in two steps.

\vspace{.1in}
\noindent
{\bf Website content.} Suppose individuals see all websites ranking-free, that is,  each one with equal probability (fixing content). 
As mentioned, each individual receives the signals $(x_n, z_n) \in \{ 0 , 1 \}^2$. 
We use these signals to define {\em attributes} (see G\''{u}l et al., 2014) that yield value to the individual in the following sense. 
Reading a website $m$, with $y_m=x_n$, yields value $\gamma \in [0,1]$, where the parameter $\gamma$ calibrates the preference for reading like-minded  news (Mullainathan and Shleifer, 2005; Gentzkow and
Shapiro, 2010).\footnote{See Yom-Tov et al. (2013); Flaxman et al. (2016); White and Horvitz (2015) for evidence of individual preference for like-minded news in the context of search engines.}
At the same time, reading a website $m$, with $y_m=z_n$, yields value $1-\gamma$, which reflects the desirability associated with reading the ex-ante most informative signal.\footnote{As mentioned above in Section~\ref{sect:infostrct}, given the assumptions on $p, q$ and $\mu$, a rational agent who derives utility from reading a website reporting a correct signal ($y_m=\omega$) but who cannot use the ranking $r_n$ to update her information about the true state of the world, prefers to ignore her signal $x_n$ and choose a website $m$ with $y_m=z_n$.  
In Section~\ref{sect:soph}, we briefly discuss some implications of more sophisticated learning in our model.} 
Using $x_n$ and $z_n$ to define attributes (with a normalization such that the value of having both attributes is 1), we obtain the {\em ranking-free values} from selecting website $m$,
\begin{equation}\label{eq:rank_free}
v_{n,m}^* =
\left\{
\begin{array}
[c]{cl}
\frac{1}{[m]} & \text{ if }x_n=y_m=z_n \\ \vspace{.02in}
\frac{\gamma}{[m]} & \text{ if }x_n=y_m \ne z_n \\ \vspace{.02in}
\frac{1 - \gamma}{[m]} & \text{ if }x_n \ne y_m=z_n \\ \vspace{.02in}
0 & \text{ if }x_n \ne y_m \ne z_n 
\end{array}
\right. , 
\end{equation}
where $[m]=\# \{ m' \in M | y_{m'} = y_{m} \}$ is the number of all websites with the same signal as $m$. The division by $[m]$ avoids the usual Luce effect when dealing with duplicate alternatives. It also allows us to interpret the $v_{m,n}^*$'s both as values as well as the individual stochastic \textit{ranking-free website choices}. In particular, $v_{n}^* : \{ 0, 1 \}^2 \rightarrow \Delta(M)$ depends only on the private signals $(x_n, z_n) \in \{ 0, 1 \}^2$.

\vspace{.1in}
\noindent
{\bf Website rankings.} Let $r_{n} \in\Delta(M)$ be the \emph{ranking} of the $M$ websites at time $t \equiv n$;
the ranking is provided by the search engine (as discussed in the next section). 
Individuals ignore who searched before them and do not update their beliefs about the true state of the world $\omega$ after observing the ranking. 
Moreover, we assume they can process the ranked list, subject to limitations of attention, meaning they implicitly favor higher ranked websites due to attention bias. 
Indeed, as shown by Pan et al. (2007);
Glick et al. (2011); Yom-Tov et al. (2013); Epstein and Robertson (2015), keeping all other things equal (e.g., the fit of a
given website with respect to the individual's preferences), highest ranked websites tend to receive significantly more \textquotedblleft
attention\textquotedblright\ by users than lower ranked ones.\footnote{More generally, when faced with ordinal lists, individuals often show a disproportionate tendency to select options that are placed at the top (Novarese and Wilson, 2013).} Since we model the ranking ($r_n$) as a probability distribution over the outlets that is multiplicatively separable from the ranking-free values ($v_{n}^*$), we can write final values as ranking-weighted values:
\begin{equation}\label{eq:rank_weight}
 v_{n,m} = r_{n,m} \cdot v_{n,m}^* .
\end{equation}
Websites with a higher ranking receive a higher weight. In Section~\ref{sect:attbias}, we introduce a parameter ($\alpha$) that further calibrates the degree of attention bias.

\vspace{.1in}
\noindent
{\bf Individual stochastic choice.} We can think of the values $v_{n,m}$ defined in (\ref{eq:rank_weight}) as representing desirability levels for the different ranked alternatives, which yields the stochastic choices defined in (\ref{RUM}), as in a standard random utility model, and thus give us the  \emph{website choice function} $\rho_{n} : \{ 0, 1 \}^2 \times \Delta(M) \rightarrow \Delta(M)$, defined by:\footnote{Fortunato et al. (2006) and Demange (2014a) use related models of individuals' choices over ranked items. 
}
\begin{equation} \label{eq:rho_idbias}
\rho_{n,m}
=  \frac{r_{n,m}  \cdot v_{n,m}^*}{\sum_{m' \in M} r_{n,m'}  \cdot v_{n,m'}^*}  . 
\end{equation}
The website choices $\rho_{n,m}$ reflect, through the ranking-free values ($v_{n,m}^*$), two \textit{attributes} of website $m$ from individual $n$'s perspective, namely, $(i)$ whether website $m$ reports a signal corresponding to the individual's prior ($y_m=x_n$), which gives desirability $\gamma$, and $(ii)$ whether website $m$ is identified as carrying a website majority signal ($y_m=z_n$), which gives desirability $1-\gamma$. At the same time, website $m$'s rank ($r_{n,m}$) re-scales its final desirability in a multiplicatively separable way, and reflects individuals' tendency to devote more attention to higher ranked websites.
Importantly, the multiplicative form leads individuals to trade off attributes and hence content, when making non-degenerate stochastic choices ($0 < \gamma < 1$). When $\gamma=0$ or $\gamma=1$ only one of the attributes matters.

Overall, $\rho_{n,m}$ summarizes some intuitive properties, namely,  individuals tend to choose websites that are $(i)$  higher ranked, $(ii)$ ex-ante more informative, $(iii)$ reflect their own prior information; and $(iv)$ they may also trade off content and ranking. Such a stochastic choice specification is consistent with agents maximizing the probability of reading a website reporting a correct signal (thereby following their signal $z_n$), while, at the same time, holding a preference for reading like-minded news (thereby following their signal $x_n$), and also being subject to attention bias (thereby implicitly weighting the alternatives by the ranking $r_n$).

\subsection{Search Engine and Ranking Algorithm}

\label{sect:ranking}

The \emph{ranking algorithm} used by the search engine is at the
center of our model. At any point in time, $t =1, 2, \ldots, N$, it gives a
\emph{ranking} $r_{t} \in\Delta(M)$ of the $M$ websites, where an element
$r_{t,m}$ is the probability that the individual searching at time $t$ is
directed to website $m$ in the absence of other factors.\footnote{For the sake of tractability, the ranking is considered as a cardinal score attached by the search engine to each given website. As pointed out by Demange (2014b, p.~918) a ranking is meant to measure the ``relative strength of $n$ items, meaning that the values taken by the scores matter up to a multiplicative constant''.}

Concretely, we assume that, given an initial ranking $r_{1}
\in\Delta(M)$, the ranking $r_{t}$ at subsequent periods, $t=2, 3, \ldots, N$,
is defined by,
\begin{equation}
\label{eq:rt}
r_{t,m} = \nu r_{t-1,m} + (1-\nu) \rho_{t-1,m},
\end{equation}
where $\nu\in(0,1)$. This says that the ranking at time $t$, $r_{t}$, is
determined by the ranking of the previous period $r_{t-1}$ and partly also by
the individual's choice over websites in the previous period $\rho_{t-1,m}$ as
defined in (\ref{eq:rho_idbias}).\footnote{See Demange (2012) for a similar specification of the popularity-ranking algorithm.}
 Such ranking dynamic reflects the algorithm used by search engines to update their rankings according to how \textquotedblleft popular\textquotedblright\ a webpage is (Dean, 2013; MOZ, 2013; Vaughn, 2014).\footnote{This
updating algorithm via the \textquotedblleft popularity\textquotedblright\ of
a website may be interpreted both in a strict sense (e.g., direct effect of the
actual clicks on the website in the search result page) and in a broad sense 
(e.g., a website that receives more clicks is also more likely to be more popular in other online platforms and vice versa). The relevance of the popularity component of algorithmic rankings is further confirmed by the empirical evidence provided by Kulshrestha et al. (2018).}
Accordingly, we will sometimes refer to the ranking as {\em popularity-based} or simply {\em popularity ranking}. The weight that is put on each term depends
on the parameter $\nu$, where for convenience we write,
\begin{equation}
(\nu, 1-\nu) = \left(  \frac{\kappa}{\kappa+1}, \frac{1}{\kappa+1} \right)  ,
\end{equation}
and where in turn $\kappa\in\mathbb{N}$ is a \emph{persistence} parameter of the
ranking algorithm. The larger $\kappa$, the more persistent the search engine's ranking is. In Section~\ref{section:welfare}, when studying asymptotic behavior, we will let $\kappa \rightarrow \infty$ as $N \rightarrow\infty$. 
This ensures that the effect of any given individual's search on the ranking becomes vanishingly small and allows the dynamic process to converge to its unique limit.


\subsection{Search Environments}
\label{sect:search_env}
A search environment is an \emph{ex ante} notion that fixes the ranking algorithm, information structure and
characteristics of individuals and websites, before they receive their signals
and before they perform their search. 
More formally, we define a \emph{search environment} $\mathcal{E}$ as a list of variables,
\begin{equation}
\mathcal{E}=\left\langle (p,q,\mu);(N,\gamma);M;\kappa\right\rangle ,
\end{equation}
where $(p,q,\mu)$ describes the information structure, $(N,\gamma)$
describes the individuals, $M$ describes the websites, and $\kappa$
describes the ranking algorithm.
Given a search environment $\mathcal{E}$, 
we refer to an \emph{(interim) realization} of $\mathcal{E}$ as to the tuple 
$\left\langle \omega;\left(L,(y_{m})_{m \in M} \right); \left((x_n)_{n \in N};(z_n)_{n \in N} \right) \right\rangle $, 
where the true state of the world and the
signals of the websites are fixed; $L$ denotes the set and the number of websites with the
correct signal $y_{m}=\omega$; individuals with signals ($x_{n};z_{n}$) enter sequentially, 
one at a time, in a random order.

We let $r_{1}$ denote the \emph{initial ranking}.
Unless otherwise specified, we assume that $r_1$ is interior, that is, 
$r_{1,m}>0$ for all $m \in M$. 
For $J \subset M$, let $r_{t,J}=\sum_{m \in J} r_{t,m}$ and $\rho_{t,J}=\sum_{m \in J} \rho_{t,m}$ denote respectively total ranking and total clicking probability at time $t$ on all websites in $J$; we will be particularly interested in the case where $J=L$.
We also often talk about the {\em expected} probability of individual $n$ accessing website $m$, 
$\widehat{\rho}_{t,m} = \E [\rho_{t,m} ]$, where the expectation is taken over the private signals of agent $n$ 
that enters to perform a search at time $t \equiv n$. 
(These expected probabilities are discussed in more detail in Appendix~\ref{Appendix-MeanDynamics}.)


\section{Popularity Ranking and the Advantage of the Fewer ($AOF$)}
\label{section:results}

The following proposition states our first main result. It illustrates a rather general phenomenon induced by popularity ranking, which we refer to as the {\em advantage of the fewer} ($AOF$), whereby a  set of websites with the same signal can get a {\em greater} total clicking probability by individuals sufficiently far up in the sequence, if the set contains fewer websites than if it contains more of them (as long as it does not switch from being a set of majority to a set of minority websites).\footnote{Recall that, unless otherwise stated, our search environments always have popularity-based rankings (see Section~\ref{sect:ranking}). Note also that, throughout the paper, we use the term decreasing (and increasing) in the weak sense, that is, we say a function $f$ is decreasing (increasing) if $x \ge y$ implies $f(x) \le f(y)$ ($f(x) \ge f(y)$). When $x \in \N$, we say $f$ is decreasing (increasing) at $x$ if $f(x+1) \le f(x)$ ($f(x+1) \ge f(x+1)$.}

\begin{prop}
\label{prop:$AOF$}
Fix a search environment ${\cal E}$ with uniform initial ranking $r_{1}$, and $N$, $\kappa$ large, and consider interim realizations of ${\cal E}$ that vary in the number of outlets with correct signal. 
Then, the clicking probability ($\rho_{N,J}$) by individual $N$ on all outlets with a fixed signal, say $J$, (i.e., where $m, m' \in J \subset M$ implies $y_{m}=y_{m'}$ and similarly for $m'', m''' \in M \backslash J$) is decreasing in the number of those outlets ($\# J$), for ``interior values'' of the parameters (i.e., when $0< \gamma < 1$ and $\#J \ne 0, \frac{M-1}{2}, M-1$).
\end{prop}

This suggests that having {\em fewer} websites reporting a given signal {\em enhances} their overall traffic.  The intuition behind the $AOF$ effect is straightforward. When there are fewer websites with a given signal, the flow of individuals interested in reading about that signal are concentrated on fewer outlets, thus leading to relatively more clicks per outlet. This (trivial) static effect transforms into a dynamic, amplified one through the interaction of the popularity ranking with individuals' stochastic choices: popularity ranking induces relatively higher rankings for those fewer websites, which makes them more attractive for subsequent individuals, inducing more trade-offs in favor of those fewer higher ranked websites,  leading to even more clicks, and so on. The process, which can be seen as embodying a {\em ``few get richer''} dynamic, gets repeated until it stabilizes in the limit, generating a potentially sizable amplification of total traffic on all websites with the given signal.\footnote{Figure \ref{fig:$AOF$} in the Online Appendix, provides a graphical illustration of the \textit{amplification effect}.}

The $AOF$ implies that having a small majority (or a small minority) of websites rather than a large majority (or minority) of websites reporting a given signal actually increases their overall traffic. At the same time, the amount of traffic that such websites can attract is limited by the fact that
majority (minority) websites attract all traffic from individuals with signals $x_n=z_n=y_K$ ($x_n=z_n=y_{M \backslash K}$).
Accordingly, the results do not imply that all traffic is directed toward a single website or group of websites, but rather that there is an \textit{amplification effect} for the traffic going to both majority or minority websites triggered by a lower number of corresponding websites. The \textit{amplification effect} generated by the $AOF$ hinges upon two main components: the popularity ranking and a non-degenerate stochastic choice of the individuals, who trade off content depending on the ranking when $\gamma$ is interior ($0<\gamma<1$). Instead, when $\gamma=0$ or $\gamma=1$ such an effect is absent and total traffic directed towards websites with a given signal is constant and does not depend on the number of websites (except if $J = 0, \frac{M-1}{2}, M-1$).

Two further remarks regarding the $AOF$ effect are in order.
\begin{enumerate}
\item  $AOF$ can contribute to the understanding of the spread of misinformation, in the sense that a signal that is carried by few outlets, for example,  a controversial or ``fake news'' report, may paradoxically receive amplified traffic precisely {\em because} it is carried by few outlets.\footnote{For example, if there was a single minority ``fake news'' website (so that $M-L=1$), then the probability that that outlet would be visited by an individual in the limit is (assuming $\mu=1$ and $\gamma$ not too small) $\rho_{\infty, fake} = 1-\frac{\gamma p (M-1)}{\gamma M -1}$. Moreover, such a ``fake news'' website will be the top ranked one if $\gamma > \frac{1}{M(1-p)}$.} This is consistent with various claims that the algorithms used by Google and Facebook have apparently promoted websites reporting ``fake news''.\footnote{On Google's search algorithm prioritizing websites reporting false information, see: 
``Harsh truths about fake news for Facebook, Google and Twitter", \textit{Financial Times}, November 21, 2016;
``Google, democracy and the truth about internet search", \textit{The Observer}, December 4, 2016. 
On Facebook showing websites reporting false information in the top list of its trending topics, see: ``Three days after removing human editors, Facebook is already trending fake news," 
\textit{The Washington Post}, August 29, 2016.} 
Indeed, even in queries with a clear factual truth, such as yes-no questions within the medical domain, top-ranked results of search engines provide a correct answer less than half of the time (White, 2013).. Importantly, empirical evidence shows that websites reporting misinformation may acquire a large relevance in terms of online traffic and in turn may affect individuals' opinions and behavior (Carvalho et al., 2011; Kata, 2012; Mocanu et al., 2015; Shao et al., 2016). 
By the same token, ``authoritative'' or mainstream websites that are relatively numerous will receive {\em less} traffic overall, the more numerous they are.

\item{\em AOF}  is an intrinsic property of popularity rankings combined with the tendency of individuals to focus on higher ranked results. It can be seen as an almost mechanical property of popularity-based rankings that holds well beyond the assumptions of our model. 
To show this, the online appendix provides examples illustrating cases, where:
$(i)$ parameters are outside the assumed range, for example, when individual signals are uninformative ($p=\mu=1/2$), 
$(ii)$ individuals need not distinguish between majority and minority websites and simply choose based on independent cues ($x_n$ and $z_n$) that can be interpreted as general desirable attributes of the different websites;
$(iii)$ rankings are not stochastic but rather a list of 1$^{st}$ to $M^{th}$ ranked outlet, as a function of traffic obtained ($\sum_{n' \le n} \rho_{n'}$),
$(iv)$ individual choices are pure realizations of the stochastic choice functions ($\rho_{n}$), or
$(v)$ where the total number of websites is not necessarily fixed, thus allowing for cases, where two or more websites can merge and thereby reduce the number of overall websites ($M$).\footnote{One might expect that allowing for free entry would tend to weaken the $AOF$. However, while addressing free entry of websites and even modeling the strategic choices of websites is outside the scope of this paper, it is worth noting that the effect of allowing free entry on the $AOF$ may be limited, especially in those cases where the ``fewer websites'' are minority websites carrying ``dubious'' information. Indeed, it may not be in the interest of mainstream websites (which are likely to also care about their reputation) to report such information. Also, new minority websites may not necessarily be able to ``steal'' much traffic from the existing ones due to their lower ranking (as explained by the \textit{rich-get-richer} dynamic, see Section \ref{sect:non_uniform}). Hence, we believe the result may be particularly relevant for dubious information, carried by relatively few websites and that ``resonates'' with a significant fraction of individuals.}  These examples further suggest that $AOF$ is a potentially robust phenomenon.

\end{enumerate}


\section{Popularity Ranking and Asymptotic Learning}
\label{section:welfare}

To assess the effect of the popularity ranking on opinion dynamics, we consider search environments from an  \emph{interim} and an \emph{ex-ante}  perspective. Since our model endogenizes individual clicking probabilities, it seems natural to evaluate efficiency in terms of asymptotic probability of clicking on a website carrying the correct signal. We further interpret this notion of efficiency as asymptotic learning, under the implicit assumption that individuals assimilate the content they read.
 
Consider the probability of individual $n$ choosing a website reporting a signal corresponding to the true state of the world ($y_m=\omega$). At the interim stage, 
we can write this as the probability $\rho_{n,L}$ ($=\sum_{m \in L} \rho_{n,m}$) of individual $n$ clicking on any website $m \in L$. 
We can also define a measure of {\em interim efficiency} (${\cal P}_L$), 
conditional on interim realizations, where the total number of websites reporting the correct signal is $L$, as:
\begin{equation}
{\cal P}_{L}(\alpha, \gamma, \mu, p) = \rho_{\infty,L}=\lim_{{N}\rightarrow\infty} \rho_{N,L} .
\label{eq:rho_expected_l}
\end{equation}
This implies the following measure of {\em ex ante efficiency} (${\cal P}$):
\begin{equation}
{\cal P}(\alpha, \gamma, \mu, p,q)=\sum_{L=0}^{M}\binom{M}{L}q^{L}(1-q)^{M-L} {\cal P}_{L} (\alpha, \gamma, \mu, p),   \label{eq: welfare}
\end{equation}
which uses the accuracy of websites' signals ($q$) to weigh the different interim levels (${\cal P}_{L}$).

To better highlight the role of popularity ranking on asymptotic learning, we use as benchmark of comparison, the case where ranking is random and uniform throughout ($r_{n,m}= \frac{1}{M}$ for all $n, m$), which we refer to simply as {\em  random ranking}. 

As explained in Appendix~\ref{Appendix-MeanDynamics}, we compute the limit
ranking and limit clicking probabilities using the mean dynamics approximation 
(Norman, 1972; Izquierdo and Izquierdo, 2013). This involves fixing an
interim search environment (essentially characterized by the number of
websites $L$ carrying the correct signal) and approximating the random clicking
probabilities $\rho_{n,m}$ in Equation~(\ref{eq:rho_idbias}) for 
$m\in M$ by their expectations $\widehat{\rho}_{n,m}=\mathbb{E}[\rho_{n,m}]$, while letting $\kappa \rightarrow \infty$ as $N \rightarrow \infty$. 
This leads to deterministic recursions that are easily 
computed in the limit by means of ordinary differential equations. To obtain
the ex ante efficiency, we take expectations over all interim environments
as specified in Equation (\ref{eq: welfare}).


\subsection{Non-monotonicity of Interim Efficiency}

Before analyzing ex-ante efficiency, we  first study interim efficiency.  The following result follows from Proposition~\ref{prop:$AOF$} and illustrates how interim efficiency is a non-monotonic function of the number of websites carrying the correct signal ($L$) due to the \textit{advantage of the fewer} effect ($AOF$).

\begin{cor} 
\label{cor:rankpro}
Fix a search environment ${\cal E}$ with uniform initial ranking $r_{1}$, and consider interim realizations of ${\cal E}$ that vary only in the number of outlets with correct signal ($L$). Then ${\cal P}_L$ is non-monotonic in $L$. 
In particular, when $\gamma$ is interior ($0<\gamma<1$), then ${\cal P}_L$ is increasing in $L$ at $L = 0, \frac{M-1}{2}, M-1$, but it is decreasing in $L$ otherwise.
\end{cor}

The non-monotonicity in $L$ follows from three basic facts: $(i)$ small majorities (or minorities) of outlets with a correct signal result in higher interim efficiency (${\cal P}_L$) than large majorities (or minorities), resulting in a decreasing effect of $L$ on ${\cal P}_{L}$ induced by $AOF$,
$(ii)$ when $L$ increases from $\frac{M-1}{2}$ to $\frac{M+1}{2}$, then interim efficiency increases, since the correct signal ($y_{L}$) switches from being a minority to being a majority signal, $(iii)$ when $L$ increases from 0 to 1 or from $M-1$ to $M$, interim efficiency obviously increases.

\begin{figure}
\par
\begin{center}
\includegraphics[width=6.5cm]{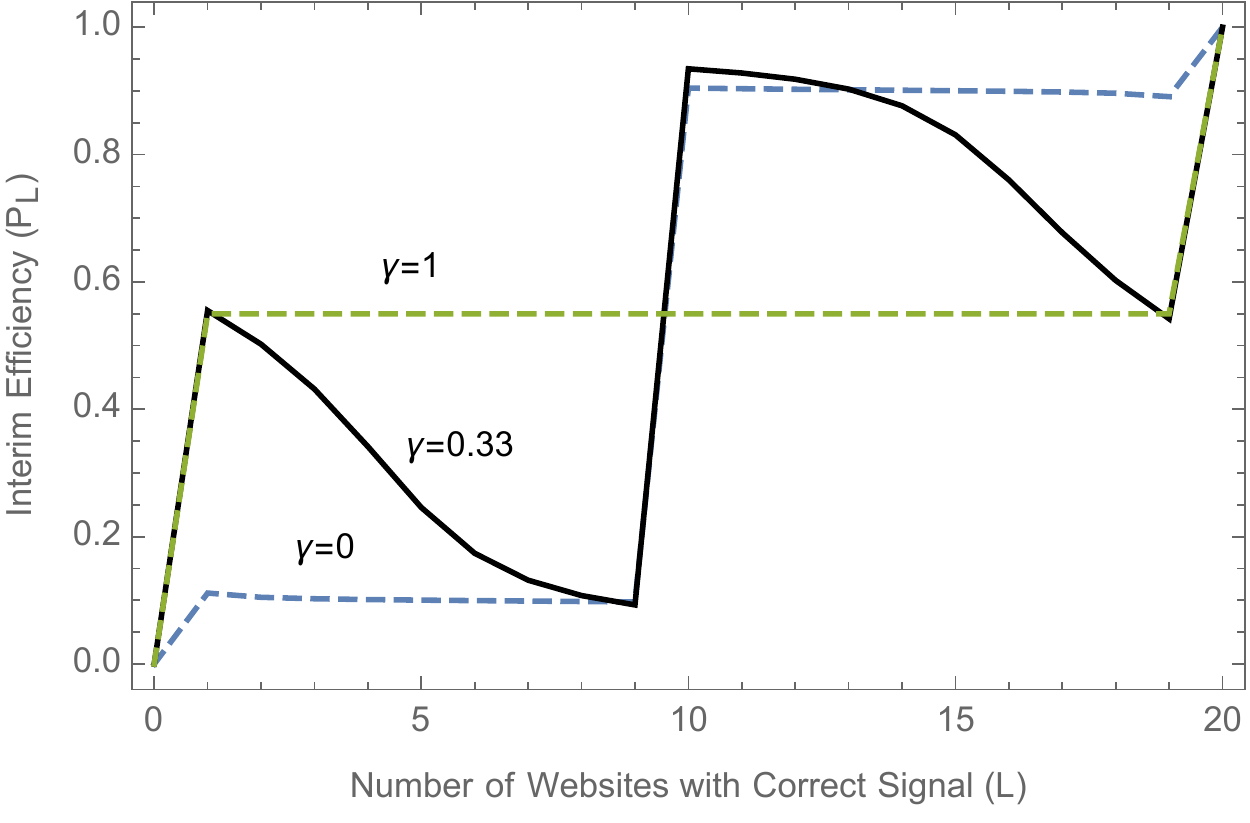} \hspace{.1in}
\includegraphics[width=6.5cm]{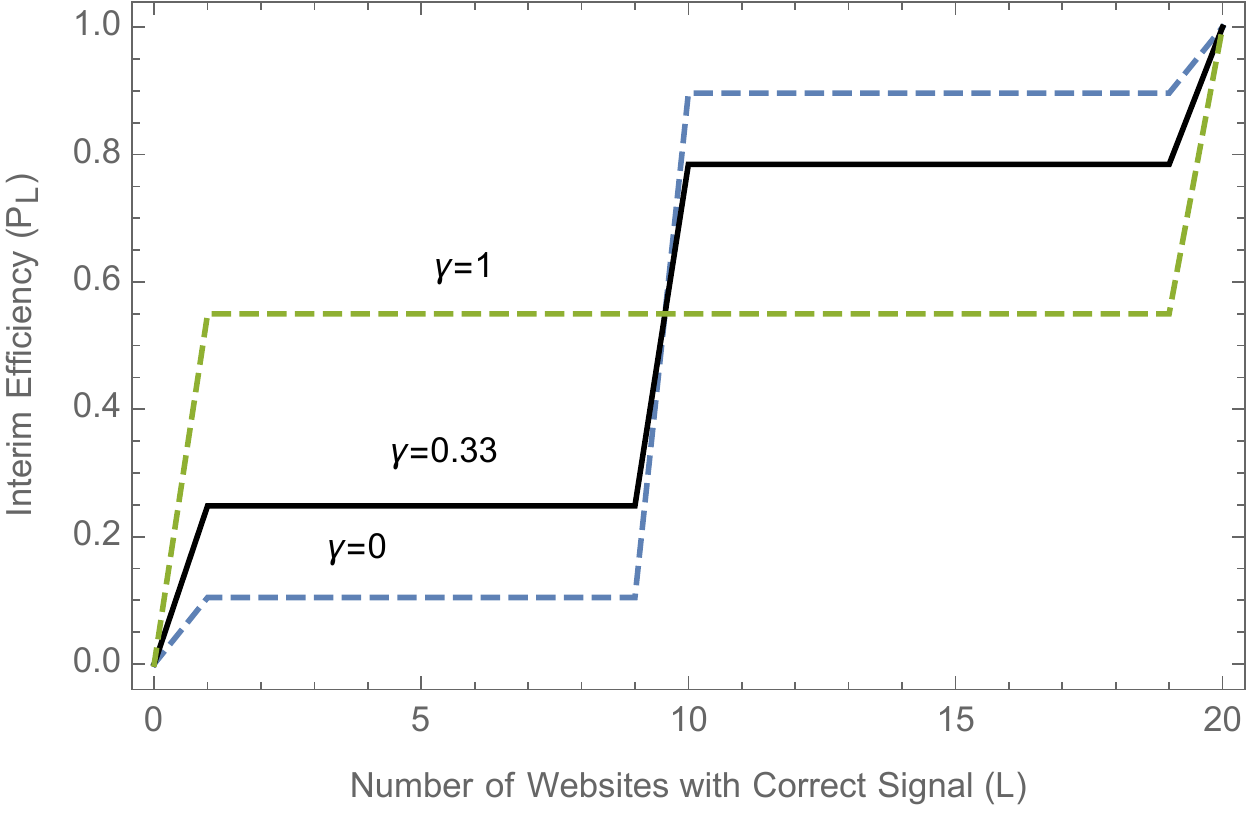} \end{center}
\par
\vspace{-15pt}\caption{\footnotesize {\em Interim} efficiency (${\cal P}_L$) as a function of the number of websites with correct signal ($L$) 
for $\gamma=0.33$ (black line) and other values of $\gamma$ (dashed lines) for the cases of  popularity ranking (left) and  random ranking (right). In both panels, $M=20$, $p=0.55$, $\mu=0.9$, and $r_1$ is uniform.}
\label{fig:ranking_L}
\end{figure}

Figure~\ref{fig:ranking_L} shows the non-monotonicity of  interim efficiency ${\cal P}_L$ in $L$, as stated in Corollary~\ref{cor:rankpro}, as well as the $AOF$ effect as discussed in  Section~\ref{section:results}, in the presence of the popularity ranking when the preference for like-minded news is interior ($\gamma=0.33$), (black line, left panel), and shows that  interim efficiency is monotonic and $AOF$ is absent when $\gamma=0, 1$ (left panel, dashed lines) or when the ranking is random (and therefore popularity ranking is switched off), (right panel).


\subsection{Ex Ante Efficiency}

\subsubsection{Comparative statics}

Our notion of ex ante efficiency (${\cal P}$) is a measure of asymptotic learning that obtains in our popularity-ranking based search environments. The following proposition describes comparative statics of ${\cal P}$ with respect to basic parameters of the model.

\begin{prop}
\label{prop:welfare} 
Let  ${\cal E}$ be a search environment with a uniform initial ranking $r_{1}$. Then:
\begin{enumerate}
\item (Individual accuracy) ${\cal P}$ is increasing in $p$ and in $\mu$.
\item (Website accuracy) ${\cal P}$ can be both increasing or decreasing in $q$.
\item (Like-mindedness) ${\cal P}$ is decreasing in $\gamma$, for $p \in \left( \frac{1}{2}, \overline{p} \right]$, for some $\overline{p}> \frac{1}{2}$.   

\end{enumerate}
\end{prop}

\begin{figure}
\par
\begin{center}
\includegraphics[width=6.5cm]{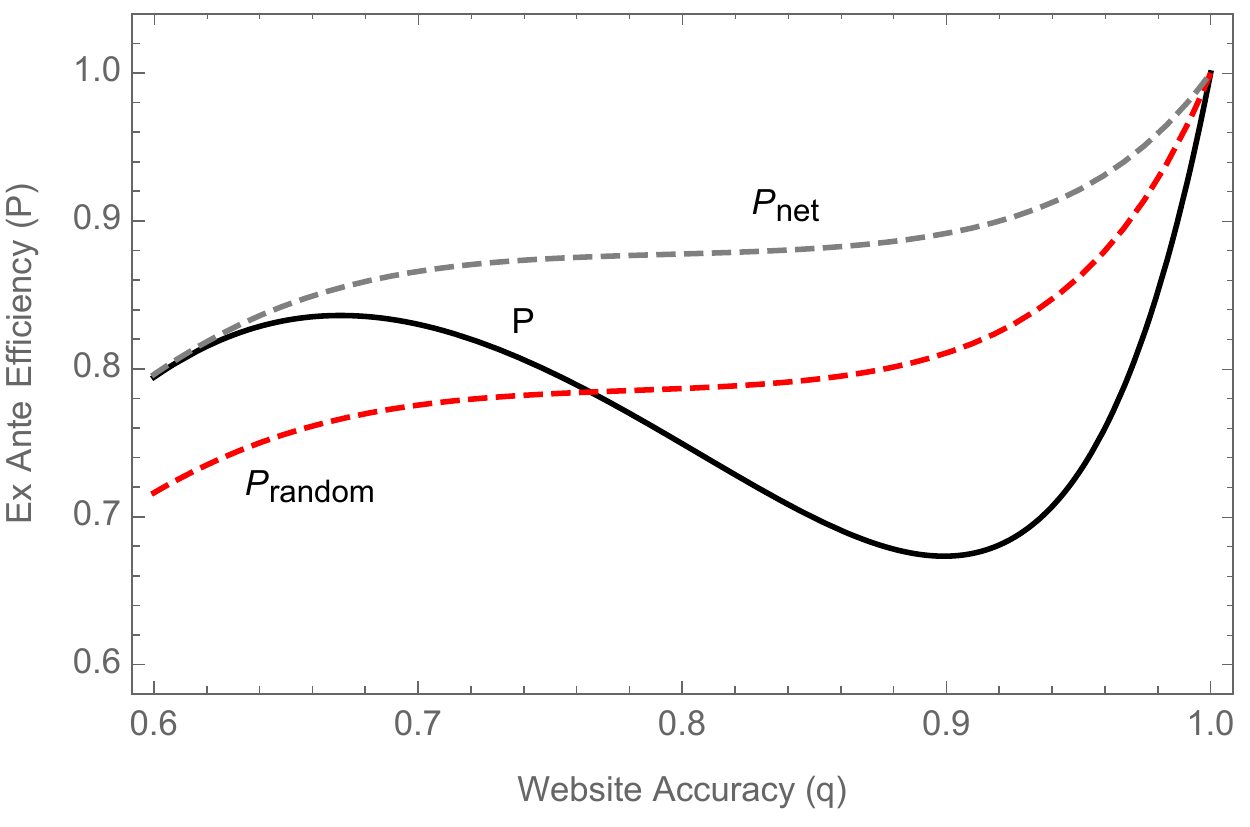} \hspace{.1in}
\end{center}
\par
\vspace{-15pt}\caption{\footnotesize  {\em Ex ante} efficiency (${\cal P}$) as a function of the accuracy of websites' signals ($q$) (right) with popularity-ranking (black line) and random-ranking (red dashed). The figure also shows net {\em ex ante} efficiency (gray dashed) with popularity-ranking. $M=20$, $p=0.55$, $\mu=0.9$,  $\gamma=0.33$ and $r_1$ is uniform.}
\label{fig:eff_alpha_q}
\end{figure}

\noindent  We briefly discuss these effects.
\begin{enumerate}
\item (\textit{Individual accuracy}) Higher levels of $p$ and $\mu$ always increase interim hence also ex ante efficiency. 
This is not just the consequence of the direct effect of more accurate private signals. The direct effect is increased by the dynamic one, since more individuals receiving a correct signal, increases the number of clicks on websites reporting that signal, which in turn increases their ranking. This further increases the probability that subsequent individuals will also choose websites reporting a correct signal so that, even though individuals are assumed to be na\"{i}ve, the higher $p$ and $\mu$ are, the more effective popularity ranking is in aggregating information and creating a positive externality that enhances asymptotic learning.
\item (\textit{Website accuracy}) A higher $q$ may increase or decrease ex ante efficiency. This  is a consequence of the non-monotonicity of the interim efficiency (Corollary~\ref{cor:rankpro}), driven by $AOF$.  Higher values of $q$ make it more likely that the number of websites with correct signal is large. Hence, due to $AOF$, a higher $q$ decreases the clicking probability on websites with a correct signal, thus reducing ex ante efficiency for intermediate values of $\gamma$.  If one could switch off $AOF$, then a higher $q$ would always increase ex-ante efficiency (i.e., efficiency ``net of $AOF$'' is always increasing in $q$).\footnote{Specifically, one can define  {\em ex ante efficiency ``net of $AOF$''} (${\cal P}_{net}$) as ex ante efficiency calculated with weighted constant ``average'' minority and majority traffic levels, respectively, ${\cal P}_{\lceil{\frac{M}{4}\rceil}}$ and ${\cal P}_{\lceil{\frac{3M}{4}\rceil}}$, formally:
\[ {\cal P}_{net} =  \sum_{k=1}^{\frac{M-1}{2}} {\cal P}_{\lceil{\frac{M}{4}\rceil}} \binom{M}{k} q^{k}(1-q)^{M-k} + 
\sum_{k=\frac{M+1}{2}}^{M-1} {\cal P}_{\lceil{\frac{3M}{4}\rceil}} \binom{M}{k} q^{k}(1-q)^{M-k} +q^M . \] 
\textcolor{black}{While, in principle, it would be possible to correct the ranking algorithm for the $AOF$ effect, we are unaware of any such correction undertaken in practice, nor have we seen the effect mentioned in the computer science or machine learning literature.}
}
This stark difference is illustrated in Figure~\ref{fig:eff_alpha_q}.
\item (\textit{Like-mindeness}) Given the assumption on the relative informativeness of private signals ($\mu q > p$), if moreover, $p$ is not too large, then a higher $\gamma$ decreases the probability of choosing a website reporting a correct signal.
\end{enumerate}

\subsubsection{Popularity-Ranking Effect ($PoR$)}

We turn to the crucial question of how well popularity ranking performs in terms of asymptotic learning. The following definition uses random ranking as a comparison benchmark.
\begin{dfn}
Let ${\cal E}$ be a search environment with a uniform  initial ranking $r_1$, and let ${\cal E}'$ be another search environment that differs from ${\cal E}$ only in that the ranking is always an uniform random ranking. Let ${\cal P}$ and ${\cal P}'$ denote ex ante efficiency of ${\cal E}$ and ${\cal E}'$ respectively, then we define the {\em popularity ranking effect} of ${\cal E}$ $($PoR$ ({\cal E}))$ as the difference:  
\[  PoR  ({\cal E})  = {\cal P} - {\cal P}'. \]
\end{dfn}
\noindent
The following result compares popularity ranking and random ranking.
\begin{prop}
\label{cor:randomrank}
Let ${\cal E}$ be a search environment with uniform initial ranking $r_1$.
Then there exist $\overline{\gamma} > 0$, $\overline{\mu} <1$, and a threshold function for $q$, $\phi: [0 , 1] \rightarrow [\frac{1}{2}, 1]$, that is decreasing in  $\hat{\gamma}$, such that $\phi(\hat{\gamma})=1$, for $\hat{\gamma} \in [0, \overline{\gamma}]$, and, moreover, for any threshold $\hat{\gamma} \in [\overline{\gamma},1]$:
\begin{enumerate}
\item  $  $PoR$ ({\cal E}) \ge 0$ for $\gamma \in [0,\hat{\gamma}]$, $\mu \in [\overline{\mu}, 1]$, provided $q \in  [\frac{p}{\overline{\mu}}, \phi(\hat{\gamma})]$.
\item  $  $PoR$ ({\cal E}) \le 0$ for $\gamma \in [\overline{\gamma},\hat{\gamma}]$, provided $q \in [\phi(\hat{\gamma}), 1]$.
\end{enumerate}
\end{prop}
\noindent
In other words, when preference for like minded news is sufficiently low ($\gamma \in [0,\overline{\gamma}]$) and the accuracy on the majority signal is sufficiently high ($\mu \in [\overline{\mu}, 1]$), then popularity ranking does better than random ranking.  That is, 
when individual clicking behavior generates sufficiently positive information externalities, popularity ranking aggregates information and increases the probability of asymptotic learning relative to random ranking. Moreover, this continues to hold for higher values of preference for like minded news ($\gamma \in [\overline{\gamma}, \hat{\gamma}]$, for $\hat{\gamma}$ that can go up to 1), provided the accuracy of websites is not too high ($q \le \phi(\hat{\gamma})$) (due to $AOF$). Put differently, due to $AOF$, an increase in the ex-ante informativeness of websites ($q$), may lead to a negative popularity ranking effect.\footnote{By contrast, ex ante efficiency ``net of $AOF$'' (${\cal P}_{net}$) is above random ranking even for large values of $q$.} 
Figures~\ref{fig:eff_alpha_q} and~\ref{fig:eff_alpha_gamma} illustrate Proposition \ref{cor:randomrank}: popularity ranking dominates random ranking in terms of ex-ante efficiency provided $q$ and $\gamma$ are not too large, and is dominated by random ranking otherwise.

\begin{figure}
\par
\begin{center}
\includegraphics[width=6.5cm]{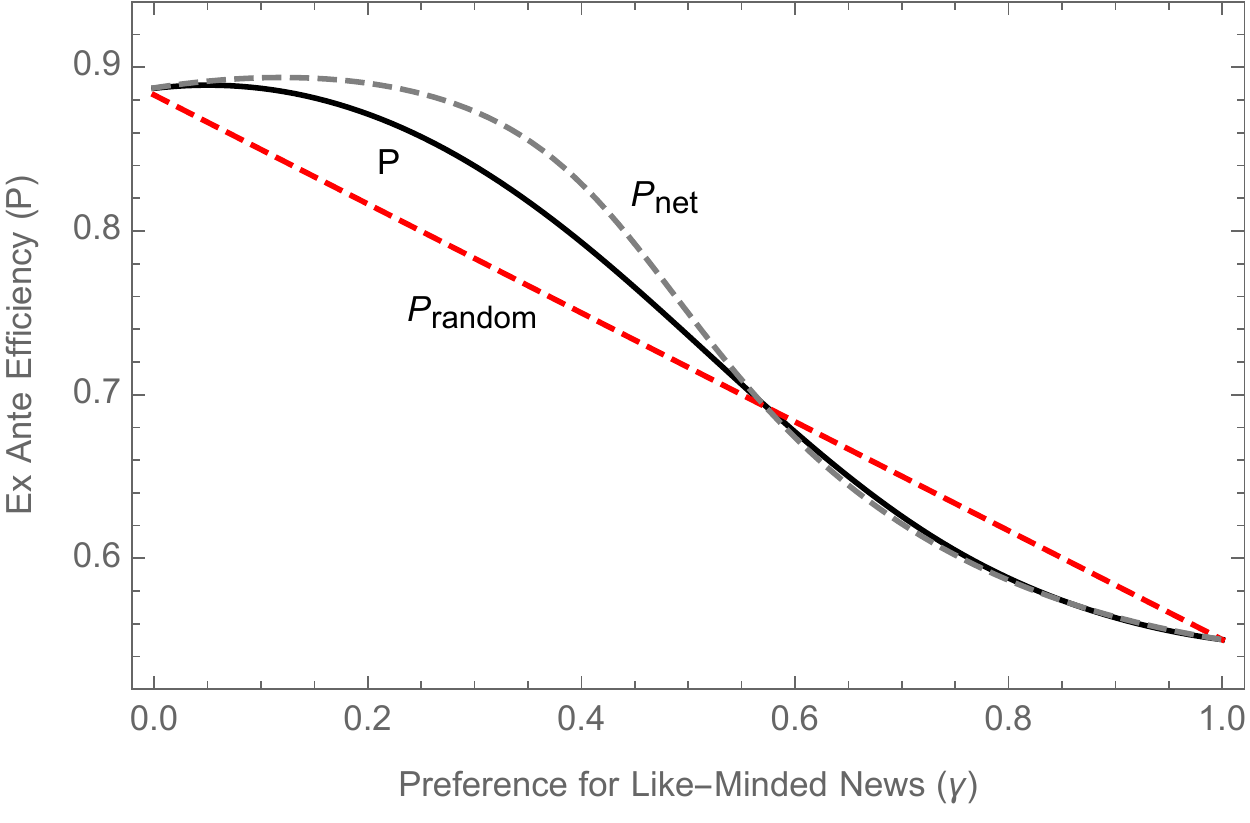} \hspace{.1in}
\end{center}
\par
\vspace{-15pt}\caption{\footnotesize  {\em Ex ante} efficiency (${\cal P}$) as a function of the preference for like-minded news  ($\gamma$) with popularity-ranking (black line) and random-ranking (red dashed). The figure also shows net {\em ex ante} efficiency (gray dashed) with popularity-ranking. The plot is drawn for $M=20$, $p=0.55$, $\mu=0.9$, $q=0.7$, and $r_1$ is uniform.}
\label{fig:eff_alpha_gamma}
\end{figure}


\section{Personalized Ranking}
\label{sect:pers_rank}

Another important question for a ranking algorithm concerns whether it should keep
track of, and use, information it has available concerning the 
individuals' identity and past searches. A search engine may want its algorithm to
condition the outcomes of searches on the geographic location of the
individuals (e.g., using the individual's \textit{IP address}) or other individual characteristics
(e.g., using the individual's search history). Accordingly, a personalized search algorithm may output different
search results to the same query performed by individuals living in different
locations and/or with different browsing histories.\footnote{Pariser (2011); Dean (2013); MOZ (2013); Vaughn (2014); Kliman-Silver et al. (2015). Hannak et al. (2013) document the presence of extensive personalization of search results. In particular, while they show that the extent of personalization varies across topics, they also point out that ``politics'' is the most personalized query category. See also Xing et al. (2014) for empirical evidence on personalization based on the \textit{Booble} extension of Chrome.}

Suppose the set of individuals $N$ is partitioned into two nonempty groups $A,B \subset N$, such that $A \cup B=N$ and $A\cap B=\emptyset$. Suppose the two groups differ in terms of the individuals' preference for like-minded news ($\gamma_{A} \neq \gamma_{B}$). In any period an individual is randomly drawn from one of the two groups, that is, from $A$ with probability $\frac{N_A}{N}$ and from $B$ with probability $\frac{N_B}{N}$, where $N_{A}=\#A>0$ and $N_{B}= \#B>0$. 
A \emph{personalized ranking algorithm} then
consists of two parallel rankings, namely, $r_{t}^{A}$ for individuals in $A$ and $r_{t}^{B}$ for individuals in $B$. 
Each one is updated as in the non-personalized case, with the difference that the weight on past choices
of individuals from the own group are possibly different than those from the other one.
Set, for any $t$ and $m$, and for $\ell=A,B$,
\begin{equation}
r_{t,m}^{\ell}=\nu_{t}^{\ell}r^{\ell}_{t-1,m}+(1-\nu_{t}^{\ell})\rho_{t-1,m}^{\ell
}\,, \label{eq:rloc}%
\end{equation}
where $\nu_{t}^{\ell}$ now depends on whether or not the individual searching
at time $t-1$ was in the same group, $t-1\in\ell$, and where the weight
$\nu_{t}^{\ell}$, for $\ell=A,B$, is given by:
\begin{equation}
(\nu_{t}^{\ell},1-\nu_{t}^{\ell})=\left( \frac{\kappa}{1-\lambda_{t}+\kappa},  \frac{1-\lambda_{t}}{1-\lambda
_{t}+\kappa} \right) ,\mbox{ where }\lambda_{t}=\left\{
\begin{array}
[c]{ll}%
0 & \text{ if }t-1\in\ell\\
\lambda & \text{ else },
\end{array}
\right.  \label{eq:nuloc}
\end{equation}
with $\kappa\in\mathbb{N}$ and $\lambda\in\lbrack0,1]$ parameters of the personalized ranking algorithm. 
This algorithm now gives different weights to past choices over websites depending on whether these choices were taken by individuals in the same group (weights $\frac{\kappa}{1+\kappa},\frac{1}{1+\kappa}$) or in the other group (weights $\frac{\kappa}{1-\lambda+\kappa},\frac{1-\lambda}{1-\lambda+\kappa}$). 
In particular, when $\lambda=1$, the ranking algorithm is fully personalized, whereas, when $\lambda=0$, it coincides with the non-personalized one previously defined.
We implicitly assume that the personalized algorithm partially separates individuals according to the individual characteristics (i.e., according to the different parameters $\gamma$). We will refer to a \emph{personalized search environment} ${\cal E}_{\lambda}$ when considering the generalization of a search environment ${\cal E}$, defined in Section \ref{sect:search_env}, to the case where the ranking is described by Equations~(\ref{eq:rloc}) and~(\ref{eq:nuloc}) with  $\lambda \ge 0$.

\subsection{Belief Polarization}

\label{section: local_search}

By introducing personalization in our model, we allow the ranking of websites to be conditioned on (observable) characteristics of the individuals such that searches performed by individuals in different groups can have different weights. 
When $\lambda=0$, there is no difference in the ranking of websites for the two groups, while as $\lambda$ increases the groups start observing potentially different rankings which may further trigger different website choices, thus leading to different opinions. In other words, increased personalization may lead to increased \textit{belief polarization}.
\begin{dfn}
Fix a personalized search environment $\mathcal{E}_{\lambda}$ with nonempty groups  of individuals $A$ and $B$. Let $K$ denote the set of websites carrying the website-majority signal, then we define the {\em degree of belief polarization} of ${\cal E}_{\lambda}$ as:
\[ {\cal BP} ({\cal E}_{\lambda}) = \left\vert \widehat{\rho}_{N,K}^{A}-\widehat{\rho}_{N,K}^{B} \right\vert  .
\]
We say environment ${\cal E}_{\lambda}$ exhibits more belief polarization than ${\cal E}_{\lambda}^{\prime}$, if
${\cal BP} ({\cal E}_{\lambda}) > {\cal BP} ({\cal E}_{\lambda}^{\prime})$.
\end{dfn}

\begin{prop}
\label{prop:polar}
Let ${\cal E}_{\lambda}$ be a personalized search environment with personalization parameter $\lambda$. Suppose there are two groups of individuals $A$ and $B$ of equal size, then ${\cal BP} ({\cal E}_{\lambda})$ is increasing in $\lambda$.
\end{prop}

Non-trivial personalization ($\lambda > 0$) can lead to different information held by relatively similar groups of individuals and thus to polarization of opinions.\footnote{Notice that, if individuals in different groups were to face also different initial rankings, ($r_{1}^{A} \ne r_{1}^{B}$), then this different rankings would clearly contribute to further accentuating the evolution of rankings seen by the two groups.}
This suggests that individuals might end up into an algorithmically-driven echo chamber. This is in line with Flaxman et al. (2016), who show that search engines can lead to a relatively high level of ideological segregation, due to web search personalization embedded in the search engine's algorithm and to individuals' preference for like-minded sources of news.  
It is also in line with existing claims and empirical evidence suggesting that the Internet---together with the online platforms embedded in it---generally contributes towards increasing ideological segregation (Putnam, 2001; Sunstein, 2009; Pariser, 2011; Halberstam and Knight, 2016; Bessi et al., 2015; Bar-Gill
and Gandal, 2017).

Next, we study how personalization may actually hinder asymptotic learning.

\subsection{Personalized-Ranking Effect ($PeR$)}
\label{sect:efficiency_pers}

We turn to the effect of personalization on ex-ante efficiency. Personalization here can be seen as progressively ``separating'' two groups of individulas by uncoupling their rankings and thereby switching off potential externalities from one group to the other. To the extent that the group with weaker preference for like minded news exerts a positive externality on the groups' rankings and overall ex ante efficiency, increasing the personalization may inhibit total ex ante efficiency.

\begin{dfn}
Let ${\cal E}$ be a non-personalized and ${\cal E}_{\lambda}$ be a personalized search environment with personalization parameter $\lambda$, and both with a uniform initial ranking $r_{1}$. Let ${\cal P}$ and ${\cal P}_{\lambda}$ denote ex ante efficiency of ${\cal E}$ and ${\cal E}_{\lambda}$ respectively, then we can define the {\em personalized ranking effect} of ${\cal E}_{\lambda}$ $( $PeR$  ({\cal E}_{\lambda}))$ as the difference:  
\[  PeR  ({\cal E}_{\lambda}) ={\cal P}_{\lambda}-{\cal P} . \] 
\end{dfn}

\begin{prop} \label{prop:personalizedsearch}
${\cal E}_{\lambda}$ be a personalized search environment with personalization parameter $\lambda$, and with a uniform initial ranking $r_{1}$. Suppose there are two nonempty groups of individuals $A$ and $B$ with $0 \le \gamma_A < \gamma_B \le \overline{\gamma}$. Then there exist $\overline{\gamma} > 0$, $\overline{\mu} <1$, and a decreasing function of $\hat{\gamma}$, $\phi: [0 , 1] \rightarrow [\frac{1}{2}, 1]$, such that $\phi(\hat{\gamma})=1$, and, moreover, for any $\hat{\gamma} \in [\overline{\gamma},1]$:
\begin{enumerate}
\item  $  $PeR$ ({\cal E}) \le 0$ for $\gamma \in [0,\hat{\gamma}]$, $\mu \in [\overline{\mu}, 1]$, provided $q \in  [\frac{p}{\overline{\mu}}, \phi(\hat{\gamma})]$.
\item  $  $PeR$ ({\cal E}) \ge 0$ for $\gamma \in [\overline{\gamma},\hat{\gamma}]$, provided $q \in [\phi(\hat{\gamma}), 1]$.
\end{enumerate}
\end{prop}

 Although the exact cutoff values for the parameters need not coincide, the parallels between the popularity ranking effect and the personalized ranking effect are stark. That is, when the preference for like minded news is sufficiently low ($\gamma \in [0,\overline{\gamma}]$) and the accuracy on the majority signal is sufficiently high ($\mu \in [\overline{\mu}, 1]$), then personalized ranking does worse than non-personalized (popularity) ranking; moreover, this continues to hold for higher values of preference for like minded news ($\hat{\gamma} \in [\overline{\gamma}, 1]$), provided the accuracy of websites is not too high ($q \le \phi(\hat{\gamma})$) (due to $AOF$); on the other hand for the higher values of preference for like-minded news ($\hat{\gamma} \in [\overline{\gamma},1]$) personalized ranking will perform better than non-personalized (popularity) ranking, provided accuracy of websites is sufficiently high ($q \ge \phi(\hat{\gamma})$). In other words, the forces that lead to a positive (negative) popularity effect are similar to the ones that lead to a negative (positive) personalization effect.

The intuition for the negative relation between the popularity ranking and the personalized ranking effects ($PoR$ and  $PeR$) is due to the fact that a positive popularity ranking effect occurs when individuals' parameters generate positive feedback into the dynamics (high accuracy $\mu$ and low preference for like-minded news $\gamma$); the same forces favor non-personalization and hence tend to generate a negative personalized ranking effect. This is because personalization can be seen as limiting the feedback between individuals in the dynamics, and so, when individuals' signals tend to generate positive feedback, it is better not to limit the feedback and hence it is better not to personalize the ranking and {\em vice versa} when individuals' signals generate negative feedback. The similarity also with respect to the website accuracy parameter ($q$) is due to the $AOF$ effect that is present with or without personalization.

\begin{figure}

\par
\begin{center} 
\includegraphics[width=6.5cm]{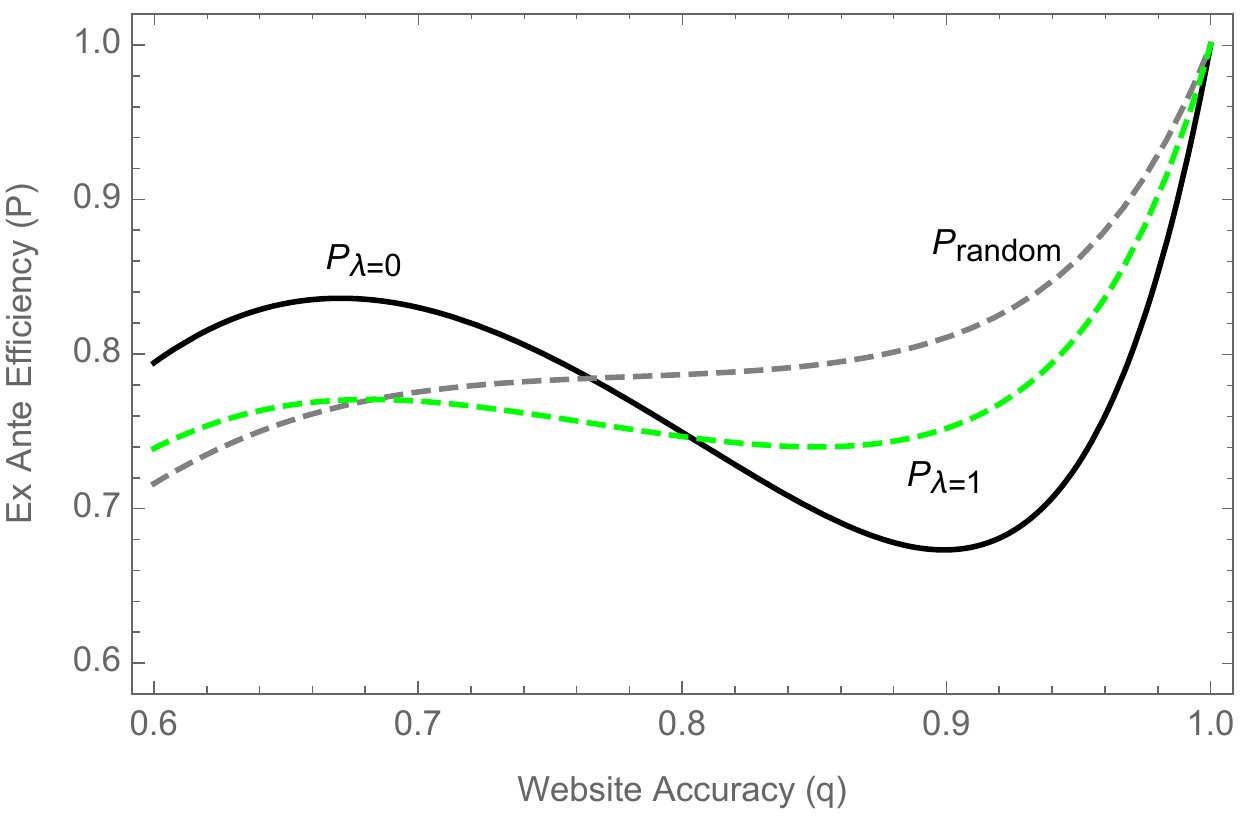}  \hspace{.1in}
\includegraphics[width=6.5cm]{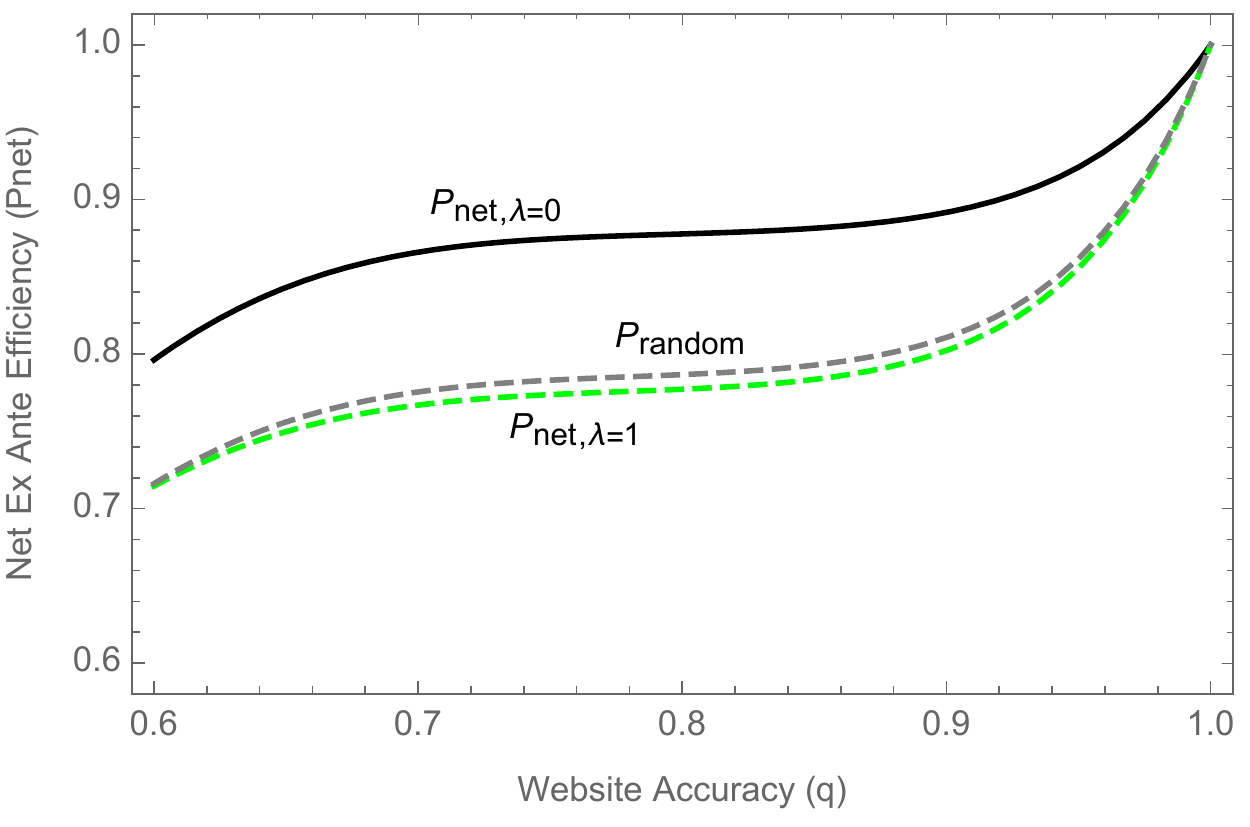}  \hspace{.1in}
\end{center}
\par
\vspace{-15pt}
\caption{\footnotesize {\em Ex ante} efficiency with no personalization (${\cal P}_{\lambda=0}$) (black) and with personalization w.r.t.~$\gamma$ (${\cal P}_{\lambda=1}$) (green dashed); and with random ranking (gray dashed). The right panel illustrates the same for ex-ante efficiency ``net of $AOF$''. In both panels, $M=20$,  $q=0.7$, $p=0.55$, $\mu=0.9$, $\gamma=0.33$, and $r_1$ uniform.}
\label{fig:personalizationeffect}
\end{figure}

Figure \ref{fig:personalizationeffect} illustrates the negative relationship between  $PoR$ and  $PeR$ . In this case, personalized ranking (green dashed) tends to be dominated (in terms of asymptotic learning) by either non-personalized ranking (black) or by random  ranking (gray dashed). The right-hand panel indicates that when $AOF$ is switched off, the corresponding ``net''  $PeR$  effect tends to be negative for any level of ex-ante accuracy of websites. 

It is important to note that our searches are common value searches in the sense that all individuals want to read outlets carrying the (same) correct signal. When individuals have private values, personalization may be an important tool that actually favors asymptotic learning of typically multiple and distinct signals.


\section{ Robustness }
\label{section: Extensions}

\subsection{Attention bias} \label{sect:attbias}

In our basic model of how individuals respond to website rankings, we assumed multiplicative separability between the ranking ($r_n$) and the ranking-free values ($v_{n}^*$), such that the ranking entered directly as a weighting function for the ranking-free-values. We now generalize the weighting function by introducing an \textit{attention bias} parameter $\alpha \ge 0$, which yields the following more general ranking-weighted values:
\[ v_{n,m} = (r_{n,m})^{\alpha} \cdot v_{n,m}^* .
\]
The parameter $\alpha$ calibrates the individual's attention bias in the following sense: 
$\alpha=1$ is a neutral benchmark in that it maintains the weight differences already present in the entries of the ranking $r_{n}$;  
$\alpha>1$ magnifies the differences in the entries of $r_{n}$, and in particular also the ones present in the
initial ranking $r_{1}$; 
$\alpha<1$ reduces the differences in the entries of $r_{n}$; in the limit, as $\alpha \rightarrow 0$, all entries have the same weight, which represents the case with \textit{no} attention bias, where all websites that provide the same signal yield the same value.
Given the values $v_{n,m}$, we can write the \emph{website choice function} with attention bias $\alpha \ge 0$, $\rho_{n} : \{ 0, 1 \}^2 \times \Delta(M) \rightarrow \Delta(M)$, as
\begin{equation} \label{eq:rho_attbias}
\rho_{n,m} 
=  \frac{(r_{n,m} )^{\alpha} \cdot v_{n,m}^*}{\sum_{m' \in M} (r_{n,m'} )^{\alpha} \cdot v_{n,m'}^*}  .
\end{equation}
When $\alpha=1$, this coincides with the choice function studied so far (Equation~(\ref{eq:rho_idbias})).
When $\alpha=0$, it yields choices that coincide with the ranking-free choices $v_{n,m}^{*}$, since all outlets have the same weight $(r_{n,m})^{0} = 1$. For this same reason, it is easy to see that, in terms of website choices and hence interim and ex ante efficiency (${\cal P}_{L}$ and ${\cal P}$), an environment with popularity ranking $r_{n,m}$ but attention bias $\alpha=0$ is outcome-equivalent to an environment with  {\em random ranking}, regardless of what $\alpha$ might be in that environment.\footnote{When there is zero attention bias ($\alpha=0$), all websites receive equal weights in the website values ($v_{n,m}$) that determine the website choice probabilities ($\rho_{n,m}$). 
Similarly, when there is random ranking ($r_1$ and every subsequent ranking $r_n$ is uniform, so that $r_{n,m}=\frac{1}{M}$ regardless of the parameter $\alpha$), then all websites will also receive equal weights in the website values. Thus website choice probabilities ($\rho_{n,m}$) always coincide in the two cases.}

The $AOF$ effect stated in Proposition~\ref{prop:$AOF$} (for $\alpha=1$) carries over {\em verbatim} to the general case of $\alpha \ge 0$;\footnote{The proof of Proposition~\ref{prop:$AOF$} in Appendix~\ref{Appendix-Proofs} is directly given for the case of $\alpha \ge 0$. It is important to note that, while, for $0 \le \alpha \le 1$, the unique limit always satisfies $AOF$; for $\alpha >1$, depending on the initial condition (i.e., the initial ranking; see also Section~\ref{sect:non_uniform}), there can be multiple limits, of which the asymptotically stable ones, that is, the only ones that can be reached by our dynamic process, also always satisfy $AOF$.} and similarly its implications for asymptotic learning or interim efficiency (Corollary~\ref{cor:rankpro}), as summarized by the following Corollary:
\begin{cor} 
\label{cor:rankpro_alpha}
Fix a search environment ${\cal E}$ with attention bias parameter $\alpha \ge 0$ and with uniform initial ranking $r_{1}$ and $\kappa$ large, and consider interim realizations of ${\cal E}$ that vary only in the number of outlets with correct signal ($L$). Then:
\begin{enumerate}
\item For $\alpha > 0$, ${\cal P}_L$ is non-monotonic in $L$. 
\item For $\alpha =0$, ${\cal P}_L$ is monotonically increasing in $L$.
\end{enumerate}
In particular, when $\alpha > 0$ and $0<\gamma<1$, then ${\cal P}_L$ is increasing in $L$ at $L = 0, L=M-1$, as well as at $L=\frac{M-1}{2}$, but it is decreasing in $L$ otherwise.
\end{cor}
In particular, this shows that $AOF$ applies when ranking matters ($\alpha>0$) and disappears when there is no attention bias and ranking does not matter  ($\alpha=0$). Thus $AOF$ disappears with random ranking. In terms of comparative statics of ex ante efficiency, the results of Proposition \ref{prop:welfare} also carry over {\em verbatim} to the general case of $\alpha \ge 0$, with the only difference that, for $\alpha=0$, ${\cal P}$ is always increasing in $q$, while, for $\alpha >0$, it can be both increasing or decreasing in $q$; again this is because $AOF$ kicks in when $\alpha >0$ and vanishes when $\alpha=0$. 

\subsection{Non-Uniform Initial Ranking and the Rich Get Richer}
\label{sect:non_uniform}
Many of the propositions stated in the paper assumed a uniform initial ranking. We now study this assumption in more detail, looking at environments with general attention bias ($\alpha \ge 0$). As it turns out, the expected limit clicking probabilities $\widehat{\rho}_{\infty, m}$ only depend on the initial ranking when $\alpha > 1$. In this case, there is also a ``rich-get-richer'' effect.

\begin{prop}\label{prop:initial_rank_0}
Let ${\cal E}$ be a search environment with attention bias parameter $\alpha \ge 0$ and with interim realization $\left\langle \omega;(L,(y_{m}))\right\rangle$, with interior initial ranking $r_{1,m}>0$ for all $m$. Then, if attention bias satisfies $0 \le \alpha \le 1$, then the expected limit clicking probabilities $\widehat{\rho}_{\infty, m}$ do not depend on $r_1$. This is not true if $\alpha>1$.
\end{prop}

The evolution of a website's ranking based on its
\textquotedblleft popularity,\textquotedblright\ interacted with a sufficiently large attention bias ($\alpha > 1$)  exhibits a \textit{rich-get-richer} dynamic, whereby the ratio of the expected clicking probabilities of two websites $m,m'$, with $r_{1,m}>r_{1,m'}>0$ increases over time as more agents perform their search. The effect is further magnified, the larger $\alpha$ is. 
Importantly, the differences in the ranking probabilities ($r_{n}$) and in the expected website choice probabilities ($\widehat{\rho}_{n}$), are driven by the initial ranking ($r_{1}$) and are amplified by the attention bias. We define this more formally. 

\begin{dfn}
Fix a search environment $\mathcal{E}$ with attention bias parameter $\alpha \ge 0$ and with interim realization $\left\langle
\omega;(L,(y_{m}))\right\rangle$. We say that $\mathcal{E}$ exhibits the \emph{rich-get-richer} dynamic if, for two websites $m, m^{\prime} \in M$
with the same signal, $y_{m} = y_{m^{\prime}}$, and different initial ranking, $r_{1, m} > r_{1,m^{\prime}}>0$, we also have $\frac{\widehat{\rho}_{n,m}}{\widehat{\rho}_{n,m^{\prime}}} > \frac{\widehat{\rho}_{n-1,m}}{\widehat{\rho}_{n-1,m^{\prime}}}$, for any $n > 0$. 
\end{dfn}

When an environment exhibits a \textit{rich-get-richer} dynamic, then the ratio of the {\em expected} probability of two websites (with the same signal but different initial ranking) being visited not only persists over time (this follows from $\kappa > 0$), but actually increases. We can state the following.

\begin{prop}
\label{prop:initial_ranking} Let $\mathcal{E}$ be a search environment with attention bias parameter $\alpha \ge 0$ and with interim realization $\left\langle \omega;(L,(y_{m}))\right\rangle$. Then we have, for any two websites $m, m^{\prime} \in M$ with $y_m = y_{m^{\prime}}$ and $r_{1, m} > r_{1,m^{\prime}}>0$, and $n>1$:
\begin{equation} \label{eq:rich gets richer}
\frac{\widehat{\rho}_{n,m}}{\widehat{\rho}_{n,m^{\prime}}} 
\hspace{.1in}
\left\{
\begin{array}
[c]{l}%
< \\
= \\
> 
\end{array}
\right	\}
\hspace{.1in}
\frac{\widehat{\rho}_{n-1,m}}{\widehat{\rho}_{n-1,m^{\prime}}}
\hspace{.1in}
\left\{
\begin{array}
[c]{l}%
\text{if }\alpha<1\\
\text{if }\alpha=1\\
\text{if }\alpha>1 \, .
\end{array}
\right	.
\end{equation}
In particular, if the attention bias is large enough ($\alpha>1$), then $\mathcal{E}$ exhibits the rich-get-richer dynamic.
\end{prop}

This proposition shows that the attention bias plays a crucial role in the evolution of website traffic.  When $0 \leq \alpha <1$, initial conditions do not matter in the limit (as $N \rightarrow \infty$), in the sense that websites with the same signal will tend to be visited with the same probability in the limit. When $\alpha=1$  the ratios of the expected clicking probabilities remain constant for websites with the same signal. 
When $\alpha>1$, initial conditions matter and the evolution of website traffic follows a rich-get-richer dynamic. Traffic concentrates on the websites that are top ranked in the initial ranking. The \textit{rich-get-richer}  dynamic is in line with the ``Googlearchy'' suggested by Hindman (2009), who argues that the dominance of popular websites via search engines is likely to be self-perpetuating. Most importantly, the \textit{rich get richer} pattern of website ranking (and traffic) via search engines is consistent with established empirical evidence (Cho
and Roy, 2004).\footnote{Indeed, even if some scholars have argued that the \textit{overall} traffic induced by search engines is less concentrated than it might appear due to the topical content of user queries (Fortunato et al., 2006), the \textit{rich-get-richer} dynamic is still present within a specific topic.}


\subsection{Sophisticated Learning}
\label{sect:soph}

The key contribution of our model is to combine endogenous ranking of websites with sequential clicking behavior of behaviorally biased and na\"{i}ve individuals. As pointed out before, such a framework reflects empirical evidence on individual preferences for confirmatory news (Gentzkow and Shapiro, 2010) and further informational and behavioral limitations in assessing the working of ranking algorithms (Granka,
2010; Eslami et al., 2016). This provides a setting where there is a certain degree of learning via the ranking but where asymptotic learning is not guaranteed.\footnote{As pointed out by Acemoglu and Ozdaglar (2011), p.~6, as disagreement on many economic, social and  political issues is ubiquitous,``useful models of learning should not always predict consensus, and certainly not the weeding out of incorrect  beliefs''.}  We now sketch how relaxing such assumptions might affect asymptotic learning. 

If individuals are not behaviorally biased ($\gamma=0$), then there will not be asymptotic learning, as individuals always click according to their signal of the website-majority signal ($z_n$).\footnote{Furthermore, if individuals are also sophisticated and can keep track of the evolution of the rankings, then they can at best learn the actual website-majority signal ($y_K$), which depending on the number of websites can be a more or less accurate signal of the true state ($\omega$).} On the other hand, if individuals are behaviorally biased ($0<\gamma<1$) and sophisticated (i.e., they know the parameters of the environment, ${\cal E}$, and can observe the evolution of the ranking, $r_t, t< n$), then they can compute the clicking probabilities ($\rho_t, t<n$). Hence, for sufficiently large $n$, they can also compute increasingly accurate estimates of the individual signals ($x_t, t<n$) and \textit{a fortiori} of their majority. The latter is an arbitrarily accurate signal of the true state ($\omega$). 
Maintaining our basic framework, we can capture an element of this ``more sophisticated'' updating by means of a further signal $\hat{z}_{n}$, that, for sufficiently large $n$, can be observed or deduced with accuracy $\hat{\mu} > \mu q > p$. %
For sufficiently large $n$, $\hat{z}_{n}$ is more informative than the  website majority signal $z_{n}$ (accordingly, 
equation (\ref{eq:rank_free}) would be defined with respect to $\hat{z}_{n}$). Figure~\ref{fig:rationalranking_L} illustrates the limiting behavior when the signal $\hat{z}_{n}$ is approximate ($\hat{\mu}=0.9$) (left) and when it is fully accurate ($\hat{\mu}=1$) (right). 
As can be seen, in both cases, when the preference for like-minded news is interior ($\gamma=0.33$, black line), the $AOF$ effect continues to hold, but {\em without} the upward jump when $L$ goes from $\frac{M-1}{2}$ to $\frac{M+1}{2}$, since the individuals' website majority signal ($z_{n}$), responsible for such a jump, no longer plays a role when the signal $\hat{z}_{n}$ is sufficiently accurate. Comparing Figure~\ref{fig:rationalranking_L} with our benchmark case with no sophisticated updating (Figure~\ref{fig:ranking_L}, left panel), it is clear that the largest discrepancy in terms of interim efficiency occurs when the website majority signal is incorrect (i.e., outlets in $L$ are a minority), in which case the na\"{i}ve individual's ``rational'' choice of following the ex-ante most informative signal ($z_{n}$) is clearly sub-optimal.

\begin{figure}
\par
\begin{center}
\includegraphics[width=6.5cm]{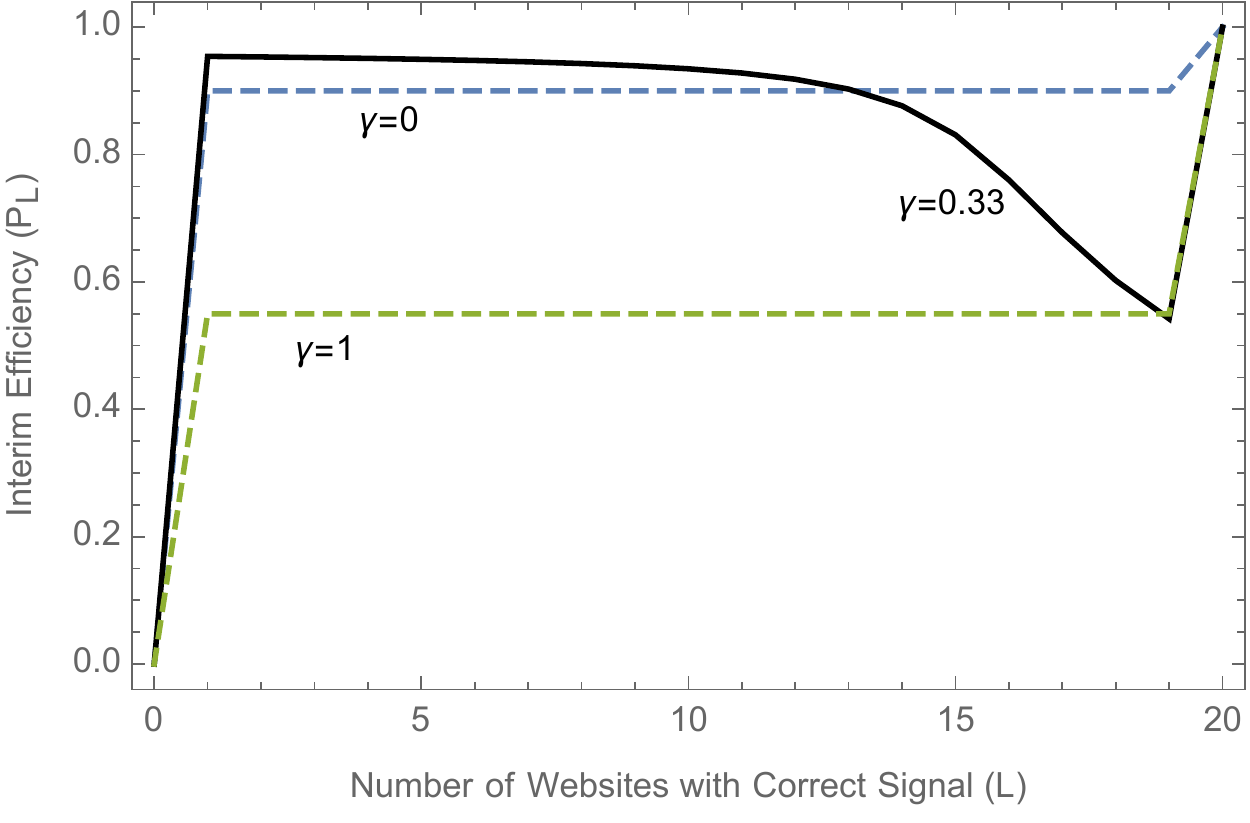} \hspace{.1in}
\includegraphics[width=6.5cm]{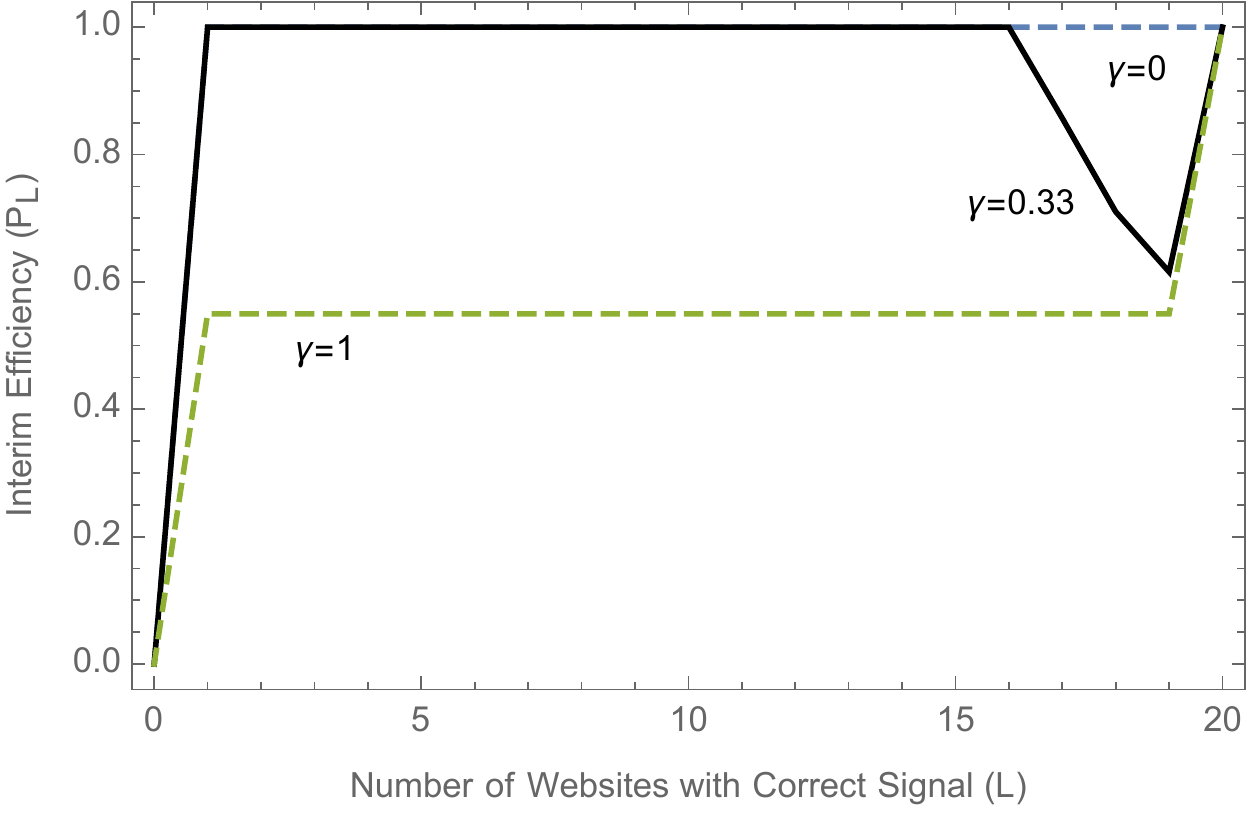} \end{center}
\par
\vspace{-15pt}\caption{\footnotesize {\em Interim} efficiency (${\cal P}_L$) with signals ($\hat{z}_{n}$) as a function of the number of websites with correct signal ($L$) 
for $\gamma=0.33$ (black line) and other values of $\gamma$ (dashed lines) for the cases of approximate signals ($\hat{\mu}=0.9$) (left) and fully accurate signals ($\hat{\mu}=1$) (right). In both panels, $M=20$, $p=0.55$ and $r_1$ is uniform.}
\label{fig:rationalranking_L}
\end{figure}

\section{\label{section: conclusions} Conclusions}

Several decades after the introduction of the Internet and the World Wide Web, there is still a vivid and growing popular and academic debate on the possible impact of digital platforms
on public opinion (e.g.,  Introna and Nissenbaum, 2000; Hargittai, 2004; Rieder, 2005; Hindman, 2009; Sunstein, 2009;
Granka, 2010; Pariser, 2011; Bakshy et al., 2015; Lazer, 2015; Tufekci, 2015). Even among regulators and policymakers, misinformation online ranks high as a key concern, leading some even to call for the direct regulation of online content (e.g., Germany and France have recently proposed laws to combat ``fake news'').\footnote{See ``How do you stop fake news? In Germany, with a law.'' \textit{Washington Post}, April 5 2017. ``Emmanuel Macron promises ban on fake news during elections.'' \textit{The Guardian}, January 3, 2018.}

Unfortunately, to understand and address these issues, it is not enough to just obtain access to the algorithm code used by digital platforms,  as the interplay between ranking algorithms and individual behavior ``yields patterns that are fundamentally emergent'' (Lazer, 2015, p.~1090). In this sense, the theoretical framework we develop seeks to inform and provide some formal guidance to the above debate, by focusing on the interaction between users' search behavior and basic and well-established aspects of ranking algorithms (popularity and personalization).  Our results uncover a rather general property of popularity-based rankings, we call the \textit{advantage of the fewer} ($AOF$) effect. Roughly speaking, it suggests that the smaller the number of websites reporting a given information, the larger the share of traffic directed to those fewer websites. 
Because dubious or particularly controversial information is often carried by a (relatively) small number of websites, the $AOF$ effect may help explain the spread of misinformation, since it shows how being small in number may actually boost overall traffic to such websites.  
Nonetheless, we find that popularity rankings---even with the $AOF$ effect---can have an overall positive effect on asymptotic learning by fostering information aggregation, as long as individuals can, on average, provide sufficiently positive feedback to the ranking algorithm. 
The model further provides insights on a controversial component of the ranking algorithm, namely personalization. While personalized rankings can clearly be efficient for search queries on private value issues (e.g., where to have dinner), we find that for queries on common value issues (e.g., whether or not to vaccinate a child) they can hinder asymptotic learning besides also deepening \textit{belief polarization}.

Understanding the role and effects of ranking algorithms, directly or indirectly used by billions of individuals daily, is a top priority for understanding the functioning of our information society. We view this paper as contributing a first step in this direction. 
\endgroup
\newpage
\noindent{\Large \textbf{References} }

\noindent

\vspace{10pt}
\noindent

\noindent Acemoglu, D. and Ozdaglar, A. (2011). Opinion Dynamics and Learning in Social Networks. \textit{Dynamic Games and Applications}, 1 (1), 3–49.

\vspace{10pt}
\noindent Acemoglu, D. and Ozdaglar, A. and ParandehGheibi, A. (2010). Spread of (mis)information in social networks. \textit{Games and Economic Behavior}, 70, 194–227.

\vspace{10pt}
\noindent Agranov, M. and Ortoleva, P. (2017). Stochastic choice and preferences for randomization. \textit{Journal of Political Economy}, 125 (1), 40–68.

\vspace{10pt}
\noindent  Allcott, H. and Gentzkow, M. (2017). Social media and fake news in the 2016 election. \textit{Journal of Economic Perspectives}, 31 (2), 211–36.

\vspace{10pt}
\noindent  Allcott, H. and Gentzkow, M. and Yu, C. (2018). Trends in the diffusion of misinformation on social media. \textit{arXiv preprint arXiv:1809.05901}.

\vspace{10pt}
\noindent  Azzimonti, M. and Fernandes, M. (2018). Social media networks, fake news, and polarization. \textit{National Bureau of Economic Research}.

\vspace{10pt}
\noindent  Bakshy, E., Messing, S. and Adamic, L. (2015). Exposure to Ideologically Diverse News and Opinion on Facebook. \textit{Science}.

\vspace{10pt}
\noindent  Bar-Gill, S. and Gandal, N. (2017). Online Exploration, Content Choice \& Echo Chambers: An Experiment., \textit{CEPR Discussion Papers}.

\vspace{10pt}
\noindent  Bessi, A., Coletto, M., Davidescu, G. A., Scala, A. and Caldarelli, G. (2015). Science vs Conspiracy: Collective Narratives in the Age of Misinformation. \textit{PloS one}, 10 (2), 2.

\vspace{10pt}
\noindent  Bikhchandani, S., Hirshleifer, D. and Welch, I. (1998). Learning From the Behavior of Others: Conformity, Fads, and Informational Cascades. \textit{The Journal of Economic Perspectives}, pp. 151–170.

\vspace{10pt}
\noindent  Block, H. D. and Marschak, J. (1960). Random orderings and stochastic theories of responses. In I. Olkin, S. Ghurye, W. Hoeffding, W. Madow and H. Mann (eds.), \textit{Contributions to probability and statistics}, vol. 2, Stanford University Press Stanford, CA, pp. 97–132.

\vspace{10pt}
\noindent  Burguet, R., Caminal, R. and Ellman, M. (2015). In Google we Trust? \textit{International Journal of Industrial Organization}, 39, 44–55.

\vspace{10pt}
\noindent  Carvalho, C., Klagge, N. and Moench, E. (2011). The Persistent Effects of a False News Shock. \textit{Journal of Empirical Finance}, 18 (4), 597–615.

\vspace{10pt}
\noindent  Cho, J. and Roy, S. (2004). Impact of search engines on page popularity. In \textit{Proceedings of the 13th international conference on World Wide Web}, ACM, pp. 20–29.

\vspace{10pt}
\noindent  Cho, J. and Roy, S. and Adams, R. E. (2005). Page Quality: In Search of an Unbiased Web Ranking. \textit{SIGMOD}, 14.

\vspace{10pt}
\noindent  De  Corniere,  A.  and  Taylor,  G.  (2014).  Integration  and  Search  Engine  Bias.  \textit{RAND  Journal  of Economics}, 45 (3), 576–597.

\vspace{10pt}
\noindent  Dean,	B.	(2013).	Google’s	200	Ranking	Factors.	\textit{Search	Engine	Journal}.

\vspace{10pt}
\noindent  DeGroot, M. H. (1974). Reaching a Consensus. \textit{Journal of the American Statistical Association}, 69 (345), 118–121.

\vspace{10pt}
\noindent  DellaVigna, S. and Gentzkow, M. (2010). Persuasion: empirical evidence. \textit{Annu. Rev. Econ.}, 2 (1), 643–669.

\vspace{10pt}
\noindent  Demange, G. (2012). On the influence of a ranking system. \textit{Social Choice and Welfare}, 39 (2-3), 431–455.

\vspace{10pt}
\noindent  Demange, G. 	(2014a). Collective Attention and Ranking Methods. \textit{Journal of Dynamics and Games}, 1 (1), 17–43.

\vspace{10pt}
\noindent  Demange, G. 	(2014b). A ranking method based on handicaps. \textit{Theoretical Economics}, 9 (3), 915–942.

\vspace{10pt}
\noindent  DeMarzo, P. M., Vayanos, D. and Zwiebel, J. (2003). Persuasion Bias, Social Influence, and Uni-Dimensional Opinions. \textit{Quarterly Journal of Economics}, 118 (3), 909–968.

\vspace{10pt}
\noindent  Epstein,  R. and Robertson,  R.  E. (2015). The Search Engine Manipulation Effect (SEME) and its Possible Impact on the Outcomes of Elections. \textit{Proceedings of the National Academy of Sciences}, 112 (33), E4512—-E4521.

\vspace{10pt}
\noindent  Eslami,  M.,  Karahalios,  K.,  Sandvig,  C.,  Vaccaro,  K.,  Rickman,  A.,  Hamilton,  K.  and Kirlik, A. (2016). First i like it, then i hide it: Folk theories of social feeds. In \textit{Proceedings of the 2016 cHI conference on human factors in computing systems}, ACM, pp. 2371–2382.

\vspace{10pt}
\noindent  Flaxman, S., Goel, S. and Rao, J. M. (2016). Filter Bubbles, Echo Chambers, and Online News Consumption,  \textit{Public Opinion Quarterly}, 80(S1), pp. 298–320.

\vspace{10pt}
\noindent  Fortunato, S., Flammini, A., Menczer, F. and Vespignani,  A. (2006). Topical  interests and the mitigation of search engine bias.  \textit{Proceedings of the National Academy of Science}, 103 (34), 12684–12689.

\vspace{10pt}
\noindent  Gentzkow, M. and Shapiro, J. (2006). Media Bias and Reputation.  \textit{Journal of Political Economy}, 114 (2), 280–316.

\vspace{10pt}
\noindent  Gentzkow, M. and Shapiro, J. (2010). What drives media slant? evidence from U.S. daily newspapers.  \textit{Econometrica}, 78 (1), 35–71.

\vspace{10pt}
\noindent  Glick, M., Richards, G., Sapozhnikov, M. and Seabright, P. (2011). How Does Page Rank Affect User Choice in Online Search?  \textit{Working Paper}.

\vspace{10pt}
\noindent  Goldman, E. (2006). Search Engine Bias and the Demise of Search Engine Utopianism.  \textit{Yale Journal of Law \& Technology}, pp. 6–8.

\vspace{10pt}
\noindent  Golub, B. and Jackson, M. O. (2010). Na\''{i}ıve Learning in Social Networks and the Wisdom of the Crowds.  \textit{American Economic Journal: Microeconomics}, 2 (1), 112–149.

\vspace{10pt}
\noindent  Granka, L. A. (2010). The Politics of Search: A Decade Retrospective.  \textit{The Information Society}, 26, 364–374.

\vspace{10pt}
\noindent  Grimmelmann, J. (2009). The Google Dilemma.  \textit{New York Law School Law Review}, 53, 939–950.

\vspace{10pt}
\noindent  G\"{u}l,  F.,  Natenzon,  P. and Pesendorfer,  W. (2014). Random choice as behavioral optimization.  \textit{Econometrica}, 82 (5), 1873–1912.

\vspace{10pt}
\noindent  Hagiu, A. and Jullien, B. (2014). Search Diversion and Platform Competition.  \textit{International Journal of Industrial Organization}, 33, 46–80.

\vspace{10pt}
\noindent  Halberstam,  Y.  and  Knight,  B.  (2016).  Homophily,  Group  Size,  and  the  Diffusion  of  Political Information in Social Networks: Evidence from Twitter.\textit{Journal of Public Economics}, 143, pp. 73-88.

\vspace{10pt}
\noindent  Hannak,  A.,  Sapiezynski,  P.,  Kakhki,  A.  M.,  Krishnamurthy,  B.,  Lazer,  D.,  Mislove,  A. and Wilson, C. (2013). Measuring Personalization of Web Search. \textit{Proceedings of the Twenty-Second International World Wide Web Conference (WWW13).}

\vspace{10pt}
\noindent  Hargittai, E. (2004). \textit{The Changing Online Landscape. Community practice in the network society: local action/global interaction.}

\vspace{10pt}
\noindent  Hazan, J. G. (2013). Stop Being Evil: A Proposal for Unbiased Google Search. \textit{Michigan Law Review}, 111 (789), 789–820.

\vspace{10pt}
\noindent  Hindman, M. (2009). \textit{The Myth of Digital Democracy}. Princeton University Press.

\vspace{10pt}
\noindent  Horrigan, J. B. (2006). The Internet as a Resource for News and Information about Science. \textit{Pew Internet \& American Life Project}.

\vspace{10pt}
\noindent  Introna, L. D. and Nissenbaum, H. (2000). Shaping the web: Why the politics of search engines matters. \textit{The information society}, 16 (3), 169–185.

\vspace{10pt}
\noindent  Izquierdo, S. S. and Izquierdo, L. R. (2013). Stochastic Approximation to Understand Simple Simulation Models. \textit{Journal of Statistical Physics}, 151, 254–276.

\vspace{10pt}
\noindent  Jansen, B. J., Booth, D. L. and Spink, A. (2008). Determining the Informational, Navigational, and Transactional Intent of Web Queries. \textit{Information Processing and Management}, (44), 1251–1266.

\vspace{10pt}
\noindent  Kata, A. (2012). Anti-Vaccine Activists, Web 2.0, and the Postmodern Paradigm–An Overview of Tactics and Tropes used Online by the Anti-Vaccination Movement. \textit{Vaccine}, 30 (25), 3778–3789.

\vspace{10pt}
\noindent  Kearney, M. S. and Levine, P. B. (2014). Media Influences on Social Outcomes: The Impact of MTV’s 16 and Pregnant on Teen Childbearing. \textit{National Bureau of Economic Research}.

\vspace{10pt}
\noindent  Kliman-Silver, C., Hannak, A., Lazer, D., Wilson, C. and Mislove, A. (2015).  Location, Location, Location: The Impact of Geolocation on Web Search Personalization. In \textit{Proceedings of the 2015 ACM Conference on Internet Measurement Conference}, ACM, pp. 121–127.

\vspace{10pt}
\noindent  Kulshrestha, J., Eslami, M., Messias, J., Zafar, M. B., Ghosh, S., Gummadi, K. P. and Karahalios, K. (2018). Search bias quantification: investigating political bias in social media and web search. \textit{Information Retrieval Journal}, pp. 1–40.

\vspace{10pt}
\noindent  Lazer, D. (2015). The Rise of the Social Algorithm. \textit{Science}, 348 (6239), 1090–1091.

\vspace{10pt}
\noindent  Luce, R. D. (1959). \textit{Individual choice behavior: A theoretical analysis}. New York: Wiley.

\vspace{10pt}
\noindent  Menczer, F., Fortunato, S., Flammini, A. and Vespignani, A. (2006). Googlearchy or Googleocracy. \textit{IEEE Spectrum Online}.

\vspace{10pt}
\noindent  Mocanu, D., Rossi, L., Zhang, Q., Karsai, M. and Quattrociocchi, W. (2015). Collective Attention in the Age of (Mis) Information. \textit{Computers in Human Behavior}, 51, 1198–1204.

\vspace{10pt}
\noindent  MOZ (2013). 2013 Search Engine Ranking Factors. 

\vspace{10pt}
\noindent  Mullainathan, S. and Shleifer, A. (2005). The Market for News. \textit{American Economic Review}, 95 (4), 1031–1053.

\vspace{10pt}
\noindent  Napoli, P. M. (2015). Social media and the public interest: Governance of news platforms in the realm of individual and algorithmic gatekeepers. \textit{Telecommunications Policy}, 39 (9), 751–760.

\vspace{10pt}
\noindent  Norman, M. F. (1972). \textit{Markov Processes and Learning Models}. Academic Press.

\vspace{10pt}
\noindent  Novarese, M. and Wilson, C. (2013). Being in the Right Place: A Natural Field Experiment on List Position and Consumer Choice. \textit{Working Paper}.

\vspace{10pt}
\noindent  Pan, B., Hembrooke, H., Joachims, T., Lorigo, L., Gay, G. and Granka, L. (2007). In Google We Trust: Users’ Decisions on Rank, Position, and Relevance. \textit{Journal of Computer-Mediated Communication}, 12, 801–823.
 
\vspace{10pt}
\noindent  Pariser, E. (2011). \textit{The Filter Bubble: How the New Personalized Web Is Changing What We Read and How We Think}. Penguin Books.

\vspace{10pt}
\noindent  Piketty, T. (1999). The Information-Aggregation Approach to Political Institutions. \textit{European Economic Review}, 43 (4-6), 791–800.

\vspace{10pt}
\noindent  Prat,  A. and Str\"{o}mberg,  D. (2013). The Political Economy of Mass Media. In: \textit{ Advances  in  Economics and Econometrics: Theory and Applications}, Tenth World Congress.

\vspace{10pt}
\noindent  Putnam, R. D. (2001). \textit{Bowling Alone:  The Collapse and Revival of American Community}. New York: Simon and Schuster.

\vspace{10pt}
\noindent  Rieder, B. (2005). Networked control: Search engines and the symmetry of confidence. \textit{International Review of Information Ethics}, 3 (1), 26–32.

\vspace{10pt}
\noindent  Rieder, B. and Sire, G. (2013). Conflict of Interest and the Incentives to Bias: A Microeconomic Critique of Google’s Tangled Position on the Web. \textit{New Media \& Society}, 0, 1–17.

\vspace{10pt}
\noindent  Shao, C., Ciampaglia, G. L., Flammini, A. and Menczer, F. (2016). Hoaxy: A Platform for Tracking Online Misinformation. In \textit{Proc. Third Workshop on Social News On the Web (WWW SNOW)}.

\vspace{10pt}
\noindent  Str\"{o}mberg, D. (2004). Mass Media Competition, Political Competition, and Public Policy. \textit{Review of Economic Studies}, 11 (1).

\vspace{10pt}
\noindent  Sunstein, C. R. (2009). \textit{Republic.com 2.0.} Princeton University Press.

\vspace{10pt}
\noindent  Taylor, G. (2013). Search Quality and Revenue Cannibalization by Competing Search Engines. \textit{Journal of Economics \& Management Strategy}, 22 (3), 445–467.

\vspace{10pt}
\noindent  Tufekci, Z. (2015). Algorithmic Harms beyond Facebook and Google: Emergent Challenges of Computational Agency. \textit{J. on Telecomm. \& High Tech. L.}, 13, 203.

\vspace{10pt}
\noindent  Vaughn, A. (2014). Google Ranking Factors. SEO Checklist. 

\vspace{10pt}
\noindent  White, R. (2013). Beliefs and Biases in Web Search. In \textit{Proceedings of the 36th international ACM SIGIR conference on Research and development in information retrieval}, ACM, pp. 3–12.

\vspace{10pt}
\noindent  White, R. W. and Horvitz, E. (2015). Belief dynamics and biases in web search. \textit{ACM Transactions on Information Systems (TOIS)}, 33 (4), 18.

\vspace{10pt}
\noindent  Xing, X., Meng, W., Doozan, D., Feamster, N., Lee, W. and Snoeren, A. C. (2014). Exposing Inconsistent Web Search Results with Bobble. In \textit{Passive and Active Measurement}, Springer, pp. 131–140.

\vspace{10pt}
\noindent  Yom-Tov, E., Dumais, S. and Guo, Q. (2013). Promoting Civil Discourse Through Search Engine Diversity. \textit{Social Science Computer Review}, pp. 1–10

\newpage

\appendix

\begingroup
\setstretch{1.06}
\noindent{\Large \textbf{APPENDIX}}

\section{Mean Dynamics Approximation}

\setcounter{table}{0}
\renewcommand{\thetable}{A.\arabic{table}}

\setcounter{figure}{0}
\renewcommand{\thefigure}{A.\arabic{figure}}

\setcounter{equation}{0}
\renewcommand{\theequation}{A.\arabic{equation}}

\label{Appendix-MeanDynamics}

In order to evaluate the actions taken by an agent in the limit as
$N\rightarrow\infty$, we use some techniques of stochastic approximation from
Norman (1972) as exposed in Izquierdo and Izquierdo (2013), which we refer to as the
\emph{mean dynamics approximation}. We here give a brief outline in order to
follow our calculations and proofs, but we refer to the latter two sources for
more details. The basic idea of the approach is to use the expected increments
to evaluate the long run behavior of a dynamic process with stochastic
increments. Rewrite the ranking probabilities as,
\begin{align*}
r_{t,m}  &  =\nu_{t}r_{t-1,m}+(1-\nu_{t})\rho_{t-1,m}\\
&  =\frac{\kappa_{t}}{1+\kappa_{t}}r_{t-1,m}+\frac{1}{1+\kappa_{t}}%
\rho_{t-1,m}\\
&  =r_{t-1,m}+\frac{1}{1+\kappa_{t}}(\rho_{t-1,m}-r_{t-1,m})
\end{align*}
Then, given Equation (\ref{eq:rho_attbias}), for search environments with attention bias ($\alpha$),
\begin{align*}
r_{t,m}&  =r_{t-1,m}+\frac{1}{1+\kappa_{t}}\left(  \frac{(r_{t-1,m})^{\alpha} \cdot v^{*}_{t-1,m}}{\sum_{m^{\prime}}(r_{t-1,m^{\prime}})^{\alpha} \cdot v^{*}_{t-1,m^{\prime}}}-r_{t-1,m}\right)  ,
\end{align*}
where in order to obtain sharper convergence results, we let $\kappa_{t}$ and
hence $\nu_{t}\in(0,1)$ vary with $t$. It is clear that the only stochastic
term is given by the expressions $v^{*}_{t-1,m}$. Replacing these
with their expectations yields the deterministic recursion in $\widehat
{r}_{t,m}$,
\begin{align*}
\widehat{r}_{t,m}  &  =\widehat{r}_{t-1,m}+\frac{1}{1+\kappa_{t}}\left(
\widehat{\rho}_{t-1,m}-\widehat{r}_{t-1,m}\right) ,
\end{align*}
where 
\begin{align} \label{eq:expclickprob}
\widehat{\rho}_{t-1,m}=\mathbb{E}[\rho_{t-1,m}] &=
p \mu \frac{(\widehat{r}_{t-1,m})^{\alpha} \cdot \widehat{v^{*}}_{m}^{00}}{\sum_{m^{\prime}}(\widehat{r}_{t-1,m^{\prime}})^{\alpha} \cdot \widehat{v^{*}}_{m^{\prime}}^{00}}
+ p (1- \mu) \frac{(\widehat{r}_{t-1,m})^{\alpha} \cdot \widehat{v^{*}}_{m}^{01}}{\sum_{m^{\prime}}(\widehat{r}_{t-1,m^{\prime}})^{\alpha} \cdot \widehat{v^{*}}_{m^{\prime}}^{01}} 
\nonumber \\
&+ (1-p) \mu \frac{(\widehat{r}_{t-1,m})^{\alpha} \cdot \widehat{v^{*}}_{m}^{10}}{\sum_{m^{\prime}}(\widehat{r}_{t-1,m^{\prime}})^{\alpha} \cdot \widehat{v^{*}}_{m^{\prime}}^{10}}
+ (1-p) (1-\mu) \frac{(\widehat{r}_{t-1,m})^{\alpha} \cdot \widehat{v^{*}}_{m}^{11}}{\sum_{m^{\prime}}(\widehat{r}_{t-1,m^{\prime}})^{\alpha} \cdot \widehat{v^{*}}_{m^{\prime}}^{11}}
\end{align}
is the expected clicking probability of the individual entering in period $t-1$, and where the expected ranking-free values $\widehat{v^{*}}_{m}^{00}, \widehat{v^{*}}_{m}^{01}, \widehat{v^{*}}_{m}^{10}, \widehat{v^{*}}_{m}^{11}$ are given by the following table, for $L\ne 0, M$:\footnote{In the extreme cases, $L=0$ or $L=M$, clearly, 
$\widehat{v^{*}}_{m}^{00}=\widehat{v^{*}}_{m}^{01}=\widehat{v^{*}}_{m}^{10}=\widehat{v^{*}}_{m}^{11}=0$ $(=1)$ if $m\in L,L=0$ or $m\notin L,L=M$ 
(if $m\in L,L=M$ or $m\notin L,L=0$).}
\begin{center}
\begin{tabular}{r|c|c|c|c|}
          &$\widehat{v^{*}}_{m}^{00}$ &$\widehat{v^{*}}_{m}^{01}$ &$\widehat{v^{*}}_{m}^{10}$ &$\widehat{v^{*}}_{m}^{11}$ \\
\cline{1-5}
$m\in L,y_L \ne y_K$&$\gamma/L$ &$1/L$&0 &$(1-\gamma)/L$ \\ 
\cline{1-5} 
$m\in L,y_L = y_K$&$1/L$&$\gamma/L$&$(1-\gamma)/L$&0 \\
\cline{1-5}
$m\notin L,y_L \ne y_K$&$(1-\gamma)/(M-L)$&$0$&$1/(M-L)$&$\gamma/(M-L)$ \\
\cline{1-5}
$m\notin L,y_L = y_K$&$0$&$(1-\gamma)/(M-L)$&$\gamma/(M-L)$&$1/(M-L)$ \\
\cline{1-5}
\end{tabular} 
\end{center}

\vspace{.1in}

Here $\widehat{v^{*}}_{m}^{00}$ represents the expected probability of an
individual choosing a website $m,$ contingent on having received two correct
signals $x_n=\omega$ and $z_n=y_K$, when $\alpha=0$ (i.e., absent attention bias);
$\widehat{v^{*}}_{m}^{01}$
represents the expected probability of an individual choosing a website $m,$
contingent on having received a correct signal on the state of the world, $x_n=\omega$, and an incorrect signal on the majority, $z_n \ne y_K$, when $\alpha=0$, and analogously for $\widehat{v^{*}}_{m}^{10}$ and $\widehat{v^{*}}_{m}^{11}$.
Importantly, $\widehat{v^{*}}_{m}^{00}, \widehat{v^{*}}_{m}^{01}, \widehat{v^{*}}_{m}^{10}$, and  $\widehat{v^{*}}_{m}^{11}$ are fixed coefficients
that do not vary with $t$. In particular, in order to apply the basic
approximation theorem we assume $\kappa_{t}$ is of the order $O(t)$ so that,
$\frac{1}{1+\kappa_{t}}\rightarrow0$ as $t \rightarrow \infty$, and, to guarantee smoothness and avoid
boundary problems, we assume there exists $\epsilon>0$ such that, each
$\widehat{r}_{t,m}\geq\epsilon$ for all $t,m$. Moreover, replacing
$\widehat{r}_{t-1,m}$ with $x_m$ (and hence the vector $\widehat{r}_{t-1}$ with the vector $x=(x_1, \ldots, x_M$)), 
we obtain a function $g:\Delta_{\epsilon
}(M)\rightarrow\mathbb{R}^{M}$, defined, for $m=1,2,\ldots,M$, by,
\[
g_{m}(x)=\mathbb{E}\left[  \rho_{t-1,m}-r_{t-1,m}\,|\,r_{t-1,m}=x_{m}%
\mbox{ for }m=1,\ldots,M\right]  =\theta_{m}(x)-x_{m},
\]
where $\theta:\Delta_{\epsilon}(M)\rightarrow\mathbb{R}^{M}$ is defined by,
\begin{eqnarray*} 
\theta_{m}(x)&=&
p \mu \frac{(x_{m})^{\alpha} \cdot \widehat{v^{*}}_{m}^{00}}{\sum_{m^{\prime}}(x_{m^{\prime}})^{\alpha} \cdot \widehat{v^{*}}_{m^{\prime}}^{00}}
+ p (1- \mu) \frac{(x_{m})^{\alpha} \cdot \widehat{v^{*}}_{m}^{01}}{\sum_{m^{\prime}}(x_{m^{\prime}})^{\alpha} \cdot \widehat{v^{*}}_{m^{\prime}}^{01}} +
\\
&&+ (1-p) \mu \frac{(x_{m})^{\alpha} \cdot \widehat{v^{*}}_{m}^{10}}{\sum_{m^{\prime}}(x_{m^{\prime}})^{\alpha} \cdot \widehat{v^{*}}_{m^{\prime}}^{10}}
+ (1-p) (1-\mu) \frac{(x_{m})^{\alpha} \cdot \widehat{v^{*}}_{m}^{11}}{\sum_{m^{\prime}}(x_{m^{\prime}})^{\alpha} \cdot \widehat{v^{*}}_{m^{\prime}}^{11}} .
\end{eqnarray*}

Given that the function $g$ is smooth in $x$ on $\Delta_{\epsilon}(M)$, it can
be shown that the expected limit of our stochastic process can be obtained by
solving the ordinary differential equation $\dot{x}=g(x)$. In particular, for
any given initial condition $r_{0}$, there is a unique limit, and for large enough
values of $\kappa$ the stochastic process $r_{t}$ (and hence also $\rho_{t}$) tends to follow the unique
solution trajectory and linger around the asymptotically stable limit point 
of the differential equation $\dot{x}=g(x)$. (See Izquierdo and Izquierdo (2013), results (i) and (iii) on p.~261).

\section{Proofs}

\setcounter{table}{0}
\renewcommand{\thetable}{B.\arabic{table}}

\setcounter{figure}{0}
\renewcommand{\thefigure}{B.\arabic{figure}}

\setcounter{equation}{0}
\renewcommand{\theequation}{B.\arabic{equation}}

\label{Appendix-Proofs} \setlength{\parindent}{0mm} \setlength{\parskip}{0mm}


\textbf{Proof of Proposition~\ref{prop:$AOF$}}. 
To avoid duplication, we prove the proposition directly for the case of $\alpha >0$ mentioned in Section~\ref{sect:attbias}, which includes the case of $\alpha=1$, stated in Proposition~\ref{prop:$AOF$}, as a special case.

We begin by characterizing the limit ranking probabilities using the differential equation from Appendix~\ref{Appendix-MeanDynamics}. 
Since the initial ranking is uniform, we have that the expected ranking probabilities are equal for websites with the same signal, that is, $\widehat{r}_{n,m}=\widehat{r}_{n,m'}$ for any two websites $m, m'$ with $y_{m}=y_{m'}$. Since we consider partitions of $M$ into sets $J$ and $M\backslash J$, and $\rho_{n,J}+\rho_{n,M\backslash J}=1$, it suffices to check the case $J=L$. 
To simplify notation, let $x=\widehat{r}_{n,L}=L \cdot \widehat{r}_{n,m}$ for $m \in L$ be the total expected ranking probability for the websites in $L$, so that $1-x$ is the total expected ranking probability for the remaining websites in $M \setminus L$. 
Using the expressions for $\widehat{v^{*}}_{m}^{00},\widehat{v^{*}}_{m}^{01},\widehat{v^{*}}_{m}^{10},\widehat{v^{*}}_{m}^{11}$ from the table in  Appendix~\ref{Appendix-MeanDynamics}, and assuming that $0\le x \le 1$, 
we have that the equations defining the limit probabilities can be reduced to two equations of the form:
{\small 
\begin{align}  \label{eqs:solsmin}
H_L^{minority}(x; \alpha, \mu, \gamma, p) & \equiv \theta_L^{minority} (x; \alpha,  \mu, \gamma, p) - x = 0,  \mbox{ for } 1 \le L \le \frac{M-1}{2} 
 \\ \label{eqs:solsmaj}
H_L^{majority}(y; \alpha, \mu, \gamma, p) & \equiv \theta_L^{majority} (y; \alpha, \mu, \gamma, p) - y = 0, \mbox{ for }  \frac{M-1}{2} \le L \le M-1, 
\end{align}
where $\theta_L^{minority}$ and $\theta_L^{majority}$ are defined as:
\begin{eqnarray}
\label{eq:thetamin}
\theta_{L}^{minority} (x; \alpha, \mu, \gamma, p) =
p(1-\mu)
+\frac{p \mu \gamma \left(\frac{x}{L}\right)^{\alpha}}{\gamma \left(\frac{x}{L}\right)^{\alpha}+(1-\gamma)\left(\frac{1-x}{M-L}\right)^{\alpha}}
+\frac{(1-p)(1-\mu)(1- \gamma) \left(\frac{x}{L}\right)^{\alpha}}{(1-\gamma) \left(\frac{x}{L}\right)^{\alpha}+\gamma\left(\frac{1-x}{M-L}\right)^{\alpha}} \end{eqnarray}
 and
\begin{eqnarray}
\label{eq:thetamaj}
\theta_{L}^{majority} (y; \alpha, \mu, \gamma, p) =
p\mu 
+ \frac{p(1-\mu)\gamma \left(\frac{y}{L}\right)^{\alpha}}{\gamma \left(\frac{y}{L}\right)^{\alpha}+(1-\gamma) \left(\frac{1-y}{M-L}\right)^{\alpha}} 
+ \frac{(1-p) \mu (1-\gamma)\left(\frac{y}{L}\right)^{\alpha}}{(1-\gamma) \left(\frac{y}{L}\right)^{\alpha}+\gamma \left(\frac{1-y}{M-L}\right)^{\alpha}} 
\end{eqnarray}
}It is easy to check that at $x=0$ and $x=1$, we have, respectively,
{\small
\[ 0 \le p(1-\mu) = \theta_{L}^{minority} (0; \alpha, \mu, \gamma, p) < \theta_{L}^{minority} (1; \alpha, \mu, \gamma, p)=p+(1-p)(1-\mu) < 1, \] 
}and similarly, at $y=0$ and $y=1$,
{\small
\[ 0 <  p\mu = \theta_{L}^{majority} (0; \alpha, \mu, \gamma, p) < \theta_{L}^{majority} (1; \alpha, \mu, \gamma, p)=p+ (1-p)\mu \le 1. \]
}In particular, $\theta_{L}^{minority}$ starts at or above the $x$-function at $x=0$ and ends below the $x$-function at $x=1$. Similarly, $\theta_{L}^{majority}$ starts above the $y$-function at $y=0$ and ends below the $y$-function at $y=1$. In Figure~\ref{fig:thetamaj}, we plot $\theta_{L}^{majority}$ for different values of $L$ and for $\alpha=1$ on the panel on the left and for $\alpha=4$ on the panel on the right. 
For most parameter values of interest (i.e., while $\alpha$ not too large) there is a unique interior solution to both Equations~(\ref{eqs:solsmin}) and ~(\ref{eqs:solsmaj}); this is the situation depicted on the left panel. However, in general there can be multiple solutions; as depicted in the right panel. Importantly, the solutions of interest are the ones where the functions $\theta_{L}^{minority}$ and $\theta_{L}^{majority}$ intersect the $x$- and $y$-functions from above.\footnote{As $\alpha \rightarrow \infty$, the stable solutions converge to the bounds $p(1-\mu)$ and $p+(1-p)(1-\mu)$ for $\theta_{L}^{minority}$ and to $p(1-\mu)$ and $p+ (1-p)\mu$ for $\theta_{L}^{majority}$, where both functions, $\theta_{L}^{minority}$ and $\theta_{L}^{majority}$, become arbitrarily flat. Because these solutions do not depend on $L$, this implies that the $AOF$ effect tends to zero in the limit. On the other hand, the unstable solutions converge to $\frac{L}{M}$, where both $\theta_{L}^{minority}$ and $\theta_{L}^{majority}$ become arbitrarily steep.} 
This is because, the process can only converge to those due to the fact that while clicking probabilities ($\theta_{L}^{minority}$ or $\theta_{L}^{majority}$) are above the respective ranking probabilities ($x$ or $y$) then the ranking probabilities will tend to increase and this will continue until the solution (at the intersection) is reached. Similarly, when clicking probabilities are below the respective ranking probabilities, then the ranking probabilities will tend to decrease until the solution is reached. In particular, only the two solutions with arrows are relevant in the panel on the right. The other interior solutions without an arrow (where the $x$- and $y$-functions are intersected from below) are unstable and are never reached by our dynamic process.\footnote{Even if the process were to start at such a solution, any extra click by an individual will move the ranking probability to the left (or to the right) thereby leading to a situation where the clicking probability is below the ranking probability, leading the ranking probability to decrease and move further to the left, again until a stable solution with an arrow is reached (or similarly, if the extra click increases the ranking probability, this will lead to a situation where the clicking probability is above the ranking probability, leading the ranking probability to increase and move further right until it also reaches a stable solution with an arrow).}
\begin{figure}
\par
\begin{center}
\includegraphics[width=6.5cm]{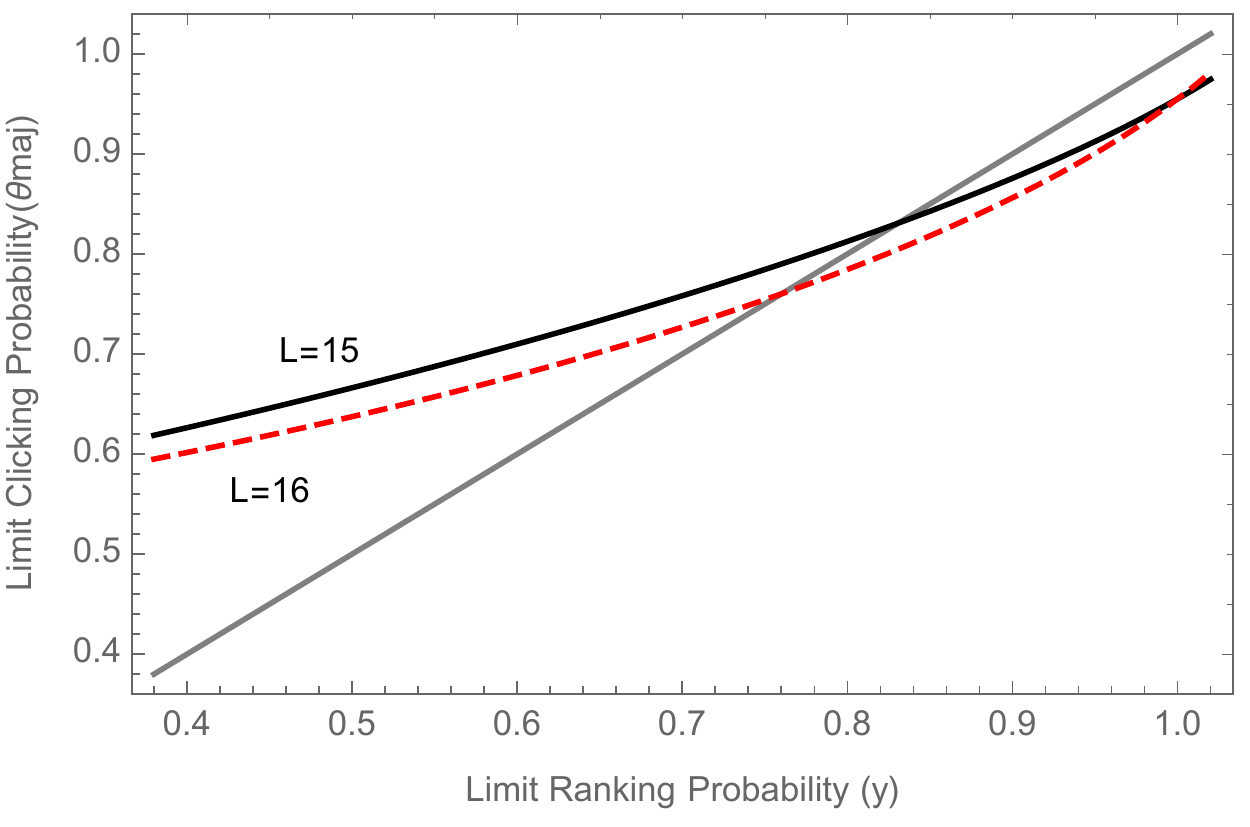} \hspace{.1in}
\includegraphics[width=6.5cm]{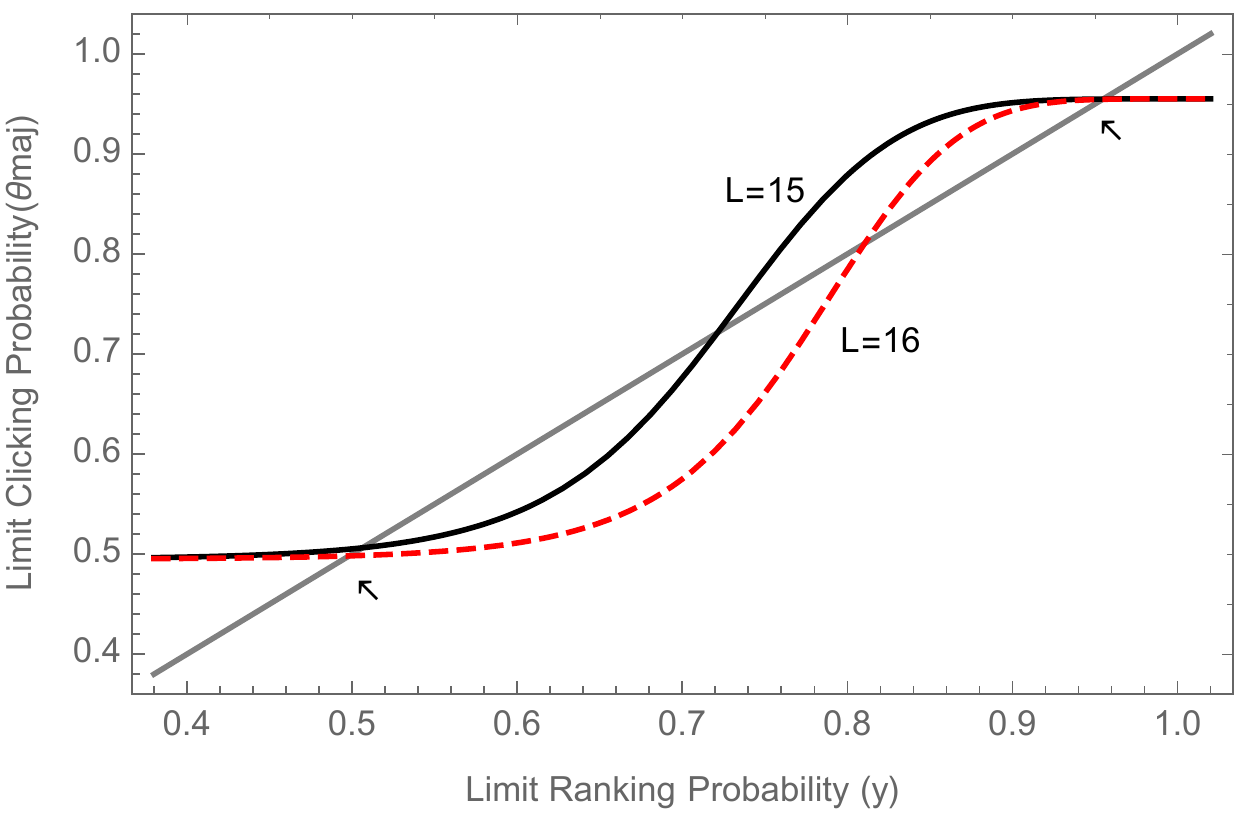} \hspace{.1in}
\end{center}
\par
\vspace{-15pt}
\caption{\footnotesize Limit clicking probability ($\theta_{L}^{majority}$) as a function of the limit ranking probability ($y$) for different values of $L$ and for $\alpha=1$ (left) and $\alpha=4$ (right). In both panels, $M=20$, $\mu=0.9$, $\gamma=0.33$, $p=0.55$, and $r_1$ is uniform.}
\label{fig:thetamaj}
\end{figure}
The key part of the proof is to show that the (stable) solutions $x=x_L^{minority}$ and $y=y_L^{majority}$ to the two equations just above are decreasing in $L$ in the corresponding ranges. 
Since at the solutions we have $H_L^{minority}(x_L^{minority}; \alpha, \mu, \gamma, p)$ $=0$ and $H_L^{majority}(y_L^{majority}; \alpha, \mu, \gamma, p)=0$,  we can write the first derivatives with respect to $L$ as:
{\small
\begin{align*}
\frac{dx_L^{minority}}{dL} &= - \frac{\partial H_L^{minority}(x_L^{minority}; \alpha, \mu, \gamma, p)/\partial L}{\partial H_L^{minority}(x_L^{minority}; \alpha, \gamma, p)/\partial x} =  \frac{\partial \theta_{L}^{minority}/\partial L}{1 - \partial \theta_{L}^{minority}/\partial x}, \\
\frac{dy_L^{majority}}{dL} &= - \frac{\partial H_L^{majority}(y_L^{majority}; \alpha, \mu, \gamma, p)/\partial L}{\partial H_L^{majority}(y_L^{majority}; \alpha, \gamma, p)/\partial y}  = \frac{\partial \theta_{L}^{majority}/\partial L}{1 - \partial \theta_{L}^{majority}/\partial y}.
\end{align*}
}To see that both are negaitive, we need to check that $\frac{\partial \theta_{L}^{minority}}{\partial L} \le 0$, $\frac{\partial \theta_{L}^{majority}}{\partial L} \le 0$ and $\frac{\partial \theta_{L}^{minority}}{\partial x}<1$, $\frac{\partial \theta_{L}^{majority}}{\partial y}<1$.
Straighforward computations yield:
{\scriptsize
\begin{eqnarray*} 
\frac{\partial \theta_{L}^{minority}}{\partial L} &=& 
- \frac{\alpha \gamma (1-\gamma) M \left( \frac{x(1-x)}{L(M-L)} \right)^{\alpha}  }{L(M-L)}  
\left( \frac{p \mu}{ \left( \gamma  \left( \frac{x}{L}\right)^{\alpha} + (1-\gamma) \left( \frac{1-x}{M-L}\right)^{\alpha} \right)^{2}} + 
\frac{(1-p)(1-\mu)}{\left( (1- \gamma)  \left( \frac{x}{L}\right)^{\alpha} + \gamma \left( \frac{1-x}{M-L}\right)^{\alpha} \right)^{2}} 
\right) 
\\
\\
\frac{\partial \theta_{L}^{majority}}{\partial L} &=& 
- \frac{\alpha \gamma (1-\gamma) M  \left( \frac{y(1-y)}{L(M-L)} \right)^{\alpha}  }{L(M-L)}  
\left( \frac{p (1-\mu)}{\left( \gamma  \left( \frac{y}{L}\right)^{\alpha} + (1-\gamma) \left( \frac{1-y}{M-L}\right)^{\alpha} \right)^{2}} + 
\frac{(1-p)\mu}{\left( (1-\gamma)  \left( \frac{x}{L}\right)^{\alpha} + \gamma \left( \frac{1-y}{M-L}\right)^{\alpha} \right)^{2}} 
\right) 
 \, , 
\end{eqnarray*}
}which are both clearly non-positive always, and negative for interior values $\alpha >0$ and $0<\gamma<1$ and $0<x<1$. 
Also,
{\scriptsize
\begin{eqnarray*} 
\frac{\partial \theta_{L}^{minority}}{\partial x} &=& 
 \frac{\alpha \gamma (1-\gamma)  \left( \frac{x(1-x)}{L(M-L)} \right)^{\alpha}  }{x(1-x)}  
\left( \frac{p \mu}{ \left( \gamma  \left( \frac{x}{L}\right)^{\alpha} + (1-\gamma) \left( \frac{1-x}{M-L}\right)^{\alpha} \right)^{2}} + 
\frac{(1-p)(1-\mu)}{\left( (1- \gamma)  \left( \frac{x}{L}\right)^{\alpha} + \gamma \left( \frac{1-x}{M-L}\right)^{\alpha} \right)^{2}} 
\right) 
\\
\\
\frac{\partial \theta_{L}^{majority}}{\partial y} &=& 
\frac{\alpha \gamma (1-\gamma)  \left( \frac{y(1-y)}{L(M-L)} \right)^{\alpha}  }{y(1-y)}  
\left( \frac{p (1-\mu)}{\left( \gamma  \left( \frac{y}{L}\right)^{\alpha} + (1-\gamma) \left( \frac{1-y}{M-L}\right)^{\alpha} \right)^{2}} + 
\frac{(1-p)\mu}{\left( (1-\gamma)  \left( \frac{y}{L}\right)^{\alpha} + \gamma \left( \frac{1-y}{M-L}\right)^{\alpha} \right)^{2}} 
\right) 
 \, .
\end{eqnarray*}
}Clearly these derivatives are all non-negative. However, we now show that they are strictly less than one or the stable solutions. 
First,  it can be checked that, for any limit clicking probability ($0\le x$ (or $y$) $\le 1$) and for any other parameter values of the model,
{\small
\[ 0 \le 
 \frac{ \gamma  \left( \frac{x}{L} \right)^{\alpha}}{\gamma  \left( \frac{x}{L}\right)^{\alpha} + (1-\gamma) \left( \frac{1-x}{M-L}\right)^{\alpha}}, 
 \frac{ (1-\gamma)  \left( \frac{1-x}{M-L} \right)^{\alpha}}{\gamma  \left( \frac{x}{L}\right)^{\alpha} + (1-\gamma) \left( \frac{1-x}{M-L}\right)^{\alpha}} ,
 \frac{ \gamma (1-\gamma)  \left( \frac{x(1-x)}{L(M-L)} \right)^{\alpha}}{\left( (1- \gamma)  \left( \frac{x}{L}\right)^{\alpha} + \gamma \left( \frac{1-x}{M-L}\right)^{\alpha} \right)^{2}}  
\le 1, \]
}which, as already seen in Figure~\ref{fig:thetamaj} above, implies that for values of $\alpha \ge 1$ there may be more than one solution to Equations~(\ref{eqs:solsmin}) and~(\ref{eqs:solsmaj}). Moreover, some of the solutions may have a slope greater or equal to one. However, as explained above, only those solutions are stable which intersect the functions $x$ (or $y$) from above. Since both $x$ and $y$ clearly have slope equal to 1 everywhere this implies that both $\theta_{L}^{minority}$ and $\theta_{L}^{majority}$ must have a slope less than one at the stable solutions and hence must satisfy $\frac{\partial \theta_{L}^{minority}}{\partial x}<1$ and $\frac{\partial \theta_{L}^{majority}}{\partial y}<1$. 
This then implies that the respective solutions $x_{L}^{minority}$ and $y_{L}^{majority}$ will satisfy $\frac{\partial \widehat{r}_{N,L}}{\partial L} \le 0$ as well as $\frac{\partial \widehat{\rho}_{N,L}}{\partial L} \le 0$ on the relevant ranges and for $N$ sufficiently large, which will also be negative for interior values of the parameters. This shows $AOF$ for both minority and majority outlets with correct signal. To see that $AOF$ also applies to outlets $J \subset M$ with incorrect signal, notice that, if increasing $L$, decreases $\widehat{\rho}_{N,L}$, then, since the total number of outlets $M$ is fixed and $\widehat{\rho}_{N,L} + \widehat{\rho}_{N,M \backslash L} = 1$, this readily implies that decreasing $M-L$, increases $\widehat{\rho}_{N,M \backslash L}$. Therefore, given the $AOF$ for outlets $L$ with the correct signal and taking $J=M \backslash L$ also shows that $AOF$ applies to minority and majority outlets with incorrect signal. This concludes the proof of Proposition~\ref{prop:$AOF$}. 
\hfill$\Box$

\bigskip
\textbf{Proof of Corollaries~\ref{cor:rankpro} and \ref{cor:rankpro_alpha}}. 
Again, to avoid duplication, we prove directly Corollary~\ref{cor:rankpro_alpha} stated in Section~\ref{sect:attbias}, which includes the case of $\alpha=1$, stated in Corollary~\ref{cor:rankpro}, as a special case.

Consider Equation (\ref{eq:rho_attbias}). Then, when $\alpha=0$ it is easy to see that:
{\small
\begin{equation*}
\rho_{N,L} =
\left\{
\begin{array}
[c]{cl}
0 & \text{ if }L=0 \\ \vspace{.02in}
(1 - \mu) p + (1 - \mu) (1 - p) (1 - \gamma) + \mu p \gamma & \text{ if }1 \le L=\frac{M-1}{2} \\ \vspace{.02in}
\mu p + \mu (1 - p) (1 - \gamma) + (1-\mu) p \gamma & \text{ if }\frac{M+1}{2} \le L \le M-1 \\ \vspace{.02in}
1 & \text{ if }L=M \, ,
\end{array} 
\right. 
\end{equation*}
} which can be checked is monotonically increasing in $L$ for $\mu \ge \frac{1}{2}$.
Now suppose $\alpha>0$. To see the non-monotonicity, notice again that at $L=0$, we have $\rho_{n,L}=0$, and at $L=M-1$, we have $\rho_{n,L}=1$, for all $n$, so that $\rho_{n,L}$ can only increase at $L=0, M-1$. For the case $L=\frac{M-1}{2}$, ($M-L=\frac{M+1}{2}$) note that the difference $\theta_{\frac{M+1}{2}}^{majority} (y; \alpha, \mu, \gamma, p) -\theta_{\frac{M-1}{2}}^{minority} (x; \alpha, \mu, \gamma, p)$ can be written as: 
{\scriptsize
\begin{align*} 
\theta_{M-L}^{majority} -\theta_{L}^{minority} =& (2\mu - 1) \left( p \left( 1 -  \frac{\gamma \left(\frac{x}{L}\right)^{\alpha}}{\gamma \left(\frac{x}{L}\right)^{\alpha}+(1-\gamma) \left(\frac{1-x}{M-L}\right)^{\alpha}} \right) + (1-p) \frac{(1-\gamma)\left(\frac{x}{L}\right)^{\alpha}}{(1-\gamma) \left(\frac{x}{L}\right)^{\alpha}+\gamma \left(\frac{1-x}{M-L}\right)^{\alpha}}  \right)  \\
&+ p(1-\mu) \left( \frac{\gamma \left(\frac{y}{M-L}\right)^{\alpha}}{\gamma \left(\frac{y}{M-L}\right)^{\alpha}+(1-\gamma) \left(\frac{1-y}{L}\right)^{\alpha}}- \frac{\gamma \left(\frac{x}{L}\right)^{\alpha}}{\gamma \left(\frac{x}{L}\right)^{\alpha}+(1-\gamma) \left(\frac{1-x}{M-L}\right)^{\alpha}}\right) \\
&+(1-p)\mu \left(  \frac{(1-\gamma)\left(\frac{y}{M-L}\right)^{\alpha}}{(1-\gamma) \left(\frac{y}{M-L}\right)^{\alpha}+\gamma \left(\frac{1-y}{L}\right)^{\alpha}}  -  \frac{(1-\gamma)\left(\frac{x}{L}\right)^{\alpha}}{(1-\gamma) \left(\frac{x}{L}\right)^{\alpha}+\gamma \left(\frac{1-x}{M-L}\right)^{\alpha}} \right) ,
\end{align*}
}where, at the relevant solutions, $x=x_L^{majority}$ and $y=y_{M-L}^{majority}$, we have $y \ge x$, such that the expressions in the second and third lines are both non-negative.
And since $\mu > \frac{p}{q} > \frac{1}{2}$, the overall difference ($\theta_{M-L}^{majority} -\theta_{L}^{minority}$) is non-negative. 
In all other cases, that is, interior values $L \ne 0, \frac{M-1}{2}, M-1$, $\rho_{N,L}$ is decreasing in $L$ by Proposition~\ref{prop:$AOF$}, for sufficiently large $N$, and hence also in the limit, for $\rho_{\infty,L}={\cal P}_L$. \hfill$\Box$

\bigskip


\textbf{Proof of Proposition~\ref{prop:welfare}}. 
Also here, to avoid duplication, we prove the proposition directly for the case of $\alpha \ge 0$ mentioned in Section~\ref{sect:attbias}, which includes the case of $\alpha=1$, stated in Proposition~\ref{prop:welfare}, as a special case.

From the definition of ${\cal P}$ in Equation~(\ref{eq: welfare}), we have:
\[ \frac{\partial {\cal P}}{\partial z} = 
\sum_{L=0}^{M}\binom{M}{L}q^{L}(1-q)^{M-L}  \frac{\partial {\cal P}_L}{\partial z} , 
\] 
for any given variable $z$. Hence, we can evaluate the comparative statics by looking at the effects on the interim efficiency, 
($ \frac{\partial {\cal P}_L}{\partial z} $).
So to see that ${\cal P}$ is weakly increasing in $p$, since the initial ranking is uniform, we can use the same reasoning as in the proof of Proposition~\ref{prop:$AOF$}. 
In particular, it suffices to consider the following derivatives for $\theta_{L}^{minority}$ and $\theta_{L}^{majority}$ defined respectively in Equations~(\ref{eq:thetamin}) and~(\ref{eq:thetamaj}) above:
{\scriptsize
\begin{eqnarray*}
\frac{\partial \theta_{L}^{minority}}{\partial p} &=& 
(1-\mu) \left( 1 - \frac{(1-\gamma)  \left( \frac{x}{L}\right)^{\alpha}}{\gamma \left( \frac{1-x}{M-L}\right)^{\alpha} + (1-\gamma)  \left( \frac{x}{L}\right)^{\alpha}} \right)
+ \mu \frac{\gamma  \left( \frac{x}{L}\right)^{\alpha}}{\gamma \left( \frac{x}{L}\right)^{\alpha} + (1-\gamma)  \left( \frac{1-x}{M-L}\right)^{\alpha}}  \ge 0 \, , \\
\frac{\partial \theta_{L}^{majority}}{\partial p} &=& 
\mu \left( 1-\frac{(1-\gamma)  \left( \frac{y}{L}\right)^{\alpha}}{\gamma \left( \frac{1-y}{M-L}\right)^{\alpha} + (1-\gamma)  \left( \frac{x}{L}\right)^{\alpha}} \right) 
+ (1-\mu) \frac{\gamma  \left( \frac{y}{L}\right)^{\alpha}}{\gamma \left( \frac{y}{L}\right)^{\alpha} + (1-\gamma)  \left( \frac{1-y}{M-L}\right)^{\alpha}}
\ge 0 \, .
\end{eqnarray*}
}This implies that $\rho_{\infty,L}={\cal P}_L$ is increasing in $p$ for all values of $L$ and hence so is ${\cal P}$. To see that ${\cal P}$ is increasing in $\mu$, notice that:
{\scriptsize
\begin{eqnarray*}
\frac{\partial \theta_{L}^{minority}}{\partial \mu} &=& 
- p \left( 1 -  \frac{\gamma  \left( \frac{x}{L}\right)^{\alpha}}{\gamma \left( \frac{x}{L}\right)^{\alpha} + (1-\gamma)  \left( \frac{1-x}{M-L}\right)^{\alpha}} \right) 
- (1-p)  \frac{(1-\gamma)  \left( \frac{x}{L}\right)^{\alpha}}{\gamma \left( \frac{1-x}{M-L}\right)^{\alpha} + (1-\gamma)  \left( \frac{x}{L}\right)^{\alpha}} 
\le 0 \, , \\
\frac{\partial \theta_{L}^{majority}}{\partial \mu} &=& 
p \left( 1 -  \frac{\gamma  \left( \frac{y}{L}\right)^{\alpha}}{\gamma \left( \frac{y}{L}\right)^{\alpha} + (1-\gamma)  \left( \frac{1-y}{M-L}\right)^{\alpha}} \right) 
+ (1-p) \frac{(1-\gamma)  \left( \frac{y}{L}\right)^{\alpha}}{\gamma \left( \frac{1-y}{M-L}\right)^{\alpha} + (1-\gamma)  \left( \frac{y}{L}\right)^{\alpha}} 
\ge 0 \, .
\end{eqnarray*}
}It can be further checked that for $\mu q > p > \frac{1}{2}$, the positive effect of when the websites in $L$ form a majority outweighs the negative effect of when they form a minority. To see this, we can rewrite the above derivatives as:
{\small 
\[
\frac{\partial \theta_{L}^{minority}}{\partial \mu} =
-\theta_{L |_{\mu=1}}^{minority} + \theta_{L |_{\mu=0}}^{minority} , \hspace{.2in}
\frac{\partial \theta_{L}^{majority}}{\partial \mu} = 
\theta_{L |_{\mu=1}}^{majority} - \theta_{L |_{\mu=0}}^{majority} \, .
\]
}Moreover, it can be checked that, at the relevant solutions $x=x_L^{minority}$ and $y=y_{M-L}^{majority}$, taking again as the relevant number of outlets with correct signal, $L$ and $M-L$ for the minority and majority case respectively, we have, 
\[ \theta_{M-L |_{\mu=1}}^{majority} (y) -  \theta_{L |_{\mu=1}}^{minority} (x) \ge 0 , \hspace{.2in} 
\theta_{M-L |_{\mu=0}}^{majority} (y) - \theta_{L |_{\mu=0}}^{minority} (x) \le 0 . \]
Together with the above equations, we can obtain the sign of $\frac{\partial {\cal P}}{\partial \mu}$ from:
{\scriptsize
\begin{align*}  
\sum_{L=1}^{\frac{M-1}{2}} & \binom{M}{L}q^{L}(1- q)^{M-L} \frac{\partial \theta_{L}^{minority}}{\partial \mu} 
+ \sum_{L=\frac{M+1}{2}}^{M-1}\binom{M}{L}q^{L}(1-q)^{M-L} \frac{\partial \theta_{L}^{majority}}{\partial \mu}  \\
 & =  \sum_{L=1}^{\frac{M-1}{2}}\binom{M}{L}q^{L}(1-q)^{M-L} (-\theta_{L |_{\mu=1}}^{minority} + \theta_{L |_{\mu=0}}^{minority}) 
+ \sum_{L=\frac{M+1}{2}}^{M-1}\binom{M}{L}q^{L}(1-q)^{M-L} (\theta_{L |_{\mu=1}}^{majority} - \theta_{L |_{\mu=0}}^{majority}) \\
 & = \sum_{L=1}^{\frac{M-1}{2}}\binom{M}{L}q^{L}(1-q)^{M-L} (-\theta_{L |_{\mu=1}}^{minority} + \theta_{L |_{\mu=0}}^{minority}) 
+ \sum_{L=1}^{\frac{M-1}{2}} \binom{M}{M-L}q^{M-L}(1-q)^{L} (\theta_{M-L |_{\mu=1}}^{majority} - \theta_{M-L |_{\mu=0}}^{majority}) \\
& \ge \sum_{L=1}^{\frac{M-1}{2}}\binom{M}{L}q^{L}(1-q)^{M-L} \left((\theta_{M-L |_{\mu=1}}^{majority} -\theta_{L |_{\mu=1}}^{minority}) 
- (\theta_{M-L |_{\mu=0}}^{majority} -  \theta_{L |_{\mu=0}}^{minority}) \right)  \ge  0,
\end{align*} 
}for $\frac{1}{2} < p < q$, which in turn implies that $\frac{\partial {\cal P}}{\partial \mu} \ge 0$.

To see point 3.~that ${\cal P}$ is decreasing in $\gamma$, provided $p$ not too large, $\mu$ not too small, notice that:
{\small
\begin{align*}
\frac{\partial \theta_{L}^{minority}}{\partial \gamma} =& 
\frac{p \mu  \left( \frac{x(1-x)}{L(M-L)} \right)^{\alpha}  }{\left( \gamma \left( \frac{x}{L}\right)^{\alpha} + (1-\gamma) \left(\frac{1-x}{M-L}\right)^{\alpha} \right)^{2} } 
-   \frac{(1- p)(1- \mu)  \left( \frac{x(1-x)}{L(M-L)} \right)^{\alpha}  }{\left( (1-\gamma) \left( \frac{x}{L}\right)^{\alpha} + \gamma \left(\frac{1-x}{M-L}\right)^{\alpha} \right)^{2} }    \ge 0 \, , \\
\frac{\partial \theta_{L}^{majority}}{\partial \gamma} = &
-\frac{(1-p) \mu  \left( \frac{y(1-y)}{L(M-L)} \right)^{\alpha}  }{\left( (1-\gamma) \left( \frac{y}{L}\right)^{\alpha} + \gamma   \left( \frac{1-y}{M-L}\right)^{\alpha} \right)^{2} } + 
\frac{p (1- \mu)  \left( \frac{y(1-y)}{L(M-L)} \right)^{\alpha}  }{\left( \gamma \left( \frac{y}{L}\right)^{\alpha} + (1-\gamma) \left(\frac{1-y}{M-L}\right)^{\alpha} \right)^{2} } 
\le 0 \, .
\end{align*}
}It can be checked that when $p = \frac{1}{2}$, the sum of the solution to Equation~(\ref{eq:thetamin}) at $L$ and the solution to Equation~(\ref{eq:thetamaj}) at $M-L$ sum to one. Thus, if $x=x_L^{minority}$ solves Equation~(\ref{eq:thetamin}) at $L$, then $y_{M-L}^{majority}=1-x_L^{minority}$ solves Equation~(\ref{eq:thetamaj}) at $M-L$. This then implies that, again at $p \approx \frac{1}{2}$,  
{\small
\[ 
 \left. \frac{\partial \theta_{M-L}^{majority}}{\partial \gamma} \right|_{1-x}
\approx
- \frac{(1-p)  \mu \left( \frac{x(1-x)}{L(M-L)} \right)^{\alpha}  }{\left( \gamma \left( \frac{x}{L}\right)^{\alpha} + (1-\gamma)   \left( \frac{1-x}{M-L}\right)^{\alpha} \right)^{2} } 
+ \frac{p (1- \mu)  \left( \frac{x(1-x)}{L(M-L)} \right)^{\alpha}  }{\left( (1-\gamma) \left( \frac{x}{L}\right)^{\alpha} + \gamma \left(\frac{1-x}{M-L}\right)^{\alpha} \right)^{2} } 
= - \left. \frac{\partial \theta_{L}^{minority}}{\partial \gamma} \right|_{x} . \]
}Summing these up, yields as the total effect on ex ante efficiency, 
{\small
\begin{align*}
\sum_{L=1}^{\frac{M-1}{2}} \binom{M}{L}q^{L}(1&-q)^{M-L}  \frac{\partial \theta_{L}^{minority}}{\partial \gamma} 
- \sum_{L=1}^{\frac{M-1}{2}} \binom{M}{M-L}q^{M-L}(1-q)^{L}  \frac{\partial \theta_{L}^{minority}}{\partial \gamma} \\
 = & \, \, \sum_{L=1}^{\frac{M-1}{2}} \binom{M}{L} \left(q^{L}(1-q)^{M-L} - q^{M-L}(1-q)^{L}  \right) \frac{\partial \theta_{L}^{minority}}{\partial \gamma}  
\le 0, 
\end{align*}
}since $q > \frac{1}{2}$, which in turn implies that $\frac{\partial {\cal P}}{\partial \gamma} \le 0$. 
But because there are always cases $\frac{M+1}{2} \le L \le M-1$, where solutions to Equation~(\ref{eq:thetamaj}) are interior,\footnote{For example, solutions always satisfy $1-\mu<x_L^{minority},y_L^{majority}<\mu$, and are interior whenever $\mu \ne 1$.} implying $\frac{\partial \theta_{L}^{majority}}{\partial \gamma}<0$ for some $L$, summing up actually implies that $\frac{\partial {\cal P}}{\partial \gamma} < 0$. This in turn allows us to extend the decreasing effect of $\gamma$ on  ${\cal P}$ to a neighborhood $[\overline{\mu} , 1] \times [ \frac{1}{2}, \overline{p}]$, for some $\overline{\mu} < 1$, $\overline{p} > \frac{1}{2}$, that exist given the continuity (in fact, linearity) of the derivatives $\frac{\partial \theta_{L}^{minority}}{\partial \gamma}$ and $\frac{\partial \theta_{L}^{majority}}{\partial \gamma}$ in both $\mu$ and $p$.

The fact that ${\cal P}$ is both increasing and decreasing in $q$ is easy to check by means of examples. As illustrated in Figure~\ref{fig:eff_alpha_q}, ${\cal P}$ is increasing (decreasing) in $q$ for low (high) values of $q$. 
\hfill$\Box$

\bigskip


\textbf{Proof of Proposition~\ref{cor:randomrank}} 
Assume first that $\mu=1$. 
Using the proof of Proposition~\ref{prop:$AOF$}, it can be checked that, for $\alpha=0$, the solutions to Equations~(\ref{eq:thetamin}) and~(\ref{eq:thetamaj}) take the form, respectively,
$x_{L|\alpha=0}^{minority} = \gamma p , \,  y_{L|\alpha=0}^{majority} = 1- \gamma (1-p)$, 
whereas, for $\alpha=1$, they take the form: 
{\small 
\[ 
x_{L|\alpha=1}^{minority} = 
\left\{
\begin{array}
[c]{cl}
\frac{(1-\gamma) L - \gamma p (M-L)}{\gamma M - L}  &\text{ if }L< \frac{\gamma p M}{1-(1-\gamma)p} \\ \vspace{.02in}
0  &\text{ if }L \ge \frac{\gamma p M}{1-(1-\gamma)p} 
\end{array}   , \right.
y_{L|\alpha=1}^{majority} = 
\left\{
\begin{array}
[c]{cl}
1  &\text{ if }L \le \frac{(1-\gamma) M}{1-\gamma p} \\ \vspace{.02in}
\frac{\gamma p L}{L - (1 - \gamma) M}  &\text{ if }L > \frac{(1-\gamma) M}{1-\gamma p}
\end{array}  . \right.
\]
}It is easy to see that there exists $\overline{\gamma}$ $(=\frac{1}{p+(1-p)M}) >0$ such that, for $\gamma \in [0, \overline{\gamma}]$, the solution for $\alpha=1$ is always $y_{L|\alpha=1}^{majority} = 1$, for $\frac{M+1}{2} \le L \le M-1$, which is greater than the corresponding solution  for $\alpha=0$, 
($y_{L|\alpha=0}^{majority} = 1- \gamma (1-p)$), for any $\frac{M+1}{2} \le L \le M-1$.  For $\gamma \in [0, \overline{\gamma}]$, the corresponding difference between the solutions $x_{L|\alpha=1}^{minority}$ and $x_{L|\alpha=0}^{minority}$ is bounded below by $-\gamma p$. Recall that the solutions coincide with our interim notion of efficiency for the given number of outlets with correct signal, $L$, so that from the definition of ${\cal P}$ in Equation~(\ref{eq: welfare}), we can write:
\begin{align*} 
 {\cal P}' &= 0 +
\sum_{L=1}^{\frac{M-1}{2}} \binom{M}{L}q^{L}(1-q)^{M-L}   x_{L|\alpha=0}^{minority} + \sum_{L=\frac{M+1}{2}}^{M-1} \binom{M}{L}q^{L}(1-q)^{M-L}   y_{L|\alpha=0}^{majority} + q^M, \\
 {\cal P} &= 0 +
\sum_{L=1}^{\frac{M-1}{2}} \binom{M}{L}q^{L}(1-q)^{M-L}   x_{L|\alpha=1}^{minority} + \sum_{L=\frac{M+1}{2}}^{M-1} \binom{M}{L}q^{L}(1-q)^{M-L}   y_{L|\alpha=1}^{majority} + q^M,
\end{align*} 
Since $q \gg p > \frac{1}{2}$, this is enough to imply that, for arbitrarily large $q$, while $\gamma \in [0, \overline{\gamma}]$, we have, ${\cal P} \ge {\cal P}'$ and hence $ $PoR$  \ge 0$. 

Suppose now $\gamma \ge \overline{\gamma}$. Again, for the given $\gamma$, there are two parts to the solution $y_{L|\alpha=1}^{majority}$ for $\alpha=1$, namely, a part which is 1 and therefore above the solution for $\alpha=0$ (for $\frac{M+1}{2} \le L \le  \frac{(1-\gamma) M}{1-\gamma p}$) and a part which is below the solution for $\alpha=0$ (for $\frac{(1-\gamma) M}{1-\gamma p} < L \le M-1$). 
In particular, there exists a level $\overline{q}$ such that, when $q \le \overline{q}$, then the cases where $L \le \frac{(1-\gamma) M}{1-\gamma p}$ and hence where the solution $y_{L|\alpha=1}^{majority}$ for $\alpha=1$ is above the solution $y_{L|\alpha=0}^{majority}$ for $\alpha=0$ will obtain sufficiently large weight, such that ex ante efficiency for $\alpha=1$ (${\cal P}$) is greater or equal to ex ante efficiency for $\alpha=0$ (${\cal P}'$), that is, such that $ $PoR$  \ge 0$.
At the same time, when $q > \overline{q}$ , then the cases where $L > \frac{(1-\gamma) M}{1-\gamma p}$ and hence where the solution $y_{L|\alpha=1}^{majority}$ for $\alpha=1$ is below the solution $y_{L|\alpha=0}^{majority}$ for $\alpha=0$ will obtain sufficiently large weight, such that ex ante efficiency for $\alpha=1$ (${\cal P}$) is less or equal to ex ante efficiency for $\alpha=0$ (${\cal P}'$), that is, such that $ $PoR$  \le 0$. 
This gives the function $\phi$, which is equal to 1 on $[0, \overline{\gamma}]$ and which is decreasing in $\gamma$ on $[\overline{\gamma}, 1]$, since the set of interim realizations, where the solution $y_{L|\alpha=1}^{majority}$ for $\alpha=1$ is above the solution $y_{L|\alpha=0}^{majority}$, is determined by the cutoff $\frac{(1-\gamma) M}{1-\gamma p}$, which is decreasing in $\gamma$. See Figure~\ref{fig:corollary2} (left panel) for an illustration of the case with $\mu=1$. 
\begin{figure}
\par
\begin{center}
\includegraphics[width=6.5cm]{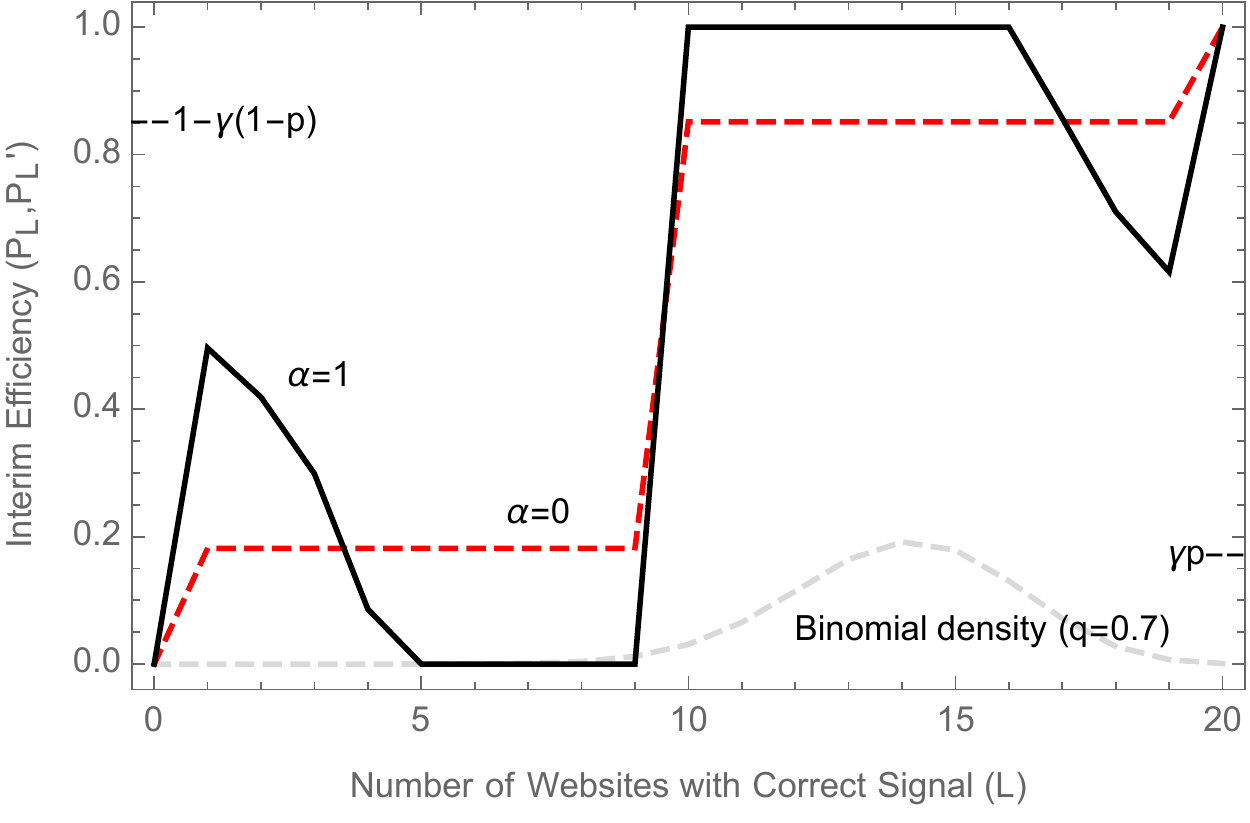} \hspace{.1in}
\includegraphics[width=6.5cm]{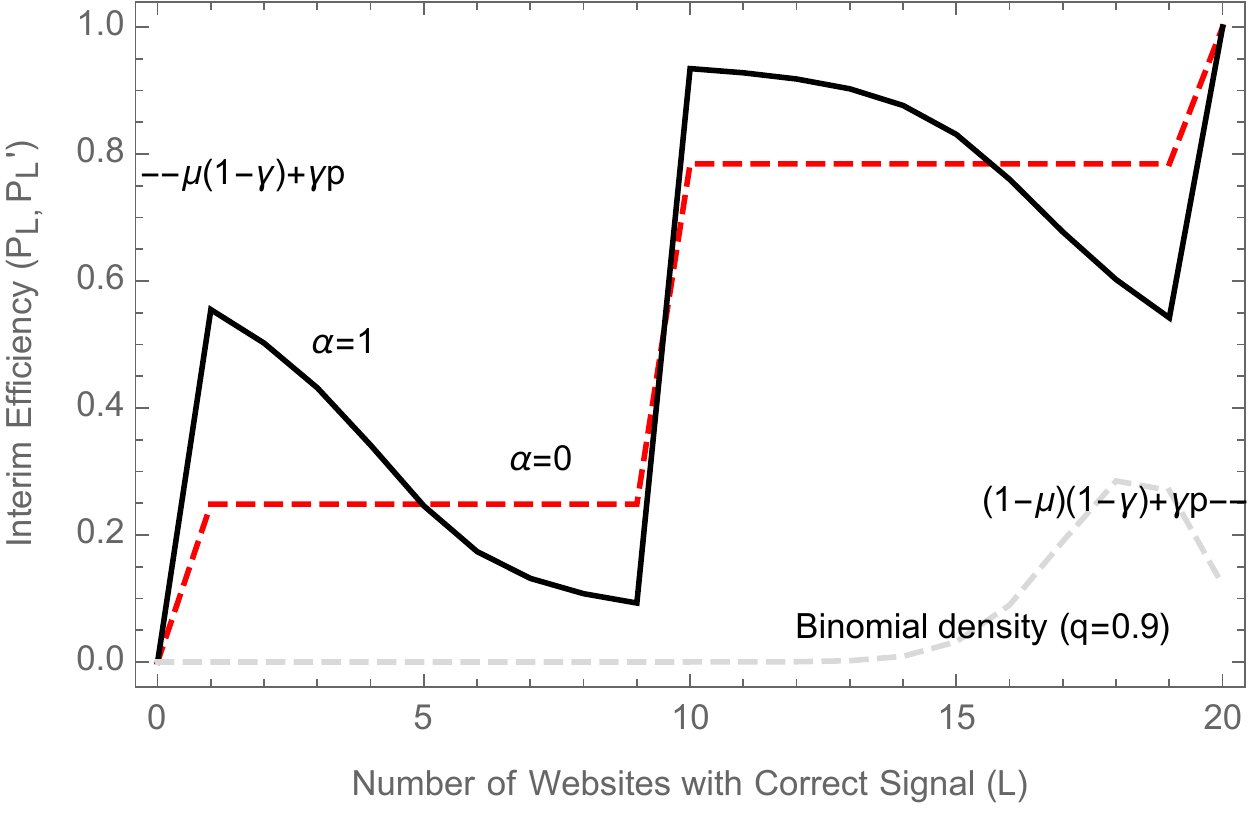} \hspace{.1in}
\end{center}
\par
\vspace{-15pt}
\caption{\footnotesize Interim efficiency (${\cal P}_{L}, {\cal P}_{L}'$) for $\alpha=0$ (black) and $\alpha=1$ (red dashed) as a function of $L$ for $\mu=1$ (left) and $\mu=0.9$ (right). In both panels, $M=20$, $\gamma=0.33$, $p=0.55$, and $r_1$ is uniform. The left panel also shows the density function for the binomial distribution for $q=0.7< \phi(\gamma)$ (light gray dashed) at which $ $PoR$  >0$, while the right panel shows the density function for the binomial distribution for $q=0.9 > \phi(\gamma)$ at which $ $PoR$  <0$.}
\label{fig:corollary2}
\end{figure}

Finally, the fact that there exists $\overline{\mu} < 1$, such that the above holds for $\mu \in [\overline{\mu}, 1]$, follows essentially from the fact that the functions $\theta_L^{minority}$ and $\theta_L^{majority}$ defining the Equations~(\ref{eq:thetamin}) and~(\ref{eq:thetamaj}) are linear in $\mu$.
This implies that the solutions are arbitrarily close to the ones computed for $\mu=1$. More specifically, it can be checked that, for general $\mu$, the solutions to Equations~(\ref{eq:thetamin}) and~(\ref{eq:thetamaj}) for $\alpha=0$ take the form, 
$x_{L|\alpha=0}^{minority} = (1-\mu)(1-\gamma)+ \gamma p , \,  y_{L|\alpha=0}^{majority} =\mu(1-\gamma) + \gamma p)$. At the same time, while the solutions for $\alpha=1$ no longer have a simple closed form, it can be checked that, for given $L$, they are above the corresponding solutions for $\alpha=0$, for
$L< ( (1-\mu)(1-\gamma)+ \gamma p)M$, if $L$ is minority, and for $L< ( \mu(1-\gamma)+ \gamma p)M$,  if $L$ is majority.
Again, checking for the levels of ex ante efficiency, this allows to compute decreasing cutoff levels for $q$ as a function of $\gamma$, $\overline{q}$, such that for $q\ge \overline{q}$, $ $PoR$ \le 0$, (provided $\gamma$ is not too small), while for $q \ge \overline{q}$, $ $PoR$  \ge 0$. See Figure~\ref{fig:corollary2} (right panel) for an illustration of the case with $\mu<1$. 
\hfill$\Box$

\bigskip


\textbf{Proof of Proposition~\ref{prop:polar}}. For simplicity, we consider the case, where one half of the population has $\gamma_A$ and the other half has $\gamma_B$, and where $\gamma_A \ne \gamma_B$. 
Recall the equation defining the ranking probabilities in the case of personalization (Equations~(\ref{eq:rloc}) and~(\ref{eq:nuloc})), for any $t$ and $m$, and for $\ell=A,B$:
\[
r_{t,m}^{\ell}=\nu_{t}^{\ell} r^{\ell}_{t-1,m}+(1-\nu_{t}^{\ell}) \rho_{t-1,m}^{\ell
}\,, 
\]
where:
\[
(\nu_{t}^{\ell},1-\nu_{t}^{\ell})=\left(  \frac{\kappa}{1-\lambda_{t}+\kappa}, \frac{1-\lambda_{t}}{1-\lambda
_{t}+\kappa}\right)
\mbox{ and where }\lambda_{t}=\left\{
\begin{array}
[c]{ll}%
0 & \text{ if }t-1\in\ell\\
\lambda & \text{ else }.
\end{array}
\right.  
\]
We can apply the mean dynamics approximation and obtain the deterministic recursions:
\begin{eqnarray} \label{eqs:persranking}
\widehat{r}_{t,m}^{A} &=& \widehat{r}_{t-1,m}^{A} + 
\frac{1}{1+\kappa_t} \left( \widehat{\rho}_{t-1,m}^A -  \widehat{r}_{t-1,m}^{A} \right) +
\frac{1 - \lambda}{1-\lambda+\kappa_t} \left( \widehat{\rho}_{t-1,m}^B -  \widehat{r}_{t-1,m}^{A} \right) 
\nonumber  \\
\widehat{r}_{t,m}^{B} &=& \widehat{r}_{t-1,m}^{B} + 
\frac{1}{1+\kappa_t} \left( \widehat{\rho}_{t-1,m}^B -  \widehat{r}_{t-1,m}^{B} \right) +
\frac{1 - \lambda}{1-\lambda+\kappa_t} \left( \widehat{\rho}_{t-1,m}^A -  \widehat{r}_{t-1,m}^{B} \right)  \, ,
\end{eqnarray}
where, following Equation~(\ref{eq:expclickprob}), we can write, for $\ell =A,B$:
\begin{align*} 
\widehat{\rho}_{t-1,m}^{\ell} = \mathbb{E}[\rho_{t-1,m}^{\ell}]  
&=
p \mu \frac{(\widehat{r}^{\ell}_{t-1,m})^{\alpha} \cdot \widehat{v^{*}}_{m}^{00}(\gamma_{\ell})}{\sum_{m^{\prime}}(\widehat{r}^{\ell}_{t-1,m^{\prime}})^{\alpha} \cdot \widehat{v^{*}}_{m^{\prime}}^{00}(\gamma_{\ell})}
+ p (1- \mu) \frac{(\widehat{r}^{\ell}_{t-1,m})^{\alpha} \cdot \widehat{v^{*}}_{m}^{01}(\gamma_{\ell})}{\sum_{m^{\prime}}(\widehat{r}^{\ell}_{t-1,m^{\prime}})^{\alpha} \cdot \widehat{v^{*}}_{m^{\prime}}^{01}(\gamma_{\ell})} 
\\
&+ (1-p) \mu \frac{(\widehat{r}^{\ell}_{t-1,m})^{\alpha} \cdot \widehat{v^{*}}_{m}^{10}(\gamma_{\ell})}{\sum_{m^{\prime}}(\widehat{r}^{\ell}_{t-1,m^{\prime}})^{\alpha} \cdot \widehat{v^{*}}_{m^{\prime}}^{10}(\gamma_{\ell})}
+ (1-p) (1-\mu) \frac{(\widehat{r}^{\ell}_{t-1,m})^{\alpha} \cdot \widehat{v^{*}}_{m}^{11}(\gamma_{\ell})}{\sum_{m^{\prime}}(\widehat{r}^{\ell}_{t-1,m^{\prime}})^{\alpha} \cdot \widehat{v^{*}}_{m^{\prime}}^{11}(\gamma_{\ell})} .
\end{align*}
Taking the limit $\kappa_{t} \rightarrow \infty$ in the equation system~(\ref{eqs:persranking}), yields the equations determining the limit clicking probabilities with personalization, which take the form:
\[ \frac{1}{2-\lambda}  \widehat{\rho}_{t,m}^A + \frac{1-\lambda}{2-\lambda}  \widehat{\rho}_{t,m}^B - r_{t,m}^A = 0 \hspace{.05in} \mbox{ and } \hspace{.05in} 
\frac{1}{2-\lambda}  \widehat{\rho}_{t,m}^B + \frac{1-\lambda}{2-\lambda}  \widehat{\rho}_{t,m}^A - r_{t,m}^B = 0 . \] 
Replacing $\widehat{r}_{t-1}^{A}$ with $x=(x_1, \ldots, x_M)$ and replacing $\widehat{r}_{t-1}^{B}$ with $y=(y_1, \ldots, y_M)$, we can study the function $g: \Delta_{\epsilon}(M) \times \Delta_{\epsilon}(M) \rightarrow \R^{2M}$, defined, for $\ell=A,B$, $m=1,\ldots, M$, by:
\[
g_{m}^{A}(x,y)=\theta_{m}^{A} (x,y) - x_{m} \hspace{.05in} \mbox{ and } \hspace{.05in}
g_{m}^{B}(x,y)=\theta_{m}^{B} (x,y) - y_{m} \]
where $\theta^{\ell}:\Delta_{\epsilon}(M) \times \Delta_{\epsilon}(M) \rightarrow\mathbb{R}^{M}$, $\ell=A,B$, is defined by, 
{\scriptsize
\begin{align*}
\hspace{-.5in}
\theta_{m}^{A}(x,y)&= \frac{1}{2-\lambda}\left( \frac{ p \mu \cdot (x_{m})^{\alpha} \cdot \widehat{v^{*}}_{m}^{00}(\gamma_{A})}
{\sum_{m^{\prime}}(x_{m^{\prime}})^{\alpha} \cdot \widehat{v^{*}}_{m^{\prime}}^{00}(\gamma_{A})}
+ \frac{p (1- \mu) \cdot (x_{m})^{\alpha} \cdot \widehat{v^{*}}_{m}^{01}(\gamma_{A})}
{\sum_{m^{\prime}}(x_{m^{\prime}})^{\alpha} \cdot \widehat{v^{*}}_{m^{\prime}}^{01}(\gamma_{A})}
+ \frac{(1-p)\mu\cdot(x_{m})^{\alpha} \cdot \widehat{v^{*}}_{m}^{10}(\gamma_{A})}{\sum_{m^{\prime}}(x_{m^{\prime}})^{\alpha} \cdot \widehat{v^{*}}_{m^{\prime}}^{10}(\gamma_{A})} 
+ \frac{(1-p)(1-\mu)\cdot(x_{m})^{\alpha} \cdot \widehat{v^{*}}_{m}^{11}(\gamma_{A})}{\sum_{m^{\prime}}(x_{m^{\prime}})^{\alpha} \cdot \widehat{v^{*}}_{m^{\prime}}^{11}(\gamma_{A})} \right)
\\
&+ \frac{1-\lambda}{2-\lambda}  \left( \frac{p \mu \cdot(y_{m})^{\alpha} \cdot  \widehat{v^{*}}_{m}^{00}(\gamma_{B})}
{\sum_{m^{\prime}} (y_{m^{\prime}})^{\alpha} \cdot \widehat{v^{*}}_{m^{\prime}}^{00}(\gamma_{B})}
+ \frac{p (1-\mu) \cdot(y_{m})^{\alpha} \cdot  \widehat{v^{*}}_{m}^{01}(\gamma_{B})}
{\sum_{m^{\prime}} (y_{m^{\prime}})^{\alpha} \cdot \widehat{v^{*}}_{m^{\prime}}^{01}(\gamma_{B})}
+  \frac{(1-p)\mu \cdot(y_{m})^{\alpha} \cdot \widehat{v^{*}}_{m}^{10}(\gamma_{B})}{\sum_{m^{\prime}
}(y_{m^{\prime}})^{\alpha} \cdot \widehat{v^{*}}_{m^{\prime}}^{10}(\gamma_{B})}
+ \frac{(1-p)(1-\mu) \cdot(y_{m})^{\alpha} \cdot \widehat{v^{*}}_{m}^{11}(\gamma_{B})}{\sum_{m^{\prime}%
}(y_{m^{\prime}})^{\alpha} \cdot \widehat{v^{*}}_{m^{\prime}}^{11}(\gamma_{B})} \right) , \\
\hspace{-.3in}
\theta_{m}^{B}(x,y)&=\frac{1-\lambda}{2-\lambda} \left(\frac{ p \mu \cdot(x_{m})^{\alpha} \cdot \widehat{v^{*}}_{m}^{00}(\gamma_{A})}
{\sum_{m^{\prime}}(x_{m^{\prime}})^{\alpha} \cdot \widehat{v^{*}}_{m^{\prime}}^{00}(\gamma_{A})}
+ \frac{ p(1- \mu) \cdot(x_{m})^{\alpha} \cdot \widehat{v^{*}}_{m}^{01}(\gamma_{A})}
{\sum_{m^{\prime}}(x_{m^{\prime}})^{\alpha} \cdot \widehat{v^{*}}_{m^{\prime}}^{01}(\gamma_{A})}
+\frac{(1-p) \mu \cdot(x_{m})^{\alpha} \cdot \widehat{v^{*}}_{m}^{10}(\gamma_{A})}{\sum_{m^{\prime}}(x_{m^{\prime}})^{\alpha} \cdot \widehat{v^{*}}_{m^{\prime}}^{10}(\gamma_{A})}
+\frac{(1-p) (1-\mu) \cdot(x_{m})^{\alpha} \cdot \widehat{v^{*}}_{m}^{11}(\gamma_{A})}{\sum_{m^{\prime}}(x_{m^{\prime}})^{\alpha} \cdot \widehat{v^{*}}_{m^{\prime}}^{11}(\gamma_{A})} \right) 
\\
&+ \frac{1}{2-\lambda}\left(\frac{ p \mu \cdot(y_{m})^{\alpha} \cdot  \widehat{v^{*}}_{m}^{00}(\gamma_{B})}
{\sum_{m^{\prime}} (y_{m^{\prime}})^{\alpha} \cdot \widehat{v^{*}}_{m^{\prime}}^{00}(\gamma_{B})}
+ \frac{ p (1-\mu) \cdot(y_{m})^{\alpha} \cdot  \widehat{v^{*}}_{m}^{01}(\gamma_{B})}
{\sum_{m^{\prime}} (y_{m^{\prime}})^{\alpha} \cdot \widehat{v^{*}}_{m^{\prime}}^{01}(\gamma_{B})}
+\frac{(1-p)\mu \cdot(y_{m})^{\alpha} \cdot \widehat{v^{*}}_{m}^{10}(\gamma_{B})}{\sum_{m^{\prime}%
}(y_{m^{\prime}})^{\alpha} \cdot \widehat{v^{*}}_{m^{\prime}}^{10}(\gamma_{B})}
+\frac{(1-p)(1-\mu) \cdot(y_{m})^{\alpha} \cdot \widehat{v^{*}}_{m}^{11}(\gamma_{B})}{\sum_{m^{\prime}%
}(y_{m^{\prime}})^{\alpha} \cdot \widehat{v^{*}}_{m^{\prime}}^{11}(\gamma_{B})} \right) .
\end{align*}
}Given that the function $g$ is smooth in $x,y$ on $\Delta_{\epsilon}(M)^2$, it can
be shown that the expected limit of our stochastic process can be obtained by
solving the system of ordinary differential equations $(\dot{x},\dot{y})=g(x,y)$ (or $(\dot{x},\dot{y})=(g^A(x,y),g^B(x,y))$) as done above in the case of $\lambda=0$. We have that if $\lambda = 0$, then it is as if there were a single group with the same (common) ranking algorithm and where the individuals' preference for like-minded news is $\gamma = \frac{\gamma_A+\gamma_B}{2}$. As $\lambda$ increases, then the rankings of the two groups 
drift apart and it is as if group $A$ had parameter $\frac{1}{2-\lambda}\gamma_A+\frac{1-\lambda}{2-\lambda}\gamma_B$ and group B had parameter $\frac{1-\lambda}{2-\lambda}\gamma_A+\frac{1}{2-\lambda}\gamma_B$ until, when $\lambda=1$, it is as if there were two separate rankings of two groups with accuracy $\gamma_A$ and $\gamma_B$ respectively. Since $\gamma_A \ne p_B$ the difference in the levels of preference for like-minded news of the two groups ($\frac{\lambda}{2-\lambda}(\gamma_A-\gamma_B)$) is increasing in $\lambda$. Finally, since clicking and ranking probabilities coincide in the limit, this translates to increasingly different probabilities of clicking on any given majority website in the two groups and hence, given the definition of ${\cal BP}$, also to a measure ${\cal BP}({\cal E}_{\lambda})$ that is increasing in $\lambda$.  \hfill$\Box$

\bigskip


\textbf{Proof of Proposition~\ref{prop:personalizedsearch}}. 
Let ${\cal P}^{\lambda}$ and ${\cal P}_{L}^{\lambda}$ denote respectively ex ante and interim efficiency as a function of the personalization parameter $\lambda$. Because the initial ranking is uniform we can look again at the solutions to the Equations~(\ref{eq:thetamin}) and~(\ref{eq:thetamaj}) in the proof of Proposition~\ref{prop:$AOF$}, where we recall again that $x_{L}^{minority}=\widehat{\rho}_{\infty,L}={\cal P}_L^{\lambda}$ when websites in $L$ have minority signal and  $y_{L}^{majority}=\widehat{\rho}_{\infty,L}={\cal P}_L^{\lambda}$ when they have majority signal.
Hence to show that ${\cal P}^{\lambda}$ is weakly decreasing in $\lambda$ we can use these solutions to study the interim efficiency levels ${\cal P}_{L}^{\lambda}$  for $\lambda \in [0,1]$ and $1 < L < M$. 
As in the proof of Proposition~\ref{cor:randomrank}, we have that, for $\alpha=1$ and $\mu=1$, the solutions to Equations~(\ref{eq:thetamin}) and~(\ref{eq:thetamaj}) take the form, respectively:
{\small 
\[ 
x_{L}^{minority} = 
\left\{
\begin{array}
[c]{cl}
\frac{(1-\gamma) L - \gamma p (M-L)}{\gamma M - L}  &\text{ if }L< \frac{\gamma p M}{1-(1-\gamma)p} \\ \vspace{.02in}
0  &\text{ if }L \ge \frac{\gamma p M}{1-(1-\gamma)p} 
\end{array}   , \right.
y_{L}^{majority} = 
\left\{
\begin{array}
[c]{cl}
1  &\text{ if }L \le \frac{(1-\gamma) M}{1-\gamma p} \\ \vspace{.02in}
\frac{\gamma p L}{L - (1 - \gamma) M}  &\text{ if }L > \frac{(1-\gamma) M}{1-\gamma p}
\end{array}  . \right.
\]
}Fix a realization of ${\cal E}$, say, parametrized by $L$.
From the proof of Proposition~\ref{prop:polar}, we have that when $\lambda=0$, it is as if there were a single group with a common ranking algorithm, with $\gamma=\frac{\gamma_A + \gamma_B}{2}$. As $\lambda$ increases, then the rankings of the two groups drift apart and are 
as if group $A$ had preference for like-minded news $\frac{1}{2-\lambda}\gamma_A + \frac{1-\lambda}{2-\lambda} \gamma_B$ 
and group $B$ had $\frac{1-\lambda}{2-\lambda}\gamma_A + \frac{1}{2-\lambda} \gamma_B$, 
until, when $\lambda = 1$, it is as if there were two separate rankings of two groups with accuracy $\gamma_A$ and $\gamma_B$, respectively. As a result interim efficiency can be written as the average of the interim efficiency of two groups, one with $\gamma_A^{\lambda}= \frac{1}{2-\lambda}\gamma_A + \frac{1-\lambda}{2-\lambda} \gamma_B$ and another with $\gamma_B^{\lambda}= \frac{1-\lambda}{2-\lambda}\gamma_A + \frac{1}{2-\lambda} \gamma_B$. 
Therefore, in the case of intermediate values of $\lambda$ interim efficiency can be seen as the average of two levels of interim efficiency corresponding to two different signal accuracies that are increasingly apart as $\lambda$ increases (that is, go from both signals corresponding to $\frac{\gamma_A+\gamma_B}{2}$ when $\lambda =0$ to being $\gamma_A$ and  $\gamma_B$ respectively when $\lambda=1$). Since we can always take $\gamma_A, \gamma_B$ to be, respectively, $\gamma_A^{\lambda}, \gamma_B^{\lambda}$, it suffices to consider directly the case of $\lambda=1$. In this case, the interim efficiency levels can be computed from the solutions evaluated at the corresponding levels of $\gamma$:
{\small 
\[ 
{\cal P}_{L}^{\lambda=0} = \left( x_{L}^{minority} \left( \frac{\gamma_A+\gamma_B}{2} \right) , y_{L}^{majority} \left( \frac{\gamma_A+\gamma_B}{2} \right) \right) , 
\]
}for the non-personalized case ($\lambda=0$), and from:
{\small
\[ 
{\cal P}_{L}^{\lambda=1} = \left( \frac{x_{L}^{minority}(\gamma_A) + x_{L}^{minority}(\gamma_B)}{2}  ,  \frac{y_{L}^{majority}(\gamma_A) + y_{L}^{majority}(\gamma_B)}{2} \right) , 
\]
}for the fully personalized case ($\lambda=1$).
\begin{figure}
\par
\begin{center}
\includegraphics[width=6.5cm]{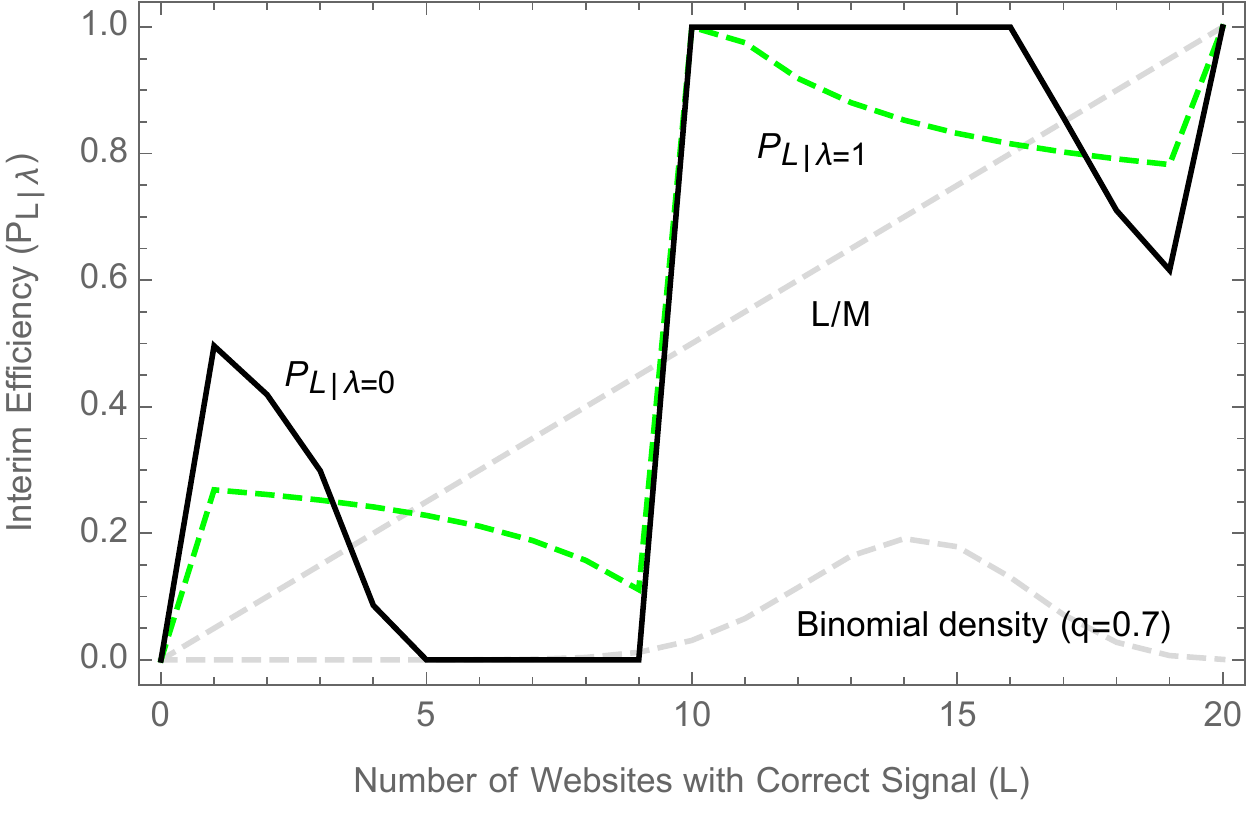} \hspace{.1in}
\includegraphics[width=6.5cm]{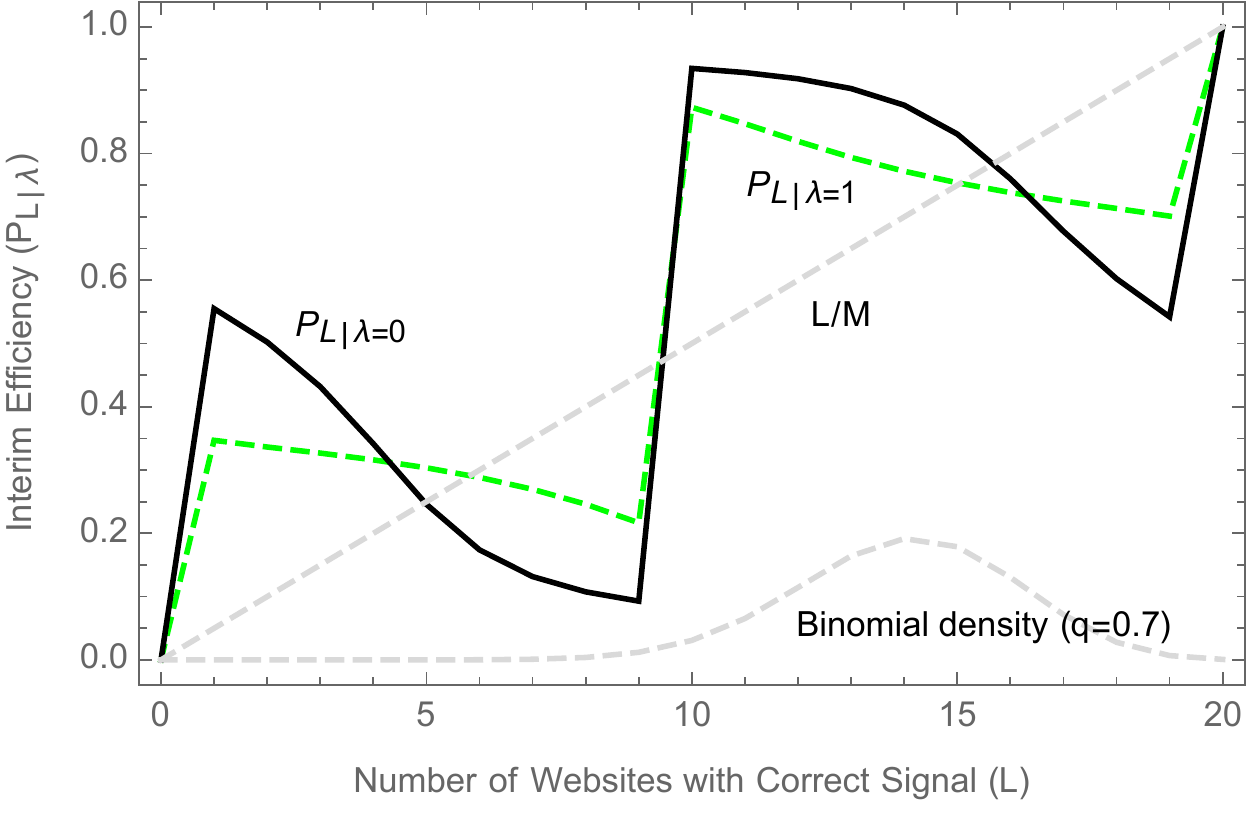} \hspace{.1in}
\end{center}
\par
\vspace{-15pt}
\caption{\footnotesize Interim efficiency with no personalization ${\cal P}_{L}^{\lambda=0}$ (black) and with full personalization ${\cal P}_{L}^{\lambda=1}$ (green dashed)  as a function of $L$ for $\mu=1$ (left) and $\mu=0.9$ (right). In both panels, $M=20$, $\gamma_A=0$, $\gamma_B=0.66$, $p=0.55$, and $r_1$ is uniform. Both panels also show (gray dashed) the function $\frac{L}{M}$ as comparison benchmark as well as the density function for the binomial distribution for $q=0.7< \phi(\gamma)$ at which $ $PeR$  >0$.}
\label{fig:prop4}
\end{figure}
Since the reasoning is very similar to that of the proof of Proposition~\ref{cor:randomrank}, we focus on the case where the outlets with correct signal are a majority. For $\lambda=0$ we have, as the level of interim efficiency, directly $y_{L}^{majority} \left( \frac{\gamma_A+\gamma_B}{2} \right)$ from above, while for $\lambda=1$, recalling $0 \le \gamma_A < \gamma_B \le 1$, and since the cutoff for $L$ ($\frac{\gamma p M}{1-(1-\gamma)p}$) is decreasing in $\gamma$, we can write this as:
{\small
\[ 
\frac{y_{L}^{majority}(\gamma_A) + y_{L}^{majority}(\gamma_B)}{2} =
\left\{
\begin{array}
[c]{ll}
1  &\text{ if } \frac{M+1}{2} \le L \le \frac{(1-\gamma_B)M}{1-\gamma_B p} \\ \vspace{.1in}
\frac{1}{2} + \frac{\gamma_B p L}{2(L - (1 - \gamma_B) M)}  &\text{ if } \frac{(1-\gamma_B) M}{1-\gamma_B p}<L \le \frac{(1-\gamma_A)M}{1-\gamma_A p}  
\\ \vspace{.1in}
\frac{\gamma_A p L}{2(L - (1 - \gamma_A) M)} + \frac{\gamma_B p L}{2(L - (1 - \gamma_B)M} &\text{ if }  \frac{(1-\gamma_A)M}{1-\gamma_A p} < L \le M-1

\end{array}  . \right.
\]
}Tedious calculations show that, when $y_{L}^{majority} \left( \frac{\gamma_A+\gamma_B}{2} \right)$ $\ge \frac{L}{M}$, we also have:
\begin{equation*}  
y_{L}^{majority} \left( \frac{\gamma_A+\gamma_B}{2} \right) \ge \frac{y_{L}^{majority}(\gamma_A) + y_{L}^{majority}(\gamma_B)}{2},
\end{equation*}
and hence ${\cal P}_{L}^{\lambda=0} \ge {\cal P}_{L}^{\lambda=1}$.\footnote{To get more intuition for the proof, notice that  ${\cal P}_{L}^{\lambda=0}$ for majority values of $L$ is a concave function of $\gamma$, when solutions satisfy $y \ge \frac{L}{M}$, and becomes convex when $y \le \frac{L}{M}$. Interim efficiency being concave implies ex ante efficiency is concave in $\gamma$ such that ex ante efficiency of an average $\gamma=\frac{\gamma_A+\gamma_B}{2}$ will be above the average ex ante efficiency of $\gamma_A$ and $\gamma_B$ (which in turn implies $PeR \le 0$, provided $q < \overline{q}$). On the other hand, the opposite holds ($PeR \ge 0$), when the function becomes convex. This shows why larger levels of $q$ ($> \overline{q}$) can reverse the result.} Using the same reasoning as in the proof of Proposition~\ref{cor:randomrank}, we can show that $y$ (or also $x$) $\ge \frac{L}{M}$ occurs, when $\mu$ is sufficiently large and $\gamma$ sufficiently small, and when $q$ is below a given threshold $\overline{q}(\gamma)$. It can further be shown that the reverse inequality holds for the same values of $\mu$ and $\gamma$ if $q$ is above the threshold $\overline{q}(\gamma)$. Finally, a similar argument as in the proof of Proposition~\ref{cor:randomrank} allows to extend to the case where $\mu > \overline{\mu}$ for some $\overline{\mu} < 1$. Figure~\ref{fig:prop4} illustrates the cases $\mu=1$ (left panel) and $\mu< 1$ (right panel). \hfill$\Box$

\bigskip


\textbf{Proof of Corollary \ref{cor:rankpro_alpha}}. This is proved above together with Corollary \ref{cor:rankpro}. \hfill$\Box$

\bigskip

\textbf{Proof of Proposition~\ref{prop:initial_rank_0} and Proposition~\ref{prop:initial_ranking}}. 
We prove directly the case with $\gamma \ge 0$.
The expected ranking probabilities are then given by:
\begin{eqnarray*}
\widehat{r}_{n,m} &=& \nu \widehat{r}_{n-1,m} + (1 - \nu) \widehat{\rho}_{n-1,m} \\
&=&  
\nu \widehat{r}_{n-1,m} + (1-\nu) \left(\frac{p \mu (\widehat{r}_{t-1,m})^{\alpha} \cdot \widehat{v^{*}}_{m}^{00}}{\sum_{m^{\prime}}(\widehat{r}_{t-1,m^{\prime}})^{\alpha} \cdot \widehat{v^{*}}_{m^{\prime}}^{00}}
+ \frac{p (1- \mu) (\widehat{r}_{t-1,m})^{\alpha} \cdot \widehat{v^{*}}_{m}^{01}}{\sum_{m^{\prime}}(\widehat{r}_{t-1,m^{\prime}})^{\alpha} \cdot \widehat{v^{*}}_{m^{\prime}}^{01}} 
\right. + \\
&& \hspace{1.5in} +   \left. \frac{(1-p) \mu (\widehat{r}_{t-1,m})^{\alpha} \cdot \widehat{v^{*}}_{m}^{10}}{\sum_{m^{\prime}}(\widehat{r}_{t-1,m^{\prime}})^{\alpha} \cdot \widehat{v^{*}}_{m^{\prime}}^{10}} 
+  \frac{(1-p) (1-\mu)(\widehat{r}_{t-1,m})^{\alpha} \cdot \widehat{v^{*}}_{m}^{11}}{\sum_{m^{\prime}}(\widehat{r}_{t-1,m^{\prime}})^{\alpha} \cdot \widehat{v^{*}}_{m^{\prime}}^{11}}\right)
\\
&=&  
\nu \widehat{r}_{n-1,m} + (1-\nu) \left(\frac{p \mu  \cdot \widehat{v^{*}}_{m}^{00}}{\sum_{m^{\prime}}(\widehat{r}_{t-1,m^{\prime}})^{\alpha} \cdot \widehat{v^{*}}_{m^{\prime}}^{00}} +
\ldots +  \frac{(1-p) (1-\mu) \cdot \widehat{v^{*}}_{m}^{11}}{\sum_{m^{\prime}}(\widehat{r}_{t-1,m^{\prime}})^{\alpha} \cdot \widehat{v^{*}}_{m^{\prime}}^{11}}\right)
\cdot (\widehat{r}_{n-1,m})^{\alpha}
\, ,
\end{eqnarray*}
where $\widehat{v^{*}}_{m}^{00}$, $\widehat{v^{*}}_{m}^{01}$, $\widehat{v^{*}}_{m}^{10}$ and $\widehat{v^{*}}_{m}^{11}$ are defined in Appendix~\ref{Appendix-MeanDynamics}. Let $m,m' \in K$ with $m \ne m'$ and $r_{1,m} > r_{1,m'} >0$. Fix $n>1$, we first show that the rich-get-richer dynamic applies to the ranking probabilities. 
To simplify notation, let $x \equiv \widehat{r}_{n-1,m}$ and $y \equiv \widehat{r}_{n-1,m'}$. Because $0 < \nu < 1$ we always have $x > y >0$ for any $n>2$, and because the two websites have the same signal they also have the equal coefficients on $(\widehat{r}_{n-1,m})^{\alpha}$ and $\widehat{r}_{n-1,m}$, say, $a$ and $b$ respectively, where $a, b>0$. Hence we can write:
\[ \widehat{r}_{n,m} = a x^{\alpha} + bx \hspace{.2in} \mbox{ and } \hspace{.2in} 
\widehat{r}_{n,m'} = a y^{\alpha} + by . \]    
But then it follows that:
\[
\frac{\widehat{r}_{n,m}}{\widehat{r}_{n,m'}} > \frac{\widehat{r}_{n-1,m}}{\widehat{r}_{n-1,m'}} 
\Longleftrightarrow \frac{ax^{\alpha} + bx}{a y^{\alpha} + by} > \frac{x}{y} 
\Longleftrightarrow ax^{\alpha}y + bxy> a x y^{\alpha} + b x y 
\Longleftrightarrow \left( \frac{x}{y} \right)^{\alpha} > \frac{x}{y}
\Longleftrightarrow \alpha > 1 
\]
and similarly
\[
\frac{\widehat{r}_{n,m}}{\widehat{r}_{n,m'}} \stackrel{(=)}{<} \frac{\widehat{r}_{n-1,m}}{\widehat{r}_{n-1,m'}} 
\Longleftrightarrow \frac{ax^{\alpha} + bx}{a y^{\alpha} + by} \stackrel{(=)}{<} \frac{x}{y} 
\Longleftrightarrow ax^{\alpha}y + bxy \stackrel{(=)}{<} a x y^{\alpha} + b x y 
\Longleftrightarrow \left( \frac{x}{y} \right)^{\alpha} \stackrel{(=)}{<} \frac{x}{y}
\Longleftrightarrow \alpha \stackrel{(=)}{<} 1 
\] 
Finally, to see the claim, notice that $\widehat{\rho}_{n,m}=\frac{a}{1-\nu}(\widehat{r}_{n,m})^{\alpha}$ and $\widehat{\rho}_{n,m'}=\frac{a}{1-\nu}(\widehat{r}_{n,m'})^{\alpha}$, where again $\frac{a}{1-\nu}>0$, so that:
\begin{eqnarray*}
\frac{\widehat{\rho}_{n,m}}{\widehat{\rho}_{n,m'}} > \frac{\widehat{\rho}_{n-1,m}}{\widehat{\rho}_{n-1,m'}}
&\Longleftrightarrow& \frac{\frac{a}{1-\nu}(\widehat{r}_{n,m})^{\alpha}}{\frac{a}{1-\nu}(\widehat{r}_{n,m'})^{\alpha}} > \frac{\frac{a}{1-\nu}(\widehat{r}_{n-1,m})^{\alpha}}{\frac{a}{1-\nu}(\widehat{r}_{n-1,m'})^{\alpha}}
\\ \\
&\Longleftrightarrow& \left( \frac{\widehat{r}_{n,m}}{\widehat{r}_{n,m'}} \right)^{\alpha} > \left( \frac{\widehat{r}_{n-1,m}}{\widehat{r}_{n-1,m'}} \right)^{\alpha} \Longleftrightarrow  \frac{\widehat{r}_{n,m}}{\widehat{r}_{n,m'}} >  \frac{\widehat{r}_{n-1,m}}{\widehat{r}_{n-1,m'}} \Longleftrightarrow \alpha > 1 ,
\end{eqnarray*}
and correspondingly for $\alpha < 1$ and $\alpha=1$.
\hfill$\Box$

\bigskip


\endgroup

\newpage

\renewcommand{\linespread}{1.1}
\appendix

\newpage\setcounter{page}{1}

\newpage

\renewcommand{\linespread}{1.1}
\appendix

\newpage\setcounter{page}{1}

\noindent{\Large \textbf{ONLINE APPENDIX} }


\bigskip

\noindent The Online Appendix presents examples that provide numerical and graphical illustrations of some results or of extensions of results presented in the paper.

\renewcommand{\ex}{}
\setcounter{ex}{0}

\setcounter{table}{0}
\renewcommand{\thetable}{A.\arabic{table}}

\setcounter{figure}{0}
\renewcommand{\thefigure}{A.\arabic{figure}}

\section*{\large A. Advantage of the Fewer ($AOF$)}

First, we provide a numerical illustration of the \textit{amplification effect} generated by the $AOF$. Then, we show that the $AOF$ effect is a rather general phenomenon that holds even outside the assumptions of the model.  In particular, we consider three cases not covered by the basic model studied in the paper, namely, (1) a case with uninformative signals, (2) a case where the ranking is an ordinal ranked list and not a probability distribution over the outlets, (3) a case where the total number of outlets is not fixed such that, for example, two outlets merging and thus reducing the total number of outlets. Finally, we illustrate the stochastic dynamic behind the $AOF$ effect.

\bigskip

\noindent
\begin{ex} 
{\bf Example A1. $AOF$ \textit{Amplification effect}}. As discussed in Section \ref{section:results}, the $AOF$ generates an \textit{amplification effect} whereby a reduction in the number of websites reporting a given signal increeases their overall share of traffic. Such an effect can be sizeable. Consider, for example, a search environment where $M=20$, $r_1$ uniform and where $p=0.55$ and $\mu=0.9$.
Figure~\ref{fig:$AOF$} illustrates the \textit{amplification effect} due to a reduction in the number of outlets reporting a correct signal, as a function of the preference for like-minded news. As the graph shows, for the case of minority outlets (e.g., a reduction in $L$ from 2 to 1), the \textit{amplification effect} may almost double the probability of clicking on a minority outlet reporting a correct signal. In the case of majority outlets (e.g., a reduction in $L$ from 16 to 15) the effect is smaller, but it may still boost the overall traffic of such outlets by up to 15\%.

\begin{figure}[H]
\par
\begin{center}

\includegraphics[width=6.5cm]{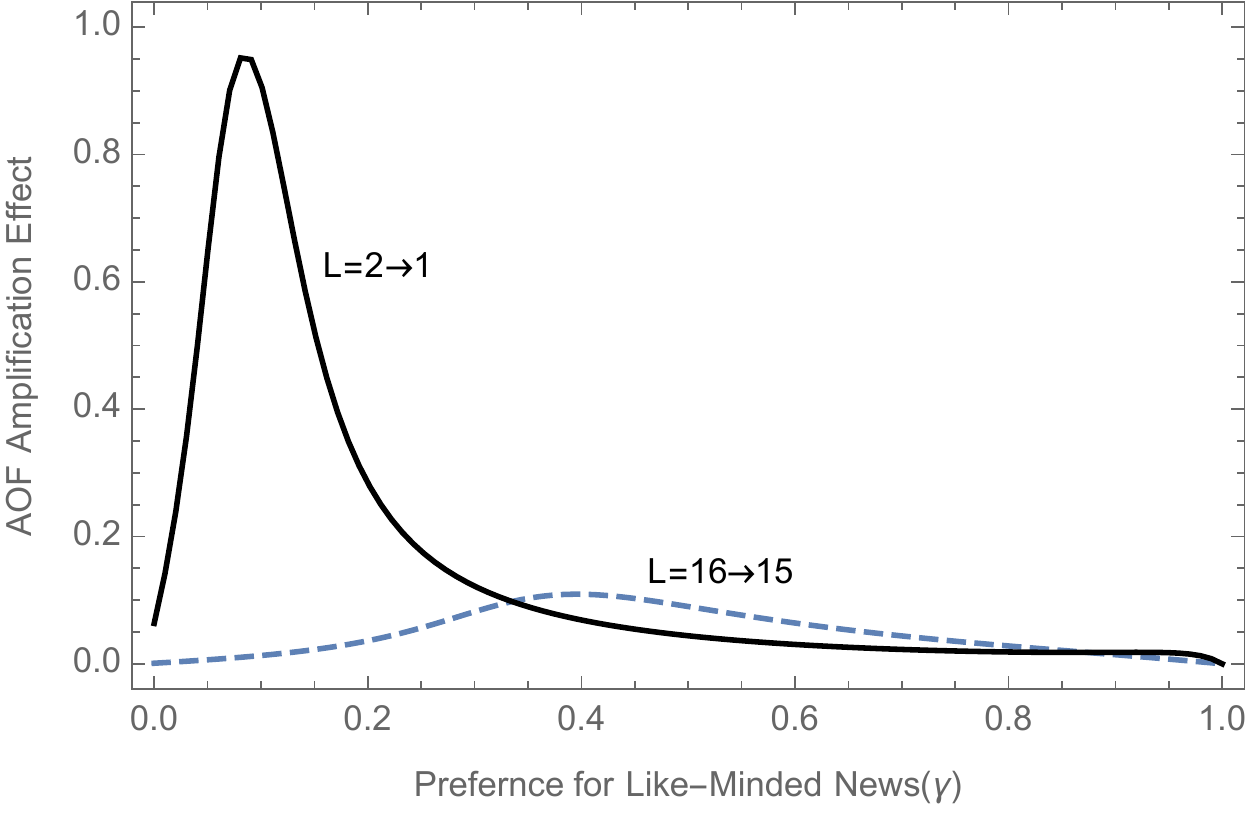} \end{center}

\par
\vspace{-15pt}\caption{\small Amplification effect of $AOF$ (as a function of the preference for like-minded news, $\gamma$) in terms of limit clicking probability of minority websites reporting a correct signal (black) when  $L$ is reduced from $2$ to $1$ (black) and of majority websites reporting a correct signal (dashed) when $L$ is reduced from $16$ to $15$.  In both panels, $M=20$, $p=0.55$, $\mu=0.9$, and $r_1$ is uniform.}
\label{fig:$AOF$}
\end{figure}
\end{ex}

\bigskip
\newpage

\noindent
\begin{ex}
{\bf Example A2. $AOF$ with uninformative signals.} Consider a search environment where $M=20$, $r_1$ uniform and where $p=\mu=\frac{1}{2}$. Then, as Figure~\ref{fig:uninformativeranking_L} shows, the limit clicking probability of choosing an outlet reporting a correct signal is monotonically decreasing in the number of websites with a correct signal (for $1\leq L<M-1$). This also shows that the $AOF$ effect can be generated with just flows of types of individuals, ones that are committed to one type of content and therefore actively search for those outlets, wherever they appear in the ranking, ones that are committed to the other type of content, and ones that are willing to trade off one content type with the other. As discussed in the text, it is the latter that generate the amplification effect.

\begin{figure}[H]
\par
\begin{center}
\includegraphics[width=6.5cm]{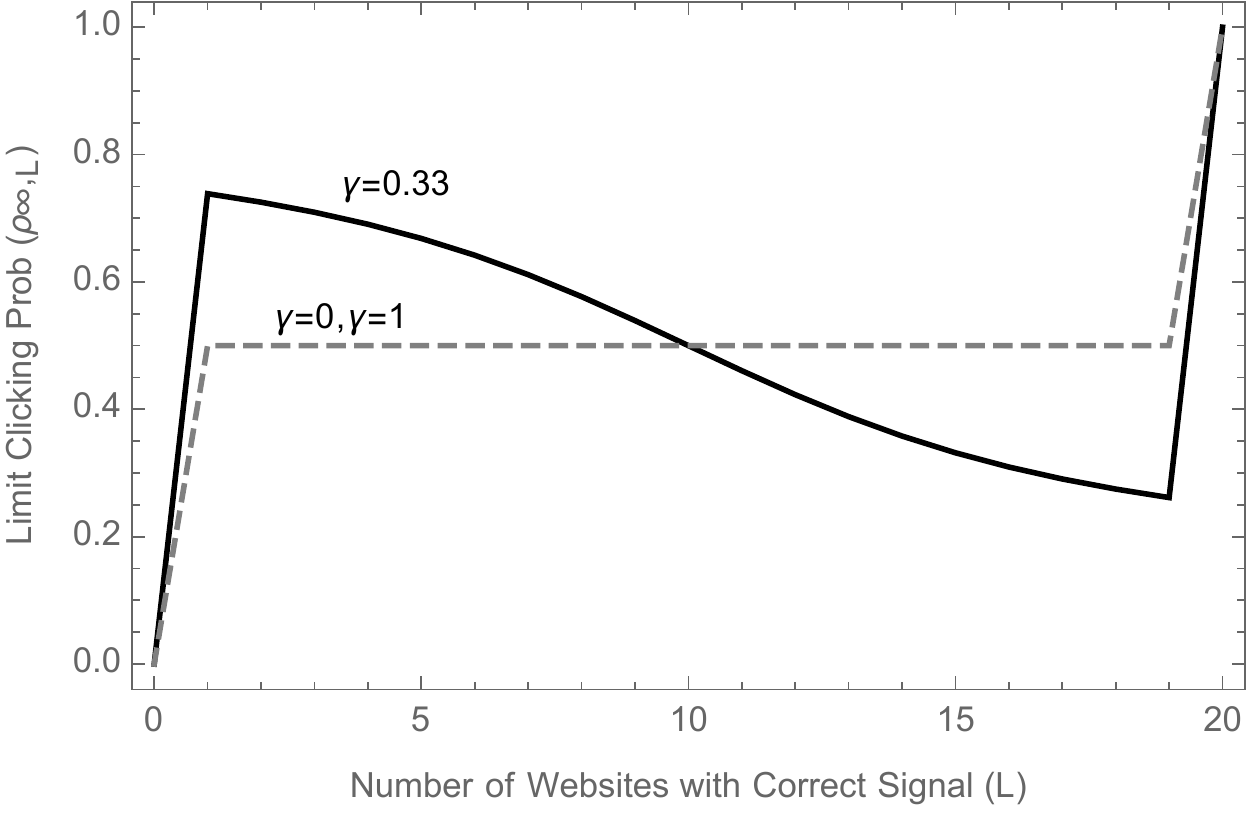}
\end{center}
\par
\caption{\small 
{\em Interim} efficiency (${\cal P}_L$) as a function of the number of websites with correct signal ($L$) 
for $\gamma=0.33$ (black line) and other values of $\gamma$ (dashed lines) with uninformative signals $p=0.5$, $\mu=0.5$. Also, $M=20$, and $r_1$ is uniform.}
\label{fig:uninformativeranking_L}
\end{figure}
\end{ex}

Importantly, this example can also be used to show that the $AOF$ effect occurs in environments, where individuals cannot distinguish between minority and majority outlets. In fact, it can be seen as illustrating a situation, where individuals simply choose between different websites based on two (neutral) independent cues ($x_n$ and $z_n$) that we interpret as general desirable attributes of the different websites.

\bigskip
\noindent 
\begin{ex}
{\bf Example A3. $AOF$ with ordinal ranking and pure clicking realizations.} Consider a search environment,  where the ranking at each time is a list from 1 to 20, and where the rank of an outlet $r_{t,m} \in \{ 1, \ldots, M\}$ depends on the number of clicks received up to $t$, such that the outlet with the greatest number of clicks has $r_{t,m}=1$, the one with the second greatest number of clicks has $r_{t,m}=2$ and so on. Suppose also that each individual chooses a single outlet according to the vector of choice probabilities $\rho_{n}$, thereby contributing a single click rather than a probability vector (as implicitly assumed in Equation (\ref{eq:rt})). In this case, we need to change the weighting function used to describe the individual stochastic choice. We use the function $\beta^{(M-r_{t,m})}$, where $\beta \ge 1$ again calibrates attention bias in a way that being one position higher corresponds to receiving $\beta$ times as much probability of being clicked. Otherwise, we use the same multinomial probabilities of choosing a website given by:
\[
\rho_{t,m} = \frac{\beta^{(M-r_{t,m})} v^*_{t,m}}{\sum_{m' \in M} \beta^{(M-r_{t,m'})} v^*_{t,m'}} .
\]
Suppose also that $r_1$ is a list that has all minority websites at the bottom, and where $M=20$, $\beta=1.5$, $p=0.55$, $\mu=0.9$.  
Then, the following figure shows that the limit clicking probability of choosing an outlet reporting a correct signal is again decreasing in the number of websites with a correct signal for $L\ne 1, \frac{M-1}{2}, M-1$. Notice that it is very similar to Figure~\ref{fig:ranking_L} obtained with the probabilistic ranking and clicking.
\begin{figure}
\par
\begin{center}
\includegraphics[width=6.5cm]{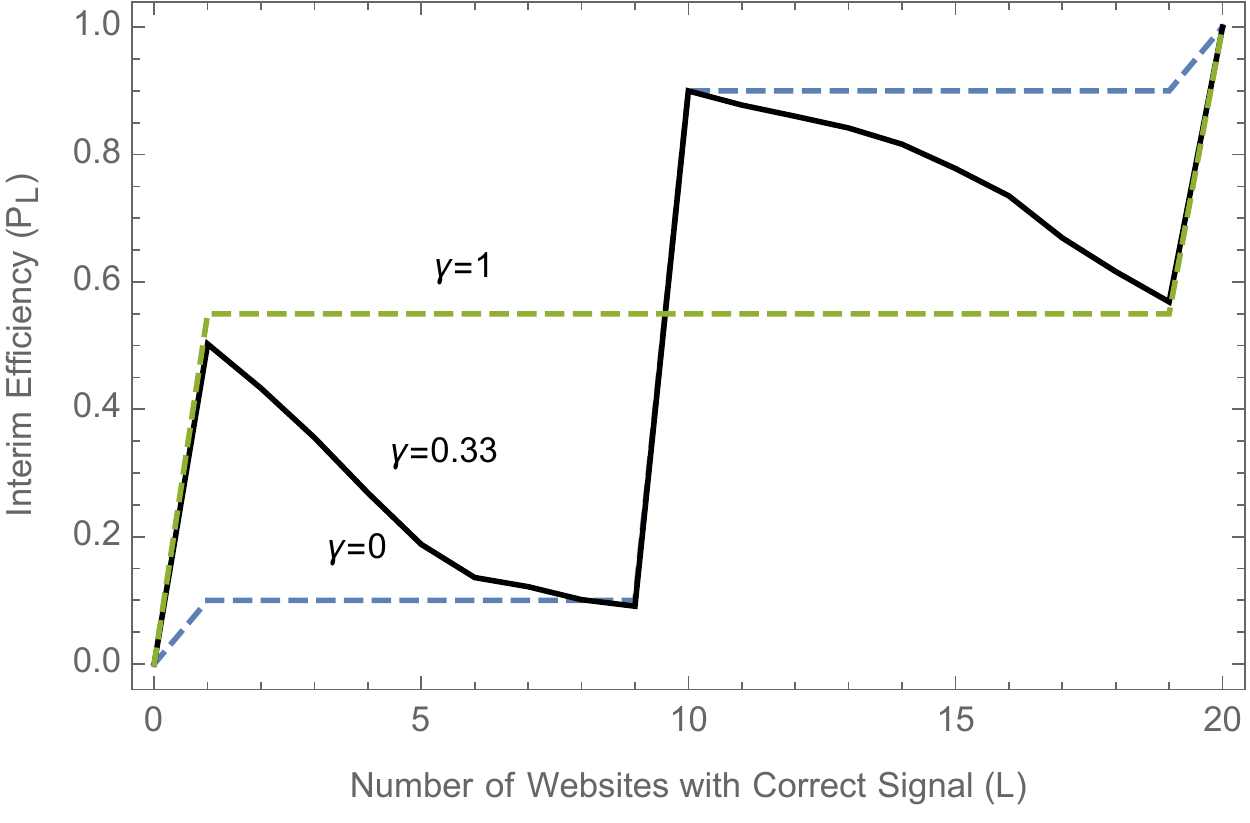} 
\end{center}
\par
\vspace{-15pt}\caption{\small  {\em Interim} efficiency (${\cal P}_L$) as a function of the number of websites with correct signal ($L$) 
for $\gamma=0.33$ (black line) and other values of $\gamma$ (dashed lines) with ordinal ranking and pure clicking realizations. $M=20$, $p=0.55$, $\mu=0.9$, $\beta=1.5$ and $r_1$ has all outlets $L$ with correct signal ranked at the bottom $L$ positions.}
\label{fig:pureranking_L}
\end{figure}
\end{ex}

\bigskip
\noindent 
\begin{ex}
{\bf Example A4. $AOF$ with merging of outlets.} Consider a search environment and consider interim realizations where the number of outlets with the correct signal varies, but where, unlike the case with a fixed total number of outlets $M$ characterizing the basic framework of the paper, the total number of outlets can vary. In particular, we want to allow two outlets with the same signal to merge such that, after they  merge, the total number of outlets decreases by one. To give an idea of what happens in this case, we show the $AOF$ effect in an environment, where there is a fixed number of outlets with the wrong signal, say $J=10$, and where the number of outlets with the correct signal varies from $L=0$ to $L=20$. In this case the total number of outlets varies from $M=10$ to $M=30$. As is illustrated in Figure~\ref{fig:mergedranking_L}. The effect is robust to considering this way of changing the number of outlets with a given signal. (Note that we could also consider the case where $L$ is fixed and the number of outlets with the wrong signal varies.) 
\begin{figure}[H]
\par
\begin{center}
\includegraphics[width=6.5cm]{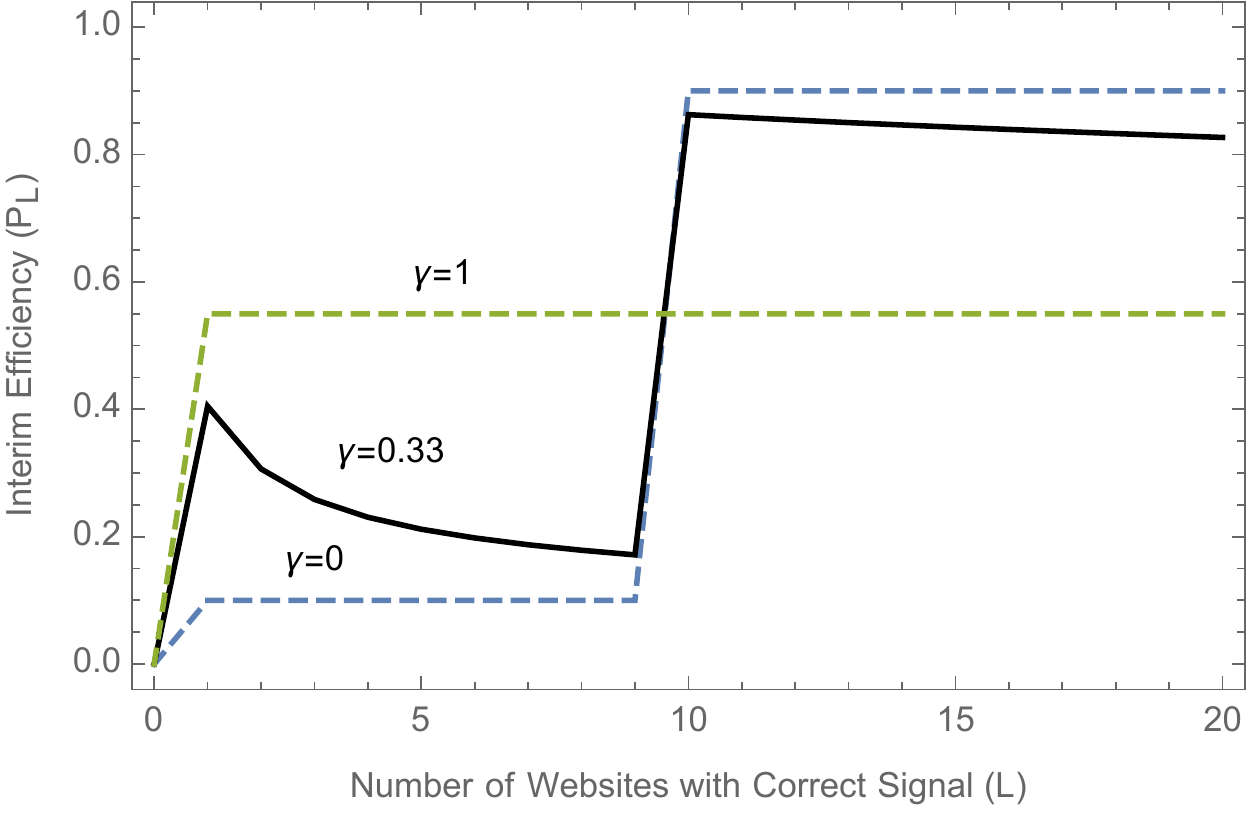}
\end{center}
\par
\caption{\small {\em Interim} efficiency (${\cal P}_L$) as a function of the number of websites with correct signal ($L$) 
for $\gamma=0.33$ (black line) and other values of $\gamma$ (dashed lines) when the number of outlets with incorrect signal is fixed at $J=10$. Also, $p=0.55$, $\mu=0.9$ and $r_1$ has all outlets $L$ with correct signal ranked at the bottom $L$ positions.}
\label{fig:mergedranking_L}
\end{figure}
\end{ex}

\bigskip

\noindent
\begin{ex} {\bf Example A5. $AOF$ and stochastic dynamics.} Consider the search environment with $M=20$, $p=0.55$, $\mu=1$, and $\gamma>0$, and assume that the initial ranking $r_1$ is uniform. Suppose there are three websites reporting a website-minority signal. Two cases may arise. The signal is the correct one ($L=3$) or is the incorrect one ($M-L=M-K=3$). In either case, the three minority websites gain from the fact that there are few websites that report the same signal. As a result they can attract a significant share of traffic even in the case where they report the incorrect signal. The share depends on $\gamma$ and can reach over 50\% total clicking probability for $\gamma$ sufficiently large. More specifically, for small values of $\gamma$ it is close to zero for $L=3$ or $M-L=3$. As $\gamma$ increases, more ranking probability is put on the minority websites, until it reaches a maximum total ranking probability for values of $\gamma$ sufficiently close to 1. 
\begin{figure}[H]
\par
\begin{center}
\includegraphics[width=6.5cm]{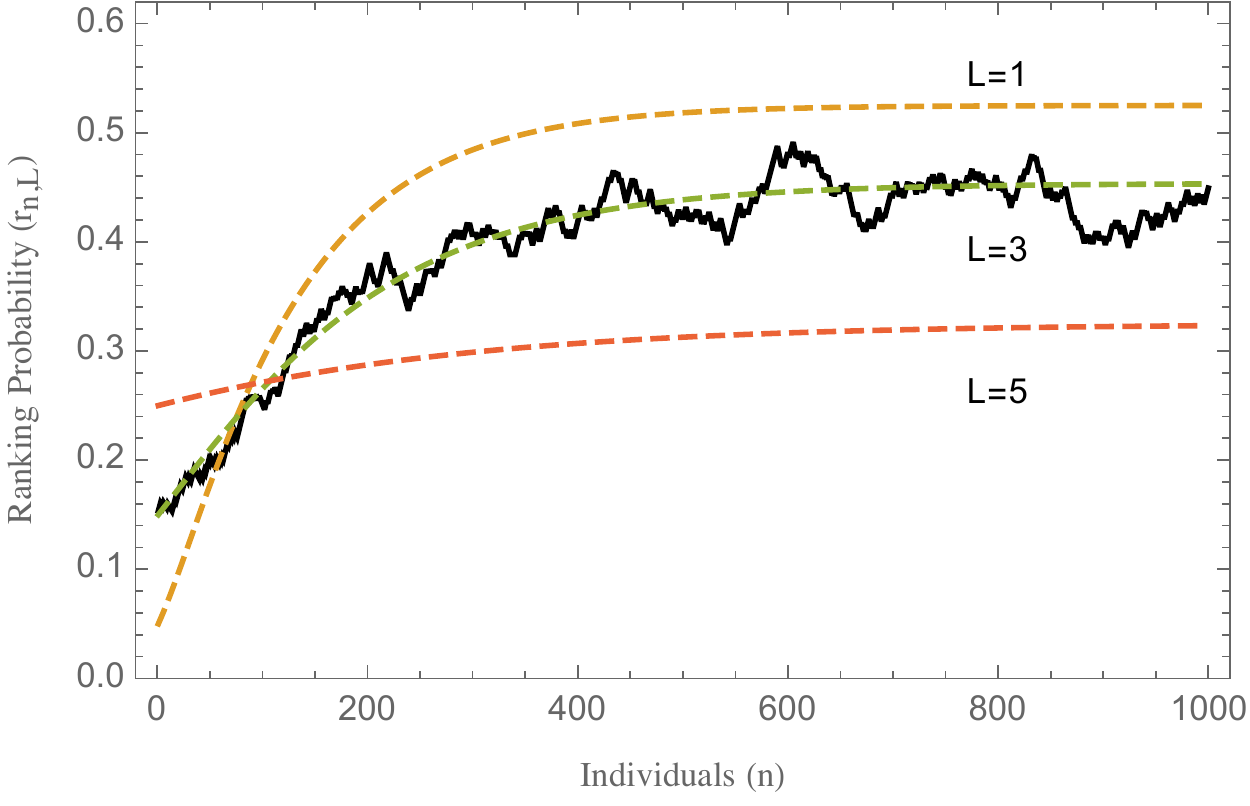}
\hspace{.1in}
\includegraphics[width=6.5cm]{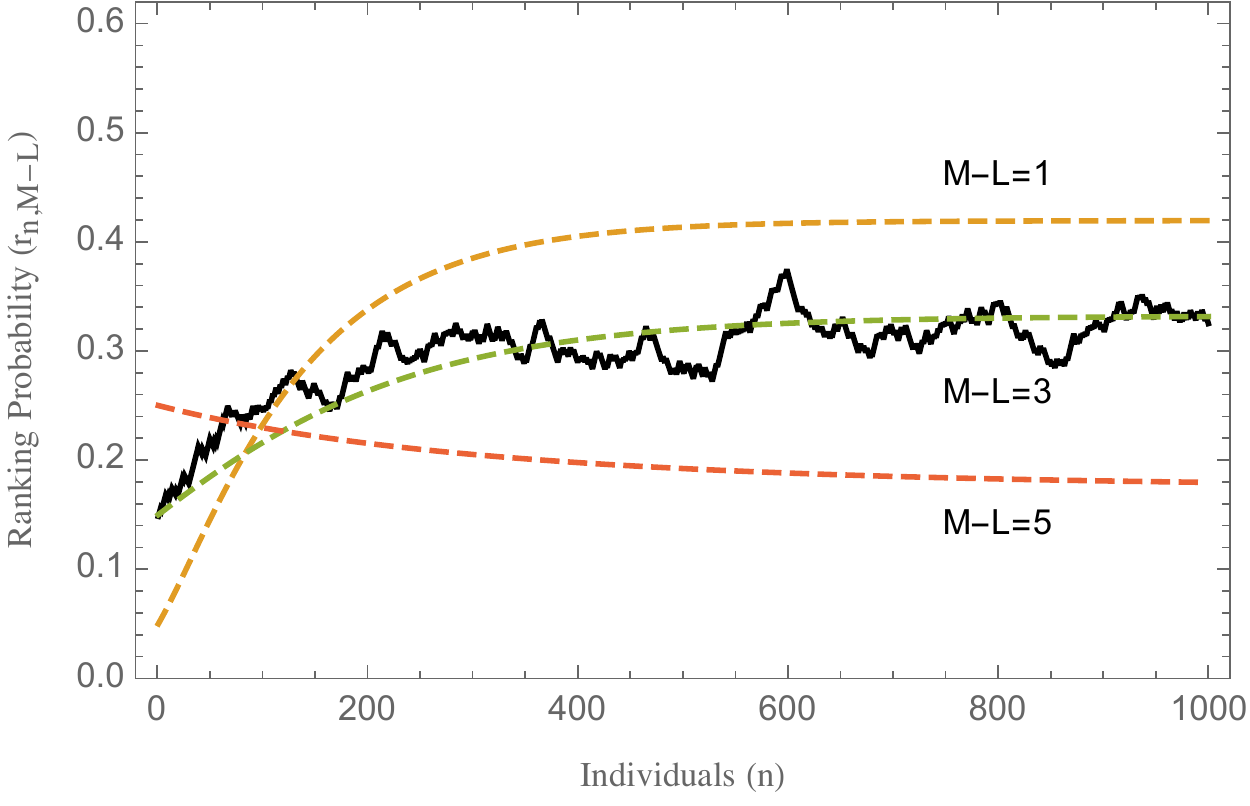}
\end{center}
\par
\vspace{-15pt}\caption{\small  Left panel: Total ranking probability on three minority
websites with correct signal ($L=M-K=3$) (black line); the dashed lines represent the mean dynamics trajectories for $L=1$ (orange), $L=3$ (green) and $L=5$ (red).
Right panel: Total ranking probability on three minority websites with wrong signal
($M-L=M-K=3$) (black line); the dashed lines represent the mean dynamics trajectories for $M-L=1$ (orange), $M-L=3$ (green) and $M-L=5$ (red) (the mean dynamics is discussed in Appendix \ref{Appendix-MeanDynamics}). In both cases $\gamma = 0.5$ and the initial ranking is uniform.}
\label{fig:concentration}
\end{figure}

Decreasing the number of minority websites to one or two ($L=1,2$ or $M-L=1,2$) {\em increases} the total ranking probability of these websites as compared to the case of three. Similarly, increasing the number of minority websites to four or five or more, {\em decreases} the total ranking probability of these websites as compared to the case of three. That is, there is  an \textit{advantage of the fewer}. Figure~\ref{fig:concentration} shows the evolution of the total clicking probability for the cases of $L=3$ (left panel) and $M-L=3$ (right panel). At the same time it shows the expected total ranking probabilities for $L=3$ ($M-L=3$) and also for $L=1$ ($M-L=1$) and for $L=5$ ($M-L=5$). The figure illustrates that decreasing the number of minority websites {\em increases} the total ranking probability of those websites regardless of whether they are reporting the correct signal or not.
\end{ex}

\section*{\large B. Rich get richer}

\setcounter{exa}{0}
\renewcommand{\exa}{}

\setcounter{table}{0}
\renewcommand{\thetable}{B.\arabic{table}}

\setcounter{figure}{0}
\renewcommand{\thefigure}{B.\arabic{figure}}

Consider now a search environment where individuals do not have any preference for like-minded news  ($\gamma=0$) and where there is no personalization ($\lambda=0$). The following numerical example shows how the attention bias ($\alpha$)  and the initial ranking ($r_1$) may create a rich-get-richer dynamic.

\bigskip

\begin{exa}
\label{ex:rich}

 \noindent\textbf{Example B1. Initial ranking, attention bias and
rich get richer.} Consider an environment with $(N,\alpha,\gamma)=(1000,\alpha,\gamma);M=20;(\kappa,\lambda)=(100,0)$.
For now, assume $\gamma=0$, $\mu=1$, and consider an interim
realization with $L=15$ and a uniform initial ranking $r_1$.
Then, agent $n$'s ranking-free website choice is:
\[
v^{*}_{n,m}=\left\{
\begin{array}
[c]{ll}
\frac{1}{15} & \text{ if }y_{m}=y_{K}\\
0 & \text{ else },
\end{array}
\right.
\]
which immediately implies $\rho_{n,m}=v^{*}_{n,m}$, for $n\in N$, $m\in M$. This means that all individuals always only read websites
with the website-majority signal $y_{L}=y_K$,
and access any one of them with the same probability $\frac{1}{15}$ so that, with $\gamma=0$ the website-majority signal prevails. Moreover, since the
initial ranking is uniform, the traffic is evenly distributed across websites
carrying such a signal.

\vspace{.05in} \noindent\emph{Non-uniform initial ranking.} Suppose now that
the initial ranking is given by:
\[
r_{1}\approx(0.06,0.059,\ldots,0.041,0.04),
\]
that is, the websites are ranked with increments of $\frac{1}{950}%
\approx0.001$ starting from 0.04 going up to 0.06. Again, since $\gamma=0$,
individuals always choose one of the fifteen websites reporting the website-majority signal $y_{L}=y_K$. At the
same time, the presence of a non-uniform initial ranking provides an advantage
to the websites with a higher initial ranking. In particular, as illustrated
in Figure~\ref{fig:rich}, increasing the attention bias parameter $\alpha$, 
makes for even larger effects of the websites' initial ranking.

\begin{figure}[H]
\par
\begin{center}
\includegraphics[width=6.5cm]{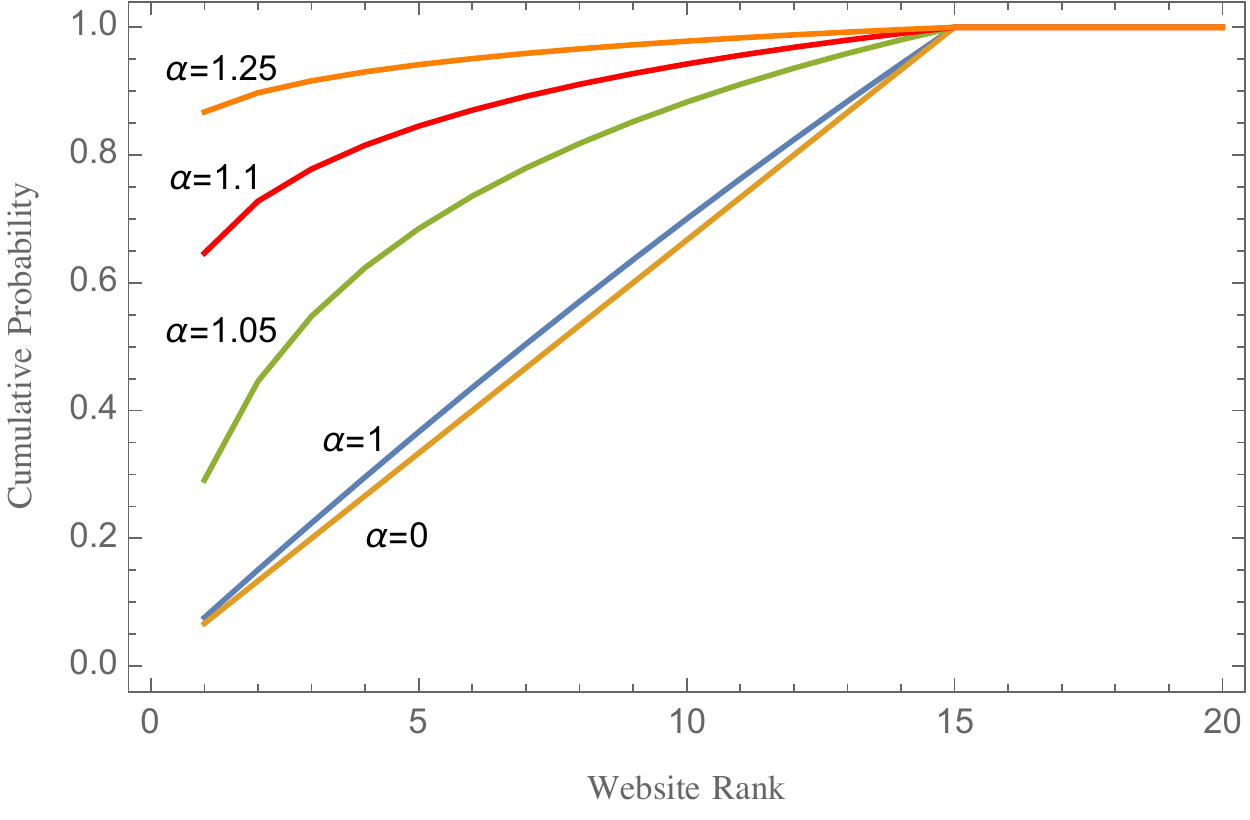}\hspace{.1in}
 \includegraphics[width=6.5cm]{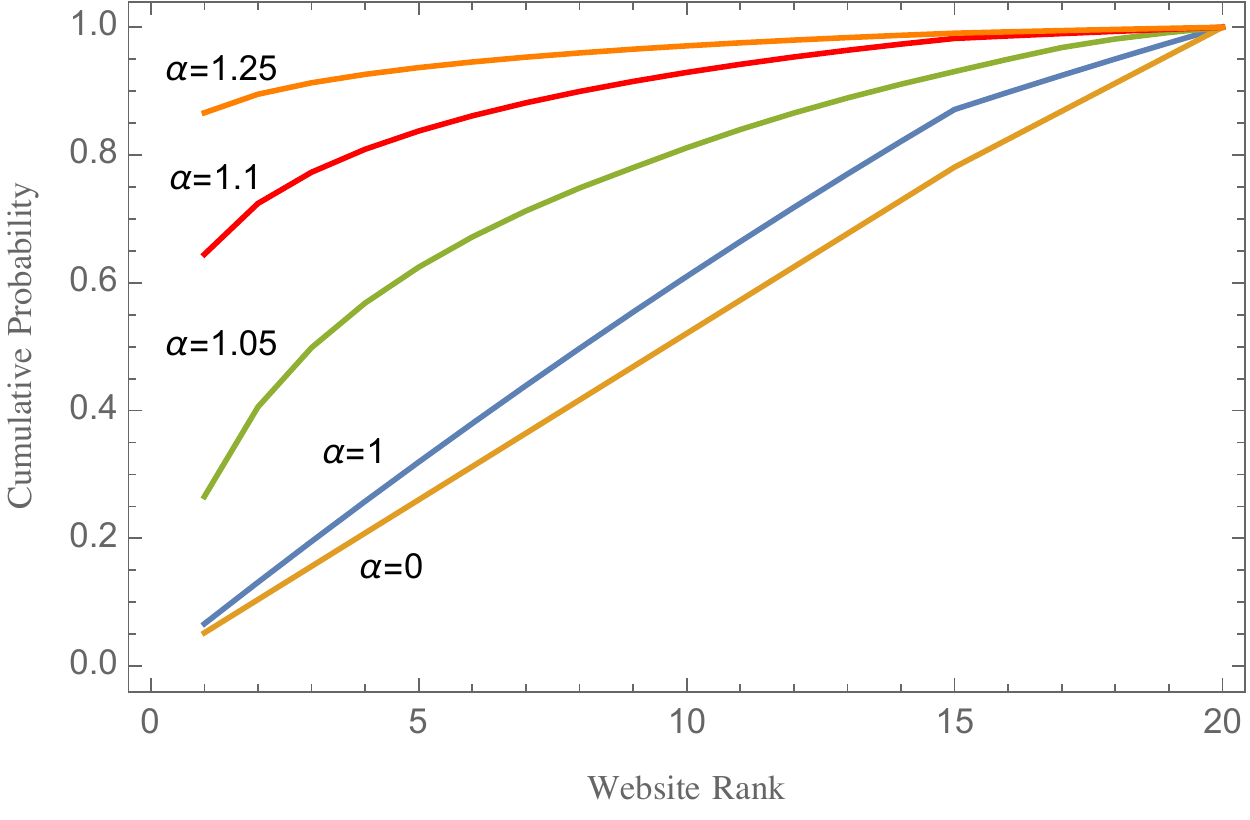}

\end{center}
\par
\vspace{-15pt}\caption{\small  Cumulative probability of clicks per website for attention bias 
$\alpha=0,1,1.05,1.1,1.25$. The graph is drawn for an initial ranking $r_{1}=(0.06,0.059,\ldots,0.041, 0.04)$ and for $\gamma=0$ (left panel) and $\gamma=0.5$ (right panel).} \label{fig:rich}%
\end{figure}

Comparing this pattern with the one arising with a uniform initial ranking, we see that, with a non-uniform initial ranking and
$\alpha > 1$, we obtain a \textit{rich-get-richer} dynamic, whereby the ratio of the expected clicking probabilities of two websites $m,m' \in K$, with the same website-majority signal and with $r_{1,m}>r_{1,m'}>0$, increases over time as more agents perform their search. The effect is further magnified, the larger $\alpha$ is. When $0 \le \alpha \le 1$ the clicking probabilities tend towards uniform ranking over the websites in $K$ (when $\alpha<1$) and a normalization of the initial ranking again over the websites in $K$ (when $\alpha=1$). A similar pattern applies when $\gamma>0$ (right panel). \hfill$\Box$

\end{exa}

This example shows that the evolution of a website's ranking based on its
\textquotedblleft popularity,\textquotedblright\ interacted with a sufficiently large
attention bias ($\alpha > 1$) of the individuals exhibits a \textit{rich-get-richer} dynamic, which in turn implies a tendency towards concentration of website traffic at the top. 
Importantly, the differences in ranking and, then, in the {\em expected} probability of accessing a given website, are driven by the initial ranking ($r_{1}$) and are amplified by the attention bias. 



\end{document}